\documentclass[journal]{IEEEtran}
\usepackage[cmex10]{amsmath}
\usepackage{eqparbox}
\usepackage[usenames, dvipsnames]{color}
\usepackage{cite}
\usepackage{mathtools,amssymb,bm,mathabx}
\usepackage{amstext}
\usepackage{amssymb}
\usepackage{graphicx}
\usepackage{color}
\usepackage{booktabs}
\usepackage{longtable}
\usepackage{multicol}
\usepackage{amsfonts}
\usepackage{dsfont}
\usepackage{array}
\usepackage{algorithmicx}
\usepackage[ruled]{algorithm}
\usepackage{algpseudocode}
\usepackage{algpascal}
\usepackage{algc}
\usepackage{cases}
\usepackage{pgfplots}
\usepackage{accents}
\usepackage{caption}
\usepackage[footnotesize]{subfigure}
\usepackage{amsthm,xparse}
\usepackage{float}
\usepackage{cite}
\usepackage [autostyle, english = american]{csquotes}
\usepackage{hyperref}
\theoremstyle{remark}

\newcommand\ASTART{\bigskip\noindent\begin{minipage}[b]{0.5\linewidth}}
	
	\newcommand\AENDSKIP{\end{minipage}\bigskip}
\newcommand\AEND{\end{minipage}}
\ifCLASSOPTIONcaptionsoff
\usepackage[nomarkers]{endfloat}
\fi
\theoremstyle{plain}
\newtheorem{thm}{\textbf{Theorem}}
\newtheorem{lem}{\textbf{Lemma}}
\newtheorem{prop}{\textbf{Proposition}}
\newtheorem{corl}{\textbf{Corollary}}

\theoremstyle{definition}
\newtheorem{defn}{\textbf{Definition}}

\newtheorem{fact}{\textbf{Fact}}

\theoremstyle{remark}
\newtheorem{rem}{\bf Remark}

\newcommand*{\rom}[1]{\expandafter\@slowromancap\romannumeral #1@}

\newcommand{\RN}[1]{%
	\textup{\uppercase\expandafter{\romannumeral#1}}%
}

\usepackage{standalone}
\graphicspath{{./Figures/}}  
\begin{document}
%
\title{Optimal Weighted Low-rank Matrix Recovery with Subspace Prior Information}
\author{Sajad~Daei, Arash~Amini, Farzan~Haddadi
}%

\maketitle

\begin{abstract}
Matrix sensing is the problem of reconstructing a low-rank matrix from a few linear measurements. In many applications such as collaborative filtering, the famous Netflix prize problem, and seismic data interpolation, there exists some prior information about the column and row spaces of the ground-truth low-rank matrix. In this paper, we exploit this prior information by proposing a weighted optimization problem where its objective function promotes both rank and prior subspace information. Using the recent results in conic integral geometry, we obtain the unique optimal weights that minimize the required number of measurements. As simulation results confirm, the proposed convex program with optimal weights requires substantially fewer measurements than the regular nuclear norm minimization.
\end{abstract}

\begin{IEEEkeywords}
Conic integral geometry, Matrix sensing, Subspace prior information, .
\end{IEEEkeywords}

%
\IEEEpeerreviewmaketitle

\section{Introduction}
\IEEEPARstart{L}{ow} rank matrix recovery (also known as matrix sensing) has appeared in numerous applications in recent years. For example, Netflix prize problem\cite{netflix,srebro2004learning}, collaborative filtering\cite{srebro2010collaborative}, seismic data interpolation\cite{oropeza2011simultaneous,aravkin2014fast}, system identification \cite{fazel2003log}, and sensor network localization\cite{so2007theory}. Mathematically, our goal is to recover a low-rank matrix $\bm{X}\in\mathbb{R}^{n_1\times n_2}$ with rank $r\ll \min\{n_1,n_2\}$ from a few linear measurements of the form $\bm{y}=\mathcal{A}(\bm{X})$, where $\mathcal{A}:\mathbb{R}^{n_1\times n_2}\mapsto\mathbb{R}^m$  is a linear operator. An idealistic approach is the following optimization problem:
\begin{align}{\label{problem.rank}}
&\min_{\bm{Z}\in\mathbb{R}^{n_1\times n_2}}{\rm rank}(\bm{Z})\nonumber\\
&\mathrm{s.t.}~\bm{y}=\mathcal{A}(\bm{Z}).
\end{align}
However, this problem is NP-hard and computationally intractable. A common alternative is to relax the objective function into the closest convex function. In fact, since rank is the number of nonzero elements of the singular value vector, its convex relaxation amounts to $\ell_1$ norm of this vector known as the nuclear norm of the matrix. Then, one may solve the following convex problem:
\begin{align}{\label{problem.nuclearnorm}}
\mathsf{P}_{\rm nuc}:~~&\min_{\bm{Z}\in\mathbb{R}^{n_1\times n_2}}\|\bm{Z}\|_*\nonumber\\
&\mathrm{s.t.}~\bm{y}=\mathcal{A}(\bm{Z}),
\end{align} 
where $\|\cdot\|_*$ computes the sum of singular values. A special case of matrix sensing known as matrix completion is to complete $\bm{X}$ from a few observed entries:
\begin{align}\label{problem.matrixcomple}
&\min_{\bm{Z}\in\mathbb{R}^{n_1\times n_2}}\|\bm{Z}\|_*\nonumber\\
&\mathrm{s.t.}~\bm{y}=\mathcal{R}_{\Omega}(\bm{Z}),
\end{align}
where $\mathcal{R}_{\Omega}:=(X_{ij})_{(i,j)\in\Omega}$ is the sampling operator that extracts the observed entries $\Omega$ of $\bm{X}$. Let us denote the column and row spaces of $\bm{X}\in\mathbb{R}^{n_1\times n_2}$ by $\mathrm{span}(\bm{X})$ and $\mathrm{span}(\bm{X}^H)$, respectively. By solving (\ref{problem.nuclearnorm}) or (\ref{problem.matrixcomple}), with high probability, one can successfully recover $\bm{X}$ by observing $\mathcal{O}(r\max\{n_1,n_2\}\log^2(n_1+n_2))$ entries\footnote{For the sake of simplicity, we investigate only square matrices in this work. The extension to non-square matrices is straightforward.}\cite{chen2015completing}. The main challenge in recovering $\bm{X}\in\mathbb{R}^{n\times n}$ in the problems (\ref{problem.nuclearnorm}) and (\ref{problem.matrixcomple}) is to identify column and row spaces of $\bm{X}$. If they are known, one can recover $\bm{X}$ from at most $r^2$ linear measurements of the form $\bm{U}_{n\times r}^{\rm H}C_{r\times r}\bm{V}_{n\times r}$ where $\bm{U}\in\mathbb{R}^{n\times r}$ and $\bm{V}\in\mathbb{R}^{n\times r}$ are orthonormal bases of $\bm{\mathcal{U}}:=\mathrm{span}(\bm{U})=\mathrm{span}(\bm{X})$ and $\bm{\mathcal{V}}:=\mathrm{span}(\bm{V})=\mathrm{span}(\bm{X}^H)$, respectively\footnote{In this paper, we occasionally use $\bm{U}$ instead of $\bm{U}_{n\times r}$ to avoid complexity.}. In this work, we consider the matrix sensing problem when prior information about the row and column spaces of $\bm{X}\in\mathbb{R}^{n\times n}$ is available.
To be precise, consider two $r$-dimensional subspaces $\widetilde{\bm{\mathcal{U}}}$ and $\widetilde{\bm{\mathcal{V}}}$ withe known principle angles\footnote{See Section \ref{section.pricipal_angles} for definition.} with the column and row subspaces of $\bm{X}$, i.e. ${\bm{\mathcal{U}}}$ and ${\bm{\mathcal{V}}}$, respectively. Intuitively, if $\widetilde{\bm{\mathcal{U}}}\approx{\bm{\mathcal{U}}}$ and $\widetilde{\bm{\mathcal{V}}}\approx{\bm{\mathcal{V}}}$ it seems that one can recover $\bm{X}$ using less measurements compared to the prior-less case. Interestingly, this case happens in many applications of interest. For example, in recommender systems, similar users share similar attributes and knowing how a particular user rates a particular item, provides some prior subspace information about the row and column spaces of the ground-truth matrix. 
\subsection{Motivations and Conjectures} 
The problem (\ref{problem.nuclearnorm}) is connected with a large body of literature known as \textit{compressed sensing} (CS) pioneered by the works in \cite{candes2005decoding} and \cite{donoho2006most}. In the same way that $\ell_1$ minimization seeks for the sparsest solution in vectors, $\mathsf{P}_{\rm nuc}$ aims at recovering the minimum rank solution under a suitable incoherent sensing operator. Naturally, there exists a parallel between CS and matrix sensing. In fact, $\ell_1$ minimization is a special case of the nuclear norm minimization in which $\bm{X}\in\mathbb{R}^{n\times n}$ is diagonal. A general question is whether the parallels between compressed sensing and matrix sensing always hold? Let us consider a relevant example. It is known that prior information about the support (non-zero locations) of a vector can be incorporated into $\ell_1$ minimization by assigning larger weights to the off-support locations than support locations\footnote{In fact, inaccurate locations are penalized more.}\cite{candes2008enhancing}, leading to a reduction in the required number of measurements. Now, let us go back to the matrix world. Consider a matrix $\bm{X}$ that lives in a union of row and column subspaces denoted by $T$. Suppose that we are given a subspace $\widetilde{T}$ that is slightly mis-aligned with $T$. Can we hope for a reduction in the required number of measurements by penalizing the orthogonal complement of $\widetilde{T}$? Are the parallels still strong?
\subsection{Notation}
Throughout the paper, scalars are denoted by lowercase letters, vectors by lowercase boldface letters, and matrices by uppercase boldface letters. The $i$th element of a vector $\bm{x}$ is shown either by ${x}(i)$ or $x_i$. $(\cdot)^\dagger$ denotes the pseudo-inverse operator. $\bm{I}_{n}$ is the identity matrix of size $n\times n$. The complement of an event $\mathcal{E}$ is shown by $\mathcal{\overline{E}}$. The nullspace of linear operators is denoted by $\mathrm{null}(\cdot)$. For a matrix $\bm{A}$, the operator norm is defined as $\|\bm{A}\|_{p\rightarrow q}=\underset{~~\|\bm{x}\|_p\le1}{\sup}\|\bm{Ax}\|_q$. The unit ball and unit sphere are shown by $\mathbb{B}^n=\{\bm{x}\in \mathbb{R}^n:~\|\bm{x}\|_2\le 1\}$ and $\mathbb{S}^{n-1}=\{\bm{x}\in \mathbb{R}^n:~\|\bm{x}\|_2=1\}$, respectively. Also, we have $\mathbb{B}_{\epsilon}^{n\times n}:=\{\bm{Z}\in\mathbb{R}^{n\times n}:~\|\bm{Z}\|_F\le \epsilon\}$ which refers to the $\epsilon$-ball of matrices according to the Frobenius norm. Consider a matrix $\bm{X}\in\mathbb{R}^{n\times n}$ with reduced SVD form $\bm{X}=\bm{U}_{n\times r}\bm{\Sigma}_{r\times r}\bm{V}_{n\times r}^{\rm H}$. Define $\bm{\mathcal{U}}:={\rm span}(\bm{U})$ and $\bm{\mathcal{V}}:={\rm span}(\bm{V})$. We denote the matrix $\bm{U}\bm{V}^{\rm H}$ by the notation $\mathrm{sgn}(\bm{X})$. Also define the \textit{support} of $\bm{X}$ by the linear subspace
\begin{align}
&T=\{\bm{Z}\in\mathbb{R}^n:~\bm{Z}=\bm{P}_{\bm{\mathcal{U}}}\bm{Z}\bm{P}_{\bm{\mathcal{V}}}+\bm{P}_{\bm{\mathcal{U}}}\bm{Z}\bm{P}_{\bm{\mathcal{V}}^\perp}+\bm{P}_{\bm{\mathcal{U}}^\perp}\bm{Z}\bm{P}_{\bm{\mathcal{V}}}\}\nonumber\\
&:=\mathrm{supp}(\bm{X}),
\end{align}
where $\bm{P}_{\bm{\mathcal{U}}}:=\bm{U}\bm{U}^{\rm H}$ and $\bm{P}_{\bm{\mathcal{V}}}:=\bm{V}\bm{V}^{\rm H}$ are unique orthogonal projections onto $\bm{\mathcal{U}}$ and $\bm{\mathcal{V}}$, respectively. $\mathcal{P}_T(\bm{Z})$ and $\mathcal{P}_{T^\perp}(\bm{Z})$ are the projection of matrix $\bm{Z}$ onto the linear subspace $T$ and $T^\perp$, respectively, and are defined as
\begin{align*}
&\mathcal{P}_{T}(\bm{Z}):=\bm{P}_{\bm{\mathcal{U}}}\bm{Z}\bm{P}_{\bm{\mathcal{V}}}+\bm{P}_{\bm{\mathcal{U}}}\bm{Z}\bm{P}_{\bm{\mathcal{V}}^\perp}+\bm{P}_{\bm{\mathcal{U}}^\perp}\bm{Z}\bm{P}_{\bm{\mathcal{V}}},\\
&\mathcal{P}_{T^\perp}(\bm{Z}):=\bm{P}_{\bm{\mathcal{U}}^\perp}\bm{Z}\bm{P}_{\bm{\mathcal{V}}^\perp}.
\end{align*}
We represent the projection onto a cone $\mathcal{C}$ with a same notation; namely,
\begin{align}
\mathcal{P}_{\mathcal{C}}(\bm{X}):=\arg\min_{\bm{Z}\in\mathcal{C}} \|\bm{Z}-\bm{X}\|_F.
\end{align}
The polar of a cone $\mathcal{C}$ is defined as $\mathcal{C}^\circ=\{\bm{z}:\langle \bm{z},\bm{x} \rangle\le 0~\forall \bm{x}\in\mathcal{C}\}$. $\bm{x}\in[a,b]^n$ for a vector $\bm{x}\in\mathbb{R}^n$ means that $a\le x_i\le b, i=1,..., n$. Also, by $\bm{x}\in(a,b]^n$, we mean $a< x_i\le b, i=1,..., n$. $\mathrm{diag}(\bm{x})$ is a diagonal matrix in which the main diagonal is determined by the elements of $\bm{x}$. For a function $f:\mathbb{R}^{n\times n}\rightarrow \mathbb{R}$, $f^*$ stands for the adjoint of the function $f$. $\bm{\sigma}(\bm{A})\in\mathbb{R}^n$ denotes the singular values of $\bm{A}$ sorted non-increasingly. $(a)_+$, $a\vee b$ and $a\wedge b$ denote $\max\{a,0\}$, $\max\{a,b\}$ and $\min\{a,b\}$, respectively. $\langle\bm{A},\bm{B} \rangle_F=tr(\bm{A}\bm{B}^{\rm H})$ denotes the Frobenius inner product of two matrices $\bm{A}$ and $\bm{B}$.
\subsection{Contributions}
In this work, we propose a new approach for exploiting the prior subspace information leading to a considerable reduction in the required number of measurements. Consider a rank $r$ matrix $\bm{X}\in\mathbb{R}^{n\times n}$ with column and row subspaces $\bm{\mathcal{U}}$ and $\bm{\mathcal{V}}$. Assume that we are given two subspaces $\widetilde{\bm{\mathcal{U}}}$ and $\widetilde{\bm{\mathcal{V}}}$, each with dimension $r'\ge r$, that have known angles from $\bm{\mathcal{U}}$ and $\bm{\mathcal{V}}$. Let $\bm{\theta}_u\in [0,90^\circ]^r$ and $\bm{\theta}_v\in[0,90^\circ]^r$\footnote{Throughout, we will occasionally exclude the symbol $\circ$ when referring to angle degree for the sake of simplicity.} represent the principle angles that $\bm{\mathcal{U}}$ and $\bm{\mathcal{V}}$ form with $\widetilde{\bm{\mathcal{U}}}$ and $\widetilde{\bm{\mathcal{V}}}$, respectively. We implicitly take these prior subspace information into account by proposing the following optimization problem.
	\begin{align}\label{eq.model1}
	&\mathsf{P}_{\bm{w},\mathrm{nuc}}: \min_{\bm{Z}\in\mathbb{R}^{n_1\times n_2}}\|h_{\bm{w}}(\bm{Z})\|_*\nonumber\\
	&~~\mathrm{s.t.}~~\bm{y}=\mathcal{A}(\bm{Z}),
	\end{align}
where,\begin{align}\label{eq.hZ}
&h_{\bm{w}}(\bm{Z})=w_1\bm{P}_{\widetilde{\bm{\mathcal{U}}}}\bm{Z}\bm{P}_{\widetilde{\bm{\mathcal{V}}}}+w_2\bm{P}_{\widetilde{\bm{\mathcal{U}}}}\bm{Z}\bm{P}_{\widetilde{\bm{\mathcal{V}}}^\perp}+w_3\bm{P}_{\widetilde{\bm{\mathcal{U}}}^\perp}\bm{Z}\bm{P}_{\widetilde{\bm{\mathcal{V}}}}\nonumber\\
&+w_4\bm{P}_{\widetilde{\bm{\mathcal{U}}}^\perp}\bm{Z}\bm{P}_{\widetilde{\bm{\mathcal{V}}}^\perp}=\nonumber\\
&\frac{1}{w_3}\Big(w_1\bm{P}_{\widetilde{\bm{\mathcal{U}}}}+w_3\bm{P}_{\widetilde{\bm{\mathcal{U}}}^\perp}\Big)\bm{Z}\Big(w_3\bm{P}_{\widetilde{\bm{\mathcal{V}}}}+w_4\bm{P}_{\widetilde{\bm{\mathcal{V}}}^\perp}\Big),\nonumber\\
&\bm{w}:=[w_1,w_2,w_3]^T,~~w_4:=\frac{w_2w_3}{w_1}.
\end{align}
The weights $w_1$ and $w_3$ reflect the uncertainty in the prior column space information. The same argument holds for $w_3$ and $w_4$ in the prior row space information. In this work, we obtain the unique weights that minimize the required number of measurements. These weights are \textit{optimal} since they minimize the number of measurements that $\mathsf{P}_{\bm{w},\mathrm{nuc}}$ needs for exact recovery of $\bm{X}$. To find optimal weights, we exploit the concept of \textit{statistical dimension} in conic integral geometry. The statistical dimension specifies the boundary of success and failure of $\mathsf{P}_{\bm{w},\mathrm{nuc}}$. To be precise, we obtain upper and lower bounds with asymptotically vanishing distances for the statistical dimension of a certain convex cone and thereby calculate a threshold $m_0(\bm{w},\bm{\theta}_u,\bm{\theta}_v)$ for the minimum required number of measurements. Then, we solve the optimization problem
\begin{align}
&\min_{\bm{w}\in\mathbb{R}_+^3} m_0(\bm{w},\bm{\theta}_u,\bm{\theta}_v),
\end{align} 
to reach the optimal weight vector $\bm{w}^*$. To better highlight our contributions, we summarize the novelties below.
\begin{enumerate}
	\item \textit{Proposing a new optimization model for matrix sensing}: We propose a new convex optimization problem in (\ref{eq.model1}) that promotes both rank and subspace information. A benefit of this model is that by suitably tuning the weights, it consistently outperforms $\mathsf{P}_{\rm nuc}$ even when the accuracy of subspace prior information is unreliable. When $\widetilde{\bm{\mathcal{U}}}\approx \bm{\mathcal{U}}$, the prior information is reliable and less penalty is assigned to $w_1$ than $w_3$. The same argument applies to $\bm{\mathcal{V}}$. If the subspace prior information is at the boundary of reliability (i.e. $\theta_u(i)=\theta_v(i)=45^\circ~\forall i=1,..., r$), then by setting $w_1=w_2=w_3$, $\mathsf{P}_{\bm{w},\rm nuc}$ reduces to $\mathsf{P}_{\rm nuc}$.
	\item \textit{Obtaining an upper-bound for the required sample complexity of $\mathsf{P}_{\rm {nuc},\bm{w}}$}: We obtain a closed-form relation for the sufficient number of measurements that $\mathsf{P}_{\rm nuc, \bm{w}}$ needs for successful recovery (denoted by $\widehat{m}_{\bm{w},\bm{\theta}_u,\bm{\theta}_v}$). This bound depends on the weights $\bm{w}$ and the principal angles $\bm{\theta}_u, \bm{\theta}_v$. By setting $w_1=w_2=w_3=1$, the bound simplifies to the required sample complexity of $\mathsf{P}_{\rm nuc}$.
	\item \textit{Obtaining an error estimate bound for $\widehat{m}_{\bm{w},\bm{\theta}_u,\bm{\theta}_v}$}: We prove that the sufficient number of measurements (i.e. $\widehat{m}_{\bm{w},\bm{\theta}_u,\bm{\theta}_v}$) is also necessary for successful recovery. To be more precise, we show that $\widehat{m}_{\bm{w},\bm{\theta}_u,\bm{\theta}_v}$ differs from the minimum required number of measurements up to an asymptotically vanishing term.
	\item \textit{Proposing a new strategy for finding optimal weights}:
	In the proposed model (\ref{eq.model1}), we obtain the weights $\bm{w}$ that minimize the required sample complexity for exact recovery. If one takes the sample complexity as the optimality criterion, then, these weights are optimal. Also, we show that, they are unique up to a positive scaling. We further propose a simple algorithm (called Optweights) that efficiently computes the unique optimal weights.
	\item \textit{Obtaining closed-form expressions for ${\rm supp}(h_{\bm{w}}(\bm{X}))$ and ${\rm sgn}(h_{\bm{w}}(\bm{X}))$}: We find that the spaces and sign of $h_{\bm{w}}(\bm{X})$ (i.e. ${\rm supp}(h_{\bm{w}}(\bm{X}))$ and ${\rm sgn}(h_{\bm{w}}(\bm{X}))$, respectively) are rotated versions of $\bm{I}_r$ spaces. More precisely,
	\begin{align}
	&\mathcal{P}_{\widehat{T}}(\bm{Z})=\bm{Q}_L\mathcal{P}_{{\rm supp}(\bm{I}_r)}(\bm{Q}_L^H\bm{Z}\bm{Q}_R)\bm{Q}_R^H,\nonumber\\
	&\mathcal{P}_{\widehat{T}^\perp}(\bm{Z})=\bm{Q}_L\mathcal{P}_{{\rm supp}(\bm{I}_r)^\perp}(\bm{Q}_L^H\bm{Z}\bm{Q}_R)\bm{Q}_R^H,\nonumber\\
	&{\rm sgn}(h_{\bm{w}}(\bm{X}))=\bm{Q}_L\mathcal{P}_{{\rm supp}(\bm{I}_r)}(\bm{I}_n)\bm{Q}_R^{\rm H},\nonumber
	\end{align}
	for any arbitrary $\bm{Z}\in\mathbb{R}^{n\times n}$, where $\bm{Q}_L\in\mathbb{R}^{n\times n}$ and $\bm{Q}_R\in \mathbb{R}^{n\times n}$ are some orthonormal bases of $\mathbb{R}^{n\times n}$, which explicitly depend on the weights $\bm{w}$ and the principal angles $\bm{\theta}_u$,  $\bm{\theta}_v$.
	\item \textit{Obtaining the limiting behavior of spectral functions}: For any non-increasingly ordered vector $f\in\mathbb{R}_{+}^{n_1}$ and Gaussian ensemble $\bm{G}\in \mathbb{R}^{n_1\times n_2}$ with $n_1\le n_2$, we obtain a closed-form relation for the limiting behavior of $$\mathds{E}\frac{1}{n_1}\sum_{i=1}^{n_1} (\sigma_i(\frac{\bm{G}}{\sqrt{n_2}})-f_i)_{+}^2.$$
\end{enumerate}
\subsection{Intuition}\label{subsec.intuition}
The optimal weights in $\mathsf{P}_{\bm{w},\mathrm{nuc}}$ depend on the orientation of the subspace $\widetilde{\bm{\mathcal{U}}}$ with respect to  $\bm{\mathcal{U}}$. Before we provide our analytical results, we intuitively describe the behavior of the weights in special cases of the relative orientation.
\begin{itemize}
\item When the principle angles $\{\theta_u(i)\}_{i=1}^{r}$ between $\bm{\mathcal{U}}$ and $\widetilde{\bm{\mathcal{U}}}$ with dimensions $r$ and $r'\ge r$, respectively, are all small (close to $0^\circ$), $\widetilde{\bm{\mathcal{U}}}$ provides a good estimate of $\bm{\mathcal{U}}$ and we expect $w_1$ to be small (penalization weight for $\widetilde{\bm{\mathcal{U}}}$). In contrast, $\widetilde{\bm{\mathcal{U}}}^{\perp}$ is a poor estimation of $\bm{\mathcal{U}}$ which should be significantly penalized (large $w_3$). Further, if $r'\approx r$, it is expected that the required number of measurements for $\mathsf{P}_{\bm{w},\mathrm{nuc}}$ approaches the optimal value $r^2$.
\item When the principle angles $\{\theta_u(i)\}_{i=1}^{r}$ are all large (close to $90^\circ$), $\widetilde{\bm{\mathcal{U}}}^{\perp}$ provides a fair estimate of $\bm{\mathcal{U}}$; therefore $w_1$ is expected to be large, while $w_3$ shall be small. Again, we predict a reduced number of required measurements for $\mathsf{P}_{\bm{w},\mathrm{nuc}}$.
\item When $\{\theta_u(i)\}_{i=1}^{r}$ are all around $45^\circ$, $\widetilde{\bm{\mathcal{U}}}$ and $\widetilde{\bm{\mathcal{U}}}^{\perp}$ are the same in terms of similarity to $\bm{\mathcal{U}}$. This means that the available data does not add any useful information for the recovery. Alternatively, all the weights become equal and the weighted problem $\mathsf{P}_{\bm{w},\mathrm{nuc}}$ simplifies to the standard problem  $\mathsf{P}_{\mathrm{nuc}}$. Hence, we expect the same number of measurements.
\item When the angles $\{\theta_u(i)\}_{i=1}^{r}$ are evenly distributed around $45^\circ$ (for instance, $\bm{\theta}_u=[25^\circ,45^\circ,75^\circ]^T$ ), we expect a similar case as if all the angles were $45^\circ$. This is partly because of the fact that both $\widetilde{\bm{\mathcal{U}}}^{\perp}$ and $\widetilde{\bm{\mathcal{U}}}$ will have the same set of principal angles with $\bm{\mathcal{U}}$, and partly because we penalize all the subspace with a single weight. In other words, the directions in $\widetilde{\bm{\mathcal{U}}}$ are on average unrelated to $\bm{\mathcal{U}}$, while some specific directions might be close to $\bm{\mathcal{U}}$. 
\end{itemize}
Similar statements also hold for row space prior information $\widetilde{\bm{\mathcal{V}}}$ and $\widetilde{\bm{\mathcal{V}}}^\perp$.

\subsection{Applications}\label{section.applications}
The application of subspace prior information in matrix sensing is very broad (seismic data interpolation \cite{aravkin2014fast}, FDD\footnote{Frequency division duplexing.} massive MIMO\footnote{Multiple input multiple output.} \cite{FDD}, Dynamic sensor network localization\cite{so2007theory}, collaborative filtering \cite{srebro2010collaborative}, Netflix problem \cite{netflix} and subspace tracking); we list some of them below.

\begin{itemize}
	\item \textit{The Netflix problem}\cite{netflix}. 
	Let us consider a large pool of movies which are seen or could potentially be seen by a pool of users. The Netflix matrix is formed by the rating of the users to the movies; the rows correspond to the movies and the columns to the users. The element in the $i$th row and $j$th column represents  the score that user $j$ gives to the movie $i$. However, many of such scores are unavailable, as not all users have seen all the films. The challenge is to estimate the unavailable scores based on the known values. It is well-studied that the Netflix matrix can be fairly approximated to be low-rank; therefore, matrix completion techniques based on relaxing the rank constraint are popular for solving the problem. In some cases, nevertheless, we might have prior information about the Netflix matrix that could improve the performance of the completion task. For instance, we might know in advance that the scores of a certain user is not much affected by the music of the movie, while the special effects significantly influences his/her scores. Another case of prior information happens in film festivals where the movies are first evaluated by professionals and critics before ordinary users. In both cases, the prior information can be translated into the angles between the columns of the Netflix matrix and some known subspaces (e.g., the subspace generated by the average score of the professionals).
	

	\item \textit{Subspace tracking}. In many setups such as in radars, it is important to estimate the subspace of the signal (e.g., to denoise the signal). Nevertheless, due to the dynamics of the system, this subspace is constantly evolving. In subspace tracking one aims at updating the estimate for the subspace based on the previous estimates and some measurements related to the recent state of the subspace. In other words, the partial similarity of the current signal subspace to its previous states is used as a key to reduce the number of required measurements.

	
	\item \textit{Dynamic sensor network localization\cite{so2007theory}}. Consider a moving network of low-power sensors scattered in an area (e.g., in the sea). The goal is to locate various objects in this area based on the observed distances to some neighboring sensors. For this purpose, the relative position of the sensors should be determined first. However, each sensor can measure its distance to only nearby sensors. It is well-known that the  matrix formed by the pairwise squared distances of the sensors is low-rank; however, only some of the elements of this matrix are measurable and the matrix is also dynamically changing. Again at each time instance, the similarity of the distance matrix to its previous versions can be employed to enhance the quality of its estimation.
	
	\item \textit{Time-varying channel estimation in FDD massive MIMO communication\cite{FDD}}. Let us imagine a multi-user wireless communication system in which the users with single-antenna transmitters are communicating with a multi-antenna base-station. The users are generally moving which makes the communication channels time-varying. It is known that due to the correlated nature of the user channels, the matrix constructed by the channel impulse responses (channel matrix) is low-rank \cite{FDD}. Besides, the physical movement of the users compared to the communication rate is rather slow; hence, the channel matrix  at each time instance can be fairly estimated using the previous time instance. Indeed, the associated channel Doppler frequency provides a maximum level of dissimilarity between the channels at consecutive time instances (could be interpreted as upper-bounds on the angles between the subspaces). By exploiting this property, one can reduce the transmission overhead reserved for channel estimation (pilots), which in turn increases the spectral efficiency. 
	
	
\end{itemize}

\subsection{Roadmap}
The paper is organized as follows. A more clear definition of principal angles between subspaces besides a few concepts from convex geometry are reviewed in Section \ref{section.convexgeometry}. Section \ref{section.mainresult} is dedicated to obtaining bounds for the required number of measurements in $\mathsf{P}_{\rm nuc}$ and $\mathsf{P}_{\bm{w},{\rm nuc}}$. Section \ref{sec.optweights} is about our strategy of finding optimal weights. In Section \ref{section.simulation}, we present some numerical experiments which validate our theory. We shall describe related works in Section \ref{sec.relatedworks}. Section \ref{sec.usefullem} is devoted to important lemmas that frequently used in our analysis. Lastly, the paper is concluded in Section \ref{section.conclusion}.
\section{Preliminaries }\label{section.convexgeometry}
\subsection{Principal angles between subspaces}\label{section.pricipal_angles}
Consider two subspaces $\bm{\mathcal{U}}$ and $\bm{\mathcal{W}}$ of an Euclidean vector space $\mathbb{R}^n$ with $\mathrm{dim}(\bm{\mathcal{U}}):=r\le \mathrm{dim}(\bm{\mathcal{W}}):=r'$. There exist $r$ non-increasingly sorted angles $\bm{\theta}:=[\theta(1),..., \theta(r)]^T\in[0^\circ,90^\circ]^r$ called the principal angles, the least one is obtained by:
\begin{align}
&\theta(r):=\min\left\{\cos^{-1}\left(\frac{|\langle \bm{u}, \bm{w}\rangle|}{\|\bm{u}\|_2\|\bm{w}\|_2}\right)~:~\bm{u}\in\bm{\mathcal{U}},\bm{w}\in\bm{\mathcal{W}}\right\}\nonumber\\
&=\angle(\bm{u}_r,\bm{w}_r).
\end{align}
The $i$th one ($i<r$) is given by:
\begin{align}
&\theta(i):=\min\Bigg\{\cos^{-1}\left(\frac{|\langle \bm{u}, \bm{w}\rangle|}{\|\bm{u}\|_2\|\bm{w}\|_2}\right)~:~\bm{u}\in\bm{\mathcal{U}},\bm{w}\in\bm{\mathcal{W}}\nonumber\\
&, \bm{u}\perp \bm{u}_j, \bm{w}\perp \bm{w}_j ~ \forall j\in \{i+1, ..., r\}\Bigg\}=\angle(\bm{u}_i,\bm{w}_i).
\end{align}
$\{\bm{u}_i,\bm{w}_i\}_{i=1}^r$ are called principal vectors. Moreover, each subspace $\bm{\mathcal{U}}, \bm{\mathcal{W}}$ is spanned by a set of linearly independent vectors. In fact, there exist orthonormal bases $\bm{U}:=[\bm{u}_1,..., \bm{u}_r]\in\mathbb{R}^{n\times r}$ and $\bm{V}:=[\bm{w}_1,..., \bm{w}_r,\bm{w}_{r+1},...,\bm{w}_{r'}]\in\mathbb{R}^{n\times r'}$ for subspaces $\bm{\mathcal{U}}$ and $\bm{\mathcal{W}}$, respectively. Also,
\begin{align}
&\bm{\mathcal{U}}=\mathrm{span}(\bm{U}),\nonumber\\
&\bm{\mathcal{W}}=\mathrm{span}(\bm{W}),\nonumber\\
&\bm{U}^{\rm H}\bm{W}=\begin{bmatrix}
\cos(\bm{\theta})&\bm{0}_{r\times r'-r}
\end{bmatrix},
\end{align}
where
\begin{align}
&\cos(\bm{\theta}):=\mathrm{diag}([\cos(\theta(r)),\cos(\theta(r-1)), ..., \cos(\theta(1))])\nonumber\\
&\in\mathbb{R}^{r\times r}.
\end{align}
In the following, basic concepts of convex geometry are reviewed.
\begin{figure}[]
	\centering
	\hspace{0cm}
	\includegraphics[scale=.44]{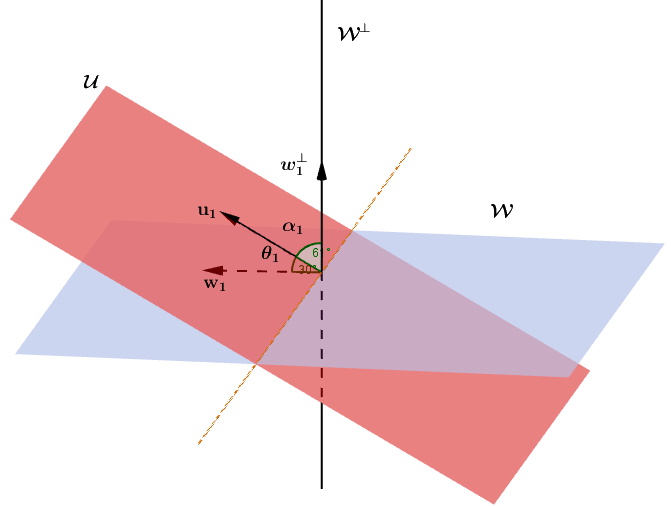}
	\caption{Principal angles and vectors in a three-dimensional Euclidean space. In this Figure, the subspaces $\bm{\mathcal{W}}$ and $\bm{\mathcal{W}}^\perp$ form principal angles $\bm{\theta}=[\theta_1,\theta_2]$ and $\alpha_1$ with the subspace $\bm{\mathcal{U}}$. $\bm{u}_1$, $\bm{w}_1$ and $\bm{w}_1^\perp$ are the corresponding principal vectors. Also, $\theta_2=0$, $\bm{u}_2$ and $\bm{w}_2$ (which are not depicted) are in the line $\bm{\mathcal{U}}\bigcap\bm{\mathcal{W}}$ in the same direction.}
	\label{fig.principalangles}
\end{figure}
\subsection{Descent Cones}
The descent cone $\mathcal{D}(f,\bm{x})$ at a point $\bm{x}\in\mathbb{R}^n$ consists of the set of directions that do not increase $f$ and is given by:
\begin{align}\label{eq.descent cone}
\mathcal{D}(f,\bm{x})=\bigcup_{t\ge0}\{\bm{z}\in\mathbb{R}^n: f(\bm{x}+t\bm{z})\le f(\bm{x})\}.
\end{align}
The descent cone reveals the local behavior of $f$ near $\bm{x}$ and is a convex set for convex functions. There is also a relationship between decent cone and subdifferential \cite[Chapter 23]{rockafellar2015convex} given by:
\begin{align}\label{eq.D(f,x)}
\mathcal{D}^{\circ}(f,\bm{x})=\mathrm{cone}(\partial f(\bm{x})):=\bigcup_{t\ge0}t.\partial f(\bm{x}).
\end{align}
\subsection{Statistical Dimension}
\begin{defn}{Statistical Dimension}\cite{amelunxen2013living}:
	Let $\mathcal{C}\subseteq\mathbb{R}^n$ be a convex closed cone. Statistical dimension of $\mathcal{C}$ is defined as:
	\begin{align}\label{eq.statisticaldimension}
	\delta(\mathcal{C}):=\mathds{E}\|\mathcal{P}_\mathcal{C}(\bm{g})\|_2^2=\mathds{E}\mathrm{dist}^2(\bm{g},\mathcal{C}^\circ),
	\end{align}
	where, $\bm{g}$ is an i.i.d. standard normal vector and $\mathcal{P}_\mathcal{C}(\bm{x})$ is the projection of $\bm{x}\in \mathbb{R}^n$ onto the set $\mathcal{C}$ defined as: $\mathcal{P}_\mathcal{C}(\bm{x})=\underset{\bm{z} \in \mathcal{C}}{\arg\min}\|\bm{z}-\bm{x}\|_2$.
\end{defn}
Statistical dimension extends the concept of linear subspaces to convex cones. Intuitively, it measures the size of a cone. Furthermore,
\begin{align}\label{eq.delta_desent_cone}
\delta(\mathcal{D}(f,\bm{x})):=\mathds{E}\inf_{t\ge 0}\inf_{\bm{z}\in\partial f(\bm{x})}\|\bm{g}-t\bm{z}\|_2^2,
\end{align}
 determines the precise number of measurements corresponding to the transition from failure to success in $\mathsf{P}_f$.
\begin{figure}[]
	\centering
	\hspace{0cm}
	\includegraphics[scale=.35]{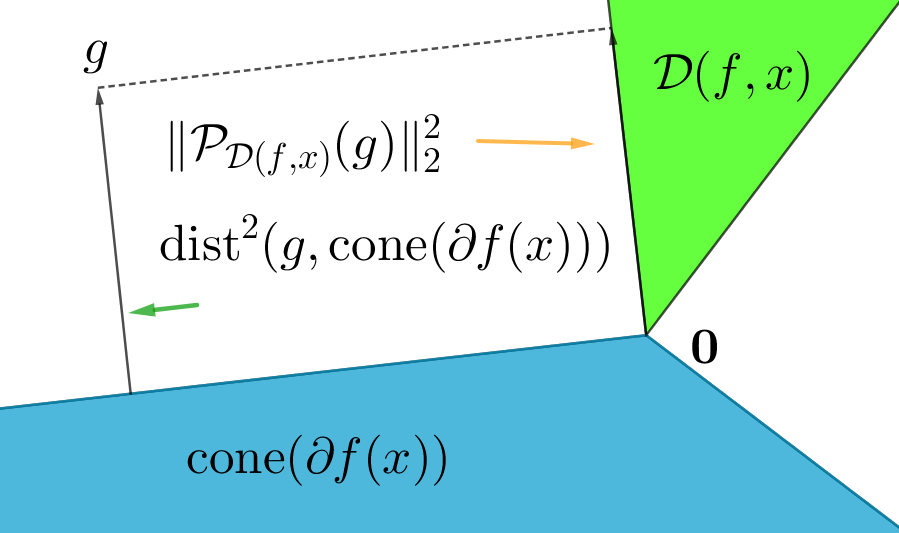}
	\caption{A schematic about the cones $\mathcal{D}(f,\bm{x})$, ${\rm cone}(\partial f(\bm{x}))$ and the equal expressions $\|\mathcal{P}_{\mathcal{D}(f,\bm{x})}(\bm{g})\|_2^2$ and ${\rm dist}^2(\bm{g},{\rm cone}(\partial f(\bm{x})))$. $\|\mathcal{P}_{\mathcal{D}(f,\bm{x})}(\bm{g})\|_2^2$ intuitively expresses the bigness of $\mathcal{D}(f,\bm{x})$.}
	\label{fig.desentcone}
\end{figure}
\subsection{Optimality Condition}
In the following, we characterize when $\mathsf{P}_f$ succeeds in the noise-free case.
\begin{prop}\cite[Proposition 2.1]{chandrasekaran2012convex} Optimality condition: Let $f$ be a proper convex function. The vector $\bm{x}\in \mathbb{R}^n$ is the unique optimal point of $\mathsf{P}_f$ if and only if $\mathcal{D}(f,\bm{x})\cap \mathrm{null}(\bm{A})=\{\bm{0}\}$.
\end{prop}
The next theorem determines the number of measurements needed for successful recovery of $\mathsf{P}_f$ for any proper convex function $f$.
\begin{thm}\label{thm.Pfmeasurement}\cite[Theorem 2]{amelunxen2013living}:
	Let $f:\mathbb{R}^n\rightarrow \mathbb{R}\cup \{\pm\infty\}$ be a proper convex function and $\bm{x}\in \mathbb{R}^n$ a fixed sparse vector. Suppose that $m$ independent Gaussian linear measurements of $\bm{x}$ are observed via the affine constraint $\bm{y}=\bm{Ax} \in \mathbb{R}^m$. Then, for a given tolerance $\eta \in [0,1]$ if
	\begin{align}
	m\ge \delta(\mathcal{D}(f,\bm{x}))+\sqrt{8\log(\frac{4}{\eta})n},\nonumber
	\end{align}
 we have
 \begin{align}
 \mathds{P}(\mathcal{D}(f,\bm{x})\cap \mathrm{null}(\bm{A})=\{\bm{0}\})\ge 1-\eta.\nonumber
 \end{align}
 Besides, if 
 \begin{align}
m\le \delta(\mathcal{D}(f,\bm{x}))-\sqrt{8\log(\frac{4}{\eta})n},\nonumber
 \end{align}
then, 
\begin{align}
\mathds{P}(\mathcal{D}(f,\bm{x})\cap \mathrm{null}(\bm{A})=\{\bm{0}\})\le \eta.\nonumber
\end{align}
\end{thm}
Also in \cite{amelunxen2013living}, the following error bound for the statistical dimension is provided:
\begin{thm}\cite[Theorem 4.3]{amelunxen2013living} For any $\bm{x}\in \mathbb{R}^n\setminus\{\bm{0}\}$:
	\begin{align}\label{eq.errorbound}
	0\le\inf_{t\ge0}\mathds{E}\mathrm{dist}^2(\bm{g},t\partial f(\bm{x}))- \delta(\mathcal{D}(f,\bm{x}))\le \frac{2\sup_{s\in \partial f(\bm{x})}\|s\|_2}{f(\frac{\bm{x}}{\|\bm{x}\|_2})}.
	\end{align}
\end{thm} 
\section{The measurement Threshold for successful recovery}\label{section.mainresult} 
Fix a probability of failure $\eta\in[0,1]$. Denote the normalized number of measurements that $\mathsf{P}_{\mathrm{nuc}}$ and $\mathsf{P}_{\bm{w},\mathrm{nuc}}$ need for exact recovery of a matrix $\bm{X}\in\mathbb{R}^{n\times n}$ by
\begin{align}
&m_{\mathrm{nuc}}:=\frac{\delta(\mathcal{D}(\|\cdot\|_*,\bm{X}))}{n^2},\nonumber\\
&m_{\bm{w},\bm{\theta}_u,\bm{\theta}_v}:=\frac{\delta(\mathcal{D}(\|h_{\bm{w}}(\cdot)\|_*,\bm{X}))}{n^2},\nonumber
\end{align}
respectively.
In \cite[Proposition 4.7]{amelunxen2013living}, an upper-bound for $m_{\mathrm{nuc}}$ is provided. To facilitate the calculations, we obtain an upper-bound for $m_{\mathrm{nuc}}$ in harmony with our strategy of finding optimal weights in this work. The proposed upper-bound asymptotically equals the upper-bound in \cite[Proposition 4.7]{amelunxen2013living}.
\begin{prop}\label{prop.mhat_nuc}
Consider a matrix $\bm{X}\in\mathbb{R}^{n\times n}$ with rank $r$. Suppose that $r,r',n\rightarrow \infty$ with limiting ratios $\sigma_1:=\frac{r}{n}$ and $\sigma_2:=\frac{r'}{n}$ with $r'\ge r$. Then, 
\begin{align}
m_{\mathrm{nuc}}\rightarrow \widehat{m}_{\rm nuc} \nonumber
\end{align}
for
\begin{align}
\widehat{m}_{\rm nuc}:=\inf_{t\ge 0}\Psi(\sigma_1,\sigma_2),
\end{align}
with
\begin{align}
&\Psi(\sigma_1,\sigma_2)=3\sigma_1^2+\tfrac{t^2\sigma_1}{n}+\sigma_1^2\phi(\tfrac{t\alpha_{22}}{\sqrt{r}},1)+2\sigma_1(\sigma_2-\sigma_1)\nonumber\\
&\phi(\tfrac{t\alpha_{23}}{r\vee (r'-r)},s_1)+2\sigma_1(1-\sigma_1-\sigma_2)\phi(\tfrac{t\alpha_{24}}{r\vee (n-r'-r)},s_2)+\nonumber\\
&2(\sigma_2-\sigma_1)(1-\sigma_1-\sigma_2)\phi(\tfrac{t\alpha_{34}}{(r'-r)\vee (n-r'-r)},s_3)+\nonumber\\
&(\sigma_2-\sigma_1)^2\phi(\tfrac{t\alpha_{33}}{\sqrt{r'-r}},1)+(1-\sigma_1-\sigma_2)^2\phi(\tfrac{t\alpha_{44}}{\sqrt{n-r-r'}},1),
\end{align}
where
\begin{align}
&\phi(\tau,s):=\int_{l_b(s)}^{u_b(s)}(u-\tau)_{+}^2\tfrac{\sqrt{(u_b(s)^2-u^2)(u^2-l_b(s)^2)}}{\pi u s}{\rm d} u,\nonumber\\
&s_1=\tfrac{r\wedge (r'-r)}{r\vee (r'-r)},s_2=\tfrac{r\wedge (n-r-r')}{r\vee (n-r-r')},s_3=\tfrac{(r'-r)\wedge (n-r-r')}{(r'-r)\vee (n-r-r')},\nonumber\\
& l_b(s)=1-\sqrt{s}, u_b(s)=1+\sqrt{s}\label{eq.ub-lb},
\end{align}
and
\begin{align}\label{eq.alphadef}
&\alpha_{22}:=\tfrac{r^2}{(n-r)^2},~\alpha_{23}:=\tfrac{r(r'-r)}{(n-r)^2},~\alpha_{24}=\tfrac{r(n-r-r')}{(n-r)^2},\nonumber\\
&\alpha_{33}=\tfrac{(r'-r)^2}{(n-r)^2},~\alpha_{34}=\tfrac{(r'-r)(n-r-r')}{(n-r)^2},\alpha_{44}=\tfrac{(n-r-r')^2}{(n-r)^2}.
\end{align}
\end{prop}
Proof. See Appendix \ref{proof.prop.mhat_nuc}.

\begin{figure}[]
	\centering
	\hspace{-0cm}
	\includegraphics[scale=.35]{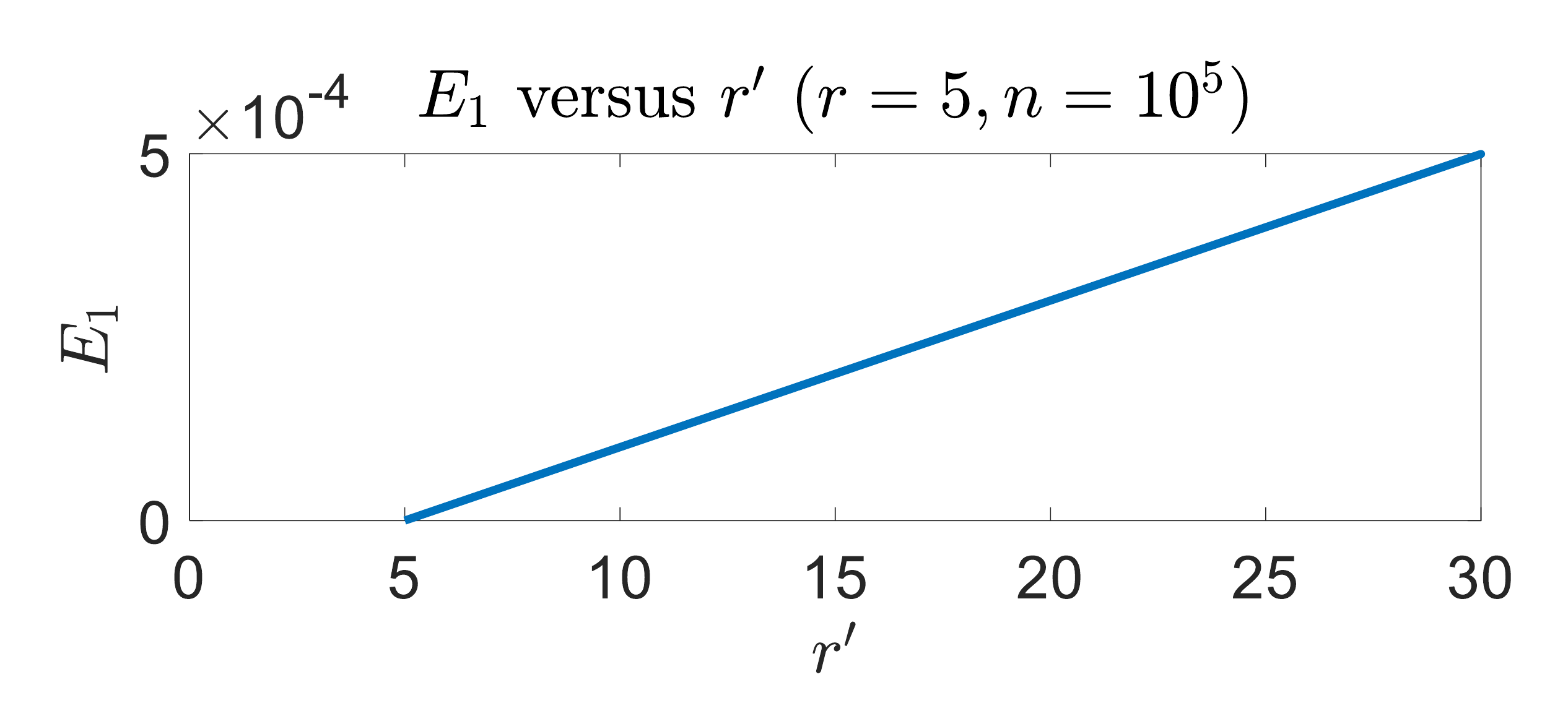}
	\caption{The difference between our upper-bound on $m_{\rm nuc}$ i.e. $\widehat{m}_{\rm nuc}$ with the one in \cite[Equation 4.8]{amelunxen2013living} (labeled $E_1$). The difference is negligible in particular when $r'$ is not far from $r$ which is practically more common.}
	\label{fig.error_vs_rprim}
\end{figure}
\begin{rem}\label{rem.ame}(Prior work)
In \cite[Equation 4.8]{amelunxen2013living} an upper-bound is derived for $m_{\rm nuc}$. Here, we compare our bound i.e. $\widehat{m}_{\rm nuc}$ with theirs. Denote the difference between $\widehat{m}_{\rm nuc}$ and the upper-bound in \cite[Equation 4.8]{amelunxen2013living} by $E_1$. From Figure \ref{fig.error_vs_rprim}, we observe that the error $E_1$ is negligible when $r'$ is not far from $r$ which is practically common. Moreover, since the upper-bound \cite[Equation 4.8]{amelunxen2013living} describes $m_{\rm nuc}$ well (the error is at most $\frac{2}{n\sqrt{nr}}$), regarding Figure \ref{fig.error_vs_rprim}, one can infer that $\widehat{m}_{\rm nuc}$ also approximates $m_{\rm nuc}$ suitably up to an asymptotically vanishing error term.
\end{rem}

In what follows, we obtain an upper-bound for $m_{\bm{w},\bm{\theta}_u,\bm{\theta}_v}$. This bound helps us to find the optimal weights later. The strategy of providing this bound is, to some extent, similar to the strategy used in Proposition \ref{prop.mhat_nuc}. However, the derivation is more elaborate; in fact, this bound, unlike $\widehat{m}_{\rm nuc}$, depends on the principal angles i.e. $\bm{\theta}_u,\bm{\theta}_v\in [0,90^\circ]^r$ and the weight vector $\bm{w}$ making it more involved.
\begin{prop}\label{prop.m_hat_nuc_weight}
Consider a rank $r$ matrix $\bm{X}\in\mathbb{R}^{n\times n}$ with column and row subspaces $\bm{\mathcal{U}}$ and $\bm{\mathcal{V}}$, respectively. Also, assume that we are given the subspaces $\widetilde{\bm{\mathcal{U}}}\subseteq\mathbb{R}^n$ and $\widetilde{\bm{\mathcal{V}}}\subseteq\mathbb{R}^n$ with dimension $r'\ge r$ that have known principal angles $\bm{\theta}_u\in[0^\circ,90^\circ]^r$ and $\bm{\theta}_v\in[0^\circ,90^\circ]^r$ with $\bm{\mathcal{U}}$ and $\bm{\mathcal{V}}$, respectively. Then,
\begin{align}
m_{\bm{w},\bm{\theta}_u,\bm{\theta}_v}\le \widehat{m}_{\bm{w},\bm{\theta}_u,\bm{\theta}_v},
\end{align}
for
\begin{align}
\widehat{m}_{\bm{w},\bm{\theta}_u,\bm{\theta}_v}:=\inf_{t\ge 0}\frac{\Psi_t(\bm{w},\bm{\theta}_u,\bm{\theta}_v)}{n^2},
\end{align}
with
\begin{align}\label{eq.Psi_w}
&\Psi_t(\bm{w},\bm{\theta}_u,\bm{\theta}_v)=3r^2+(tw_1)^2\sum_{i=1}^{r}\cos^2(\theta_u(i))\cos^2(\theta_v(i))+\nonumber\\
&(tw_2)^2\sum_{i=1}^{r}\cos^2(\theta_u(i))\sin^2(\theta_v(i))+(tw_3)^2\sum_{i=1}^{r}\sin^2(\theta_u(i))\nonumber\\
&\cos^2(\theta_v(i))+(tw_4)^2\sum_{i=1}^{r}\sin^2(\theta_u(i))\sin^2(\theta_v(i))+\nonumber\\
&(\tfrac{w_4}{w_3}-1)^2(tw_1+tw_2)^2\sum_{i=1}^{r}\Bigg\{\tfrac{w_1^2{\cos}^2(\bm{\theta}_u(i))+w_3^2{\sin}^2(\bm{\theta}_u(i))}{w_1^2{\cos}^2(\bm{\theta}_v(i))+w_2^2{\sin}^2(\bm{\theta}_v(i))}\nonumber\\
&\sin^2(\theta_v(i))\cos^2(\theta_v(i))\Bigg\}+(\tfrac{w_3}{w_1}-1)^2(tw_1+tw_3)^2\sum_{i=1}^{r}\nonumber\\
&\Bigg\{\tfrac{w_1^2{\cos}^2(\bm{\theta}_v(i))+w_2^2{\sin}^2(\bm{\theta}_v(i))}{w_1^2{\cos}^2(\bm{\theta}_u(i))+w_3^2{\sin}^2(\bm{\theta}_u(i))}\sin^2(\theta_u(i))\cos^2(\theta_u(i))\Bigg\}+\nonumber\\
&\sum_{i=1}^{r}r\phi(\tfrac{\sigma_i(\bm{E}_{22})+(tw_1)(tw_3)(tw_4)\sigma_i(\bm{C}_L^{-1})\sigma_i(\bm{C}_R^{-1})\alpha_{22}}{\sqrt{r}},1)+\nonumber\\
&r\vee (r'-r)\sum_{i=1}^{r\wedge (r'-r)}\phi(\tfrac{(tw_1)(tw_3)\sigma_i(\bm{C}_L^{-1})\alpha_{23}}{\sqrt{r\vee (r'-r)}}, s_1)+\nonumber\\
&r\vee(n-r-r')\sum_{i=1}^{r\wedge (n-r'-r)}\phi(\tfrac{(tw_1)(tw_4)\sigma_i(\bm{C}_L^{-1})\alpha_{24}}{\sqrt{r\vee (n-r-r')}},s_2)+\nonumber\\
&r\vee(r'-r)\sum_{i=1}^{r\wedge (r'-r)}\phi(\tfrac{(tw_2)(tw_3)\sigma_i(\bm{C}_R^{-1})\alpha_{32}}{\sqrt{r\vee (r'-r)}},s_1)+(r'-r)^2\nonumber\\
&\phi(\tfrac{(tw_1)\alpha_{33}}{\sqrt{(r'-r)}},1)+(r'-r)\vee (n-r-r')\nonumber\\
&\sum_{i=1}^{(r'-r)\wedge (n-r-r')}\phi(\tfrac{tw_2\alpha_{34}}{\sqrt{(r'-r)\vee (n-r-r')}}, s_3)+r\vee(n-r-r')\nonumber\\
&\sum_{i=1}^{r \wedge n-r-r'}\phi(\tfrac{(tw_3)(tw_4)\sigma_i(\bm{C}_R^{-1})\alpha _{42}}{\sqrt{r\wedge (n-r-r')}}, s_2)+(r'-r)\vee(n-r-r')\nonumber\\
&\sum_{i=1}^{(r'-r)\wedge(n-r-r')}\phi(\tfrac{tw_3\alpha_{43}}{\sqrt{(r'-r)\vee (n-r-r')}}, s_3)+(n-r-r')^2\nonumber\\
&\phi(\tfrac{tw_4\alpha_{44}}{\sqrt{n-r-r'}}, 1),
\end{align}
where
\begin{align}\label{eq.E22}
&\bm{E}_{22}=\Big((tw_4)^2-(tw_3)^2-(tw_2)^2+(tw_1)^2\Big)\nonumber\\
&\Bigg((tw_1)^2{\cos}^2(\bm{\theta}_u){\cos}^2(\bm{\theta}_v)+(tw_2)^2{\cos}^2(\bm{\theta}_u){\sin}^2(\bm{\theta}_v)+\nonumber\\
&(tw_3)^2{\sin}^2(\bm{\theta}_u){\cos}^2(\bm{\theta}_v)+(tw_4)^2{\sin}^2(\bm{\theta}_u){\sin}^2(\bm{\theta}_v)\Bigg)^{-\frac{1}{2}}
\nonumber\\
&{\sin}(\bm{\theta}_u){\cos}(\bm{\theta}_u){\sin}(\bm{\theta}_v){\cos}(\bm{\theta}_v),
\end{align}
\begin{align}\label{eq.CL}
\bm{C}_L:=\Big((tw_1)^2{\cos}^2(\bm{\theta}_u)+(tw_3)^2{\sin}^2(\bm{\theta}_u)\Big)^{\frac{1}{2}},
\end{align}
\begin{align}\label{eq.CR}
\bm{C}_R:=\Big((tw_3)^2{\cos}^2(\bm{\theta}_v)+(tw_4)^2{\sin}^2(\bm{\theta}_v)\Big)^{\frac{1}{2}}.
\end{align}
\end{prop}
Proof. See Appendix \ref{proof.prop.m_hat_nuc_weight}.

\begin{rem}\label{rem.special}(Special case of Proposition \ref{prop.m_hat_nuc_weight})
Interestingly, $\widehat{m}_{\bm{w},\bm{\theta}_u,\bm{\theta}_v}$ coincides with $\widehat{m}_{\rm nuc}$ when we set $\bm{w}=\bm{1}\in\mathbb{R}^3$. In other words, this implies the fact that the required number of measurements of $\mathsf{P}_{\bm{w},{\rm nuc}}$ in the special case of $\bm{w}=\bm{1}\in\mathbb{R}^{3}$ is the same as the number of measurements that $\mathsf{P}_{\rm nuc}$ needs for successful recovery.
\end{rem}
 From Remarks \ref{rem.ame} and \ref{rem.special}, it is obvious that $\widehat{m}_{\bm{1},\bm{\theta}_u,\bm{\theta}_v}$ is the same as $m_{\rm nuc}$. Thus, one could simply think of the following question:
\begin{itemize}
	\item Is $\widehat{m}_{\bm{w},\bm{\theta}_u,\bm{\theta}_v}$ a good description of $m_{\bm{w},\bm{\theta}_u,\bm{\theta}_v}$?
\end{itemize}
In the following Lemma, we provide a positive answer to this question. In fact, we demonstrate that the proposed upper-bound in Proposition \ref{prop.m_hat_nuc_weight} is asymptotically tight.
\begin{lem}\label{lem.errorbound}
The number of measurements that $\mathsf{P}_{\bm{w},\mathrm{nuc}}$ with parameters $\bm{\theta}_u=[\theta_u(1),..., \theta_u(r)]^T$ and $\bm{\theta}_v=[\theta_v(1), ... , \theta_v(r)]^T$, needs for exact recovery of $\bm{X}\in\mathbb{R}^{n\times n}$ satisfies the following error bound:
\begin{align}\label{eq.error_weight}
\widehat{m}_{\bm{w},\bm{\theta}_u,\bm{\theta}_v}-\frac{2}{n\sqrt{n r}c}\le m_{\bm{w},\bm{\theta}_u,\bm{\theta}_v}\le \widehat{m}_{\bm{w},\bm{\theta}_u,\bm{\theta}_v},
\end{align}
where
\begin{align*}
c=\min\{\sin(\theta_u(1)),\cos(\theta_u(r)) \} \min\{\sin(\theta_v(1)),\cos(\theta_v(r)) \}.
	\end{align*}

\end{lem}
Proof. See Appendix \ref{proof.errorbound}

It is worth mentioning that the error term is independent of $\bm{w}$, constant and vanishes asymptotically.
\section{How to find optimal weights}\label{sec.optweights}
In this section, we propose the strategy of finding the unique optimal weights. First, we present a general Lemma about the function $\delta(\mathcal{D}(\|h_{\bm{w}}(\cdot)\|_*,\bm{X}))$. Actually, this Lemma states that this function (ignoring the infimum on $t\ge0$ in the definition of statistical dimension) is strictly convex with respect to $\bm{w}\in\mathbb{R}_{+}^3$. This Lemma helps us later in proving the uniqueness of optimal weights.
\begin{lem}\label{lem.objective_convex}
Assume $\mathcal{C}:=\partial\|\cdot\|_*(h_{\bm{w}}(\bm{X}))$ does not contain the origin. Also, denote $\bm{G}\in\mathbb{R}^{n\times n}$ a random matrix with i.i.d. standard normal entries. Consider the function
\begin{align}\label{eq.Jv}
&J(\bm{v}):=\mathds{E}\mathrm{dist}^2(\bm{G},h_{\bm{v}}(\mathcal{C})):=\mathds{E}[J_{\bm{G}}(\bm{v})],\nonumber\\
&\text{with}~~ \bm{v}:=[v_1,v_2,v_3].
\end{align}
The function $J$ is strictly convex and continuous at $\bm{v}\in\mathbb{R}_{+}^{3}$. Further, it attains its minimum in the set $\Big(0,n\Big(1+\frac{(n^2+1)^{\frac{1}{4}}}{\sqrt{\sqrt{n^2+1}-n}}\Big)\Big]^3$.
\end{lem}
Proof. See Appendix \ref{proof.lem.objective_convex}.

Now, we introduce our strategy of finding the unique optimal weights. Consider the error bound in Lemma \ref{lem.errorbound}. By taking infimum from both sides, it holds that
\begin{align}
&\inf_{\bm{w}\in\mathbb{R}_{+}^3}\widehat{m}_{\bm{w},\bm{\theta}_u,\bm{\theta}_v}-\frac{2}{n\sqrt{n r}c}\le \inf_{\bm{w}\in\mathbb{R}_{+}^3}m_{\bm{w},\bm{\theta}_u,\bm{\theta}_v}\le\nonumber\\
& \inf_{\bm{w}\in\mathbb{R}_{+}^3}\widehat{m}_{\bm{w},\bm{\theta}_u,\bm{\theta}_v}.
\end{align}   
$m_{\bm{w},\bm{\theta}_u,\bm{\theta}_v}$ is surrounded by the same upper and lower-bounds up to an asymptotically vanishing constant term. We minimize this expression so as to reach the optimal weights $\bm{w}^*=[w_1^*,w_2^*,w_3^*]^T$ via
\begin{align}\label{eq.optimal weights}
&\bm{w}^*:=\mathop{\arg\min}_{\bm{w}\in\mathbb{R}_{+}^3}\widehat{m}_{\bm{w},\bm{\theta}_u,\bm{\theta}_v}=\mathop{\arg\min}_{\bm{w}\in\mathbb{R}_{+}^3}\inf_{t\ge 0}\Psi_t(\bm{w},\bm{\theta}_u,\bm{\theta}_v)=\nonumber\\
&\mathop{\arg\min}_{\bm{v}\in\mathbb{R}_{+}^3}J(\bm{v}).
\end{align}
The reason to name these weights, \textit{optimal}, lies in the fact that they asymptotically (as $n\rightarrow \infty$) minimize the required number of measurements in $\mathsf{P}_{\bm{w},{\rm nuc}}$. Note that in the second equality of (\ref{eq.optimal weights}), we converted two variables $\bm{w}$ and $t$ into a single vector variable $\bm{v}=t\bm{w}\in\mathbb{R}^3$. This is since $\bm{w}$ in  $\Psi_t(\bm{w},\bm{\theta}_u,\bm{\theta}_v)$ of (\ref{eq.Psi_w}) always appears along with the scalar $t$ (namely in the form of $t\bm{w}$). Therefore, by finding $\bm{v}^*$ (last term in (\ref{eq.optimal weights})), we can reach the optimal weights $\bm{w}^*$ up to a positive scaling factor. As a matter of fact, by the aid of Lemma \ref{lem.objective_convex}, $\bm{v}^*$ is unique and lies in the set $\Big(0,n\Big(1+\frac{(n^2+1)^{\frac{1}{4}}}{\sqrt{\sqrt{n^2+1}-n}}\Big)\Big]^3$. Hence, $\bm{w}^*$ is unique up to a positive scaling factor. Note that this scaling factor is not the case since it is effectless on $\mathsf{P}_{\bm{w},{\rm nuc}}$.
To obtain $\bm{w}^*$ in (\ref{eq.optimal weights}), we propose a simple algorithm in Algorithm \ref{alg.optweights} called Optweights. In Optweights, we solve the convex optimization problem
\begin{align}
[w_1^*,w_2^*,w_3^*]=\mathop{\rm argmin}_{\substack{w_1\ge 0\\w_2\ge 0\\w_3\ge 0}}\widehat{m}_{\bm{w},\bm{\theta}_u,\bm{\theta}_v},
\end{align}
to reach the triple $[w_1^*,w_2^*,w_3^*]$.\par
Qualitatively speaking, Algorithm \ref{alg.optweights} is based on alternating minimization (AM) approach. AM method is used to solve multivariate unconstrained optimization problems. The idea is based on optimizing each coordinate, individually. The advantages of our proposed algorithm are
\begin{itemize}
	\item Each iteration is cheap.
	\item Unlike the gradient-based algorithms, it needs no step-size tuning.
	\item  It is simple to implement.
\end{itemize}
In essence, Optweights (Algorithm \ref{alg.optweights}) converts the multivariate optimization problem into some with scalar variables. For solving scalar optimization problems in Optweights (i.e. Step \ref{step.scalar} in Algorithm \ref{alg.optweights}), we use Golden Section Search (GSS) method (Algorithm \ref{alg.GSS}) which tries to narrow the range of values ($a$ and $b$ in Algorithm \ref{alg.GSS}) inside which the minimum is known to exist. 
\alglanguage{pseudocode}
\begin{algorithm}[t]
	\caption{Optweights (Proposed algorithm for finding optimal weights)}\label{alg.optweights}
	\begin{algorithmic}[1]
		\Procedure {Optweights}{$\widehat{m}_{\bm{w},\bm{\theta}_u,\bm{\theta}_v}$, $\rm maxiter$, $\rm Tol$}
	
		\State 
		$f(w_1,w_2,w_3)=\widehat{m}_{{\bm{w},\bm{\theta}_u,\bm{\theta}_v}},$ \\\quad with $\bm{w}=[w_1,w_2,w_3]^T,$   cost function \label{step.costfun}
		\State $w_i^{1}\leftarrow 1 ~~\forall i\in\{1,2,3\},$
		\State $k\leftarrow 1,$
		\Repeat 
		
		\For{$i=1~{\rm to}~3$}
		\State Optimize the $i$th coordinate
		\State $\phi(\zeta):=f(\underbrace{w_1^{k+1},... ,w_{i-1}^{k+1}}_{\rm done} ,\underbrace{\zeta}_{\rm current},\underbrace{w_{i+1}^{k},..., w_{3}^{k}}_{\rm to~ do}),$
		\State \begin{align}
		w_i^{k+1}\leftarrow \mathop{\rm argmin}_{\zeta \in \mathbb{R}} \phi(\zeta)~~~~\text{use GSS algorithm \ref{alg.GSS}},\nonumber
		\end{align}\label{step.scalar}
		\EndFor
	\State $k\leftarrow k+1$
		\Until
		{	$\|\bm{w}^{k}-\bm{w}^{k-1}\|_2 <\rm Tol$ or $|f(\bm{w}^{k})-f(\bm{w}^{k-1})|<\rm Tol$ or $k>{\rm maxiter}$, }
		\State \textbf{Output} $\bm{w}^*\leftarrow [w_1^k,w_2^k,w_3^k]^T,$
		\EndProcedure
		\Statex
	\end{algorithmic}
\end{algorithm}
\alglanguage{pseudocode}
\begin{algorithm}[t]
	\caption{GSS (Golden Section Search)}\label{alg.GSS}
	\begin{algorithmic}[1]
		\Procedure {GSS}{$a$, $b$, $\rm maxiter$, $\rm Tol$}
		\State $a$ and $b$ are some lower and upper-bounds for $x_{\min}$
		\State 
		$\tau=\frac{\sqrt{5}-1}{2}$   golden ratio
		\State 
		$f={\rm costfun}(x)$ cost function
		\State $x_1=a+(1-\tau)(b-a)$
		\State $x_2=a+\tau(b-a)$
		\State $k\leftarrow 1$
		\Repeat 
		\State
		$k\leftarrow k+1$
	\If {$f(x_1)<f(x_2)$}
	\State $b\leftarrow x_2$
	\State $x_2\leftarrow x_1$
	\State $x_1=a+(1-\tau)(b-a)$
	\Else
	\State $a\leftarrow x_1$
	\State $x_1\leftarrow x_2$
	\State $x_2\leftarrow a+\tau(b-a)$
	\State $k\leftarrow k+1$
	\EndIf
	\Until{	$|b-a|<\rm Tol$ and $k>maxiter$ }
	\If 
	{$f(x_1)\le f(x_2)$}
	\State $x_{\min}\leftarrow x_1$
	\Else
	\State $x_{\min}\leftarrow x_2$
	\EndIf
%
%
%
%
		\State \textbf{Output} $x_{\min}$
		\EndProcedure
		\Statex
	\end{algorithmic}
\end{algorithm}
\section{Numerical experiments}\label{section.simulation}
In this section, we present the result of some computer experiments designed to evaluate the effect of optimal weighting strategy in matrix sensing given some prior subspace information. Note that the optimal weights are obtained using Algorithm \ref{alg.optweights}. First, we construct a matrix
\begin{align}
\bm{X}=\bm{U}_{n\times r}\bm{\Sigma}_{r\times r}\bm{V}_{n\times r}^{\rm H},
\end{align}
with $n=10, r=3$. Then, we construct two subspaces $\widetilde{\bm{\mathcal{U}}}$ and $\widetilde{\bm{\mathcal{V}}}$ with dimension $r'\ge r$, that have known principal angles $\bm{\theta}_u\in[0,90^\circ]^r$ and $\bm{\theta}_v\in [0,90^\circ]^r$ with column and row subspaces of the ground-truth matrix $\bm{X}$ i.e. $\bm{\mathcal{U}}={\rm span}(\bm{U})$ and $\bm{\mathcal{V}}={\rm span}(\bm{V})$, respectively. Note that, the bases $\widetilde{\bm{U}}$ and $\widetilde{\bm{V}}$ are chosen such that 
\begin{align}
&\bm{U}^{\rm H}\widetilde{\bm{U}}=\begin{bmatrix}
{\rm cos}(\bm{\theta}_u)&\bm{0}_{r\times (r'-r)}
\end{bmatrix},\nonumber\\
&\bm{V}^{\rm H}\widetilde{\bm{V}}=\begin{bmatrix}
{\rm cos}(\bm{\theta}_v)&\bm{0}_{r\times (r'-r)}
\end{bmatrix}.\nonumber
\end{align}
Next, we compute the optimal weights $\bm{w}^*$ by Algorithm \ref{alg.optweights}. We compare $\mathsf{P}_{\rm nuc}$ with $\mathsf{P}_{\bm{w}^*,{\rm nuc}}$ for different $\bm{\theta}_u$ and $\bm{\theta}_v$. Our assessment criterion is the probability of success over $50$ Monte Carlo trials. A trial is declared successful if
\begin{align}
\frac{\|\bm{X}-\widehat{\bm{X}}\|_F}{\|\bm{X}\|_F}\le 10^{-2},
\end{align}
where $\widehat{\bm{X}}$ is the solution of optimization problems provided by CVX MATLAB package \cite{cvx}. Below, we investigate different cases of principal angles.  

\begin{figure*}[t]
	\centering
	\mbox{\subfigure[]{\includegraphics[width=2.34in]{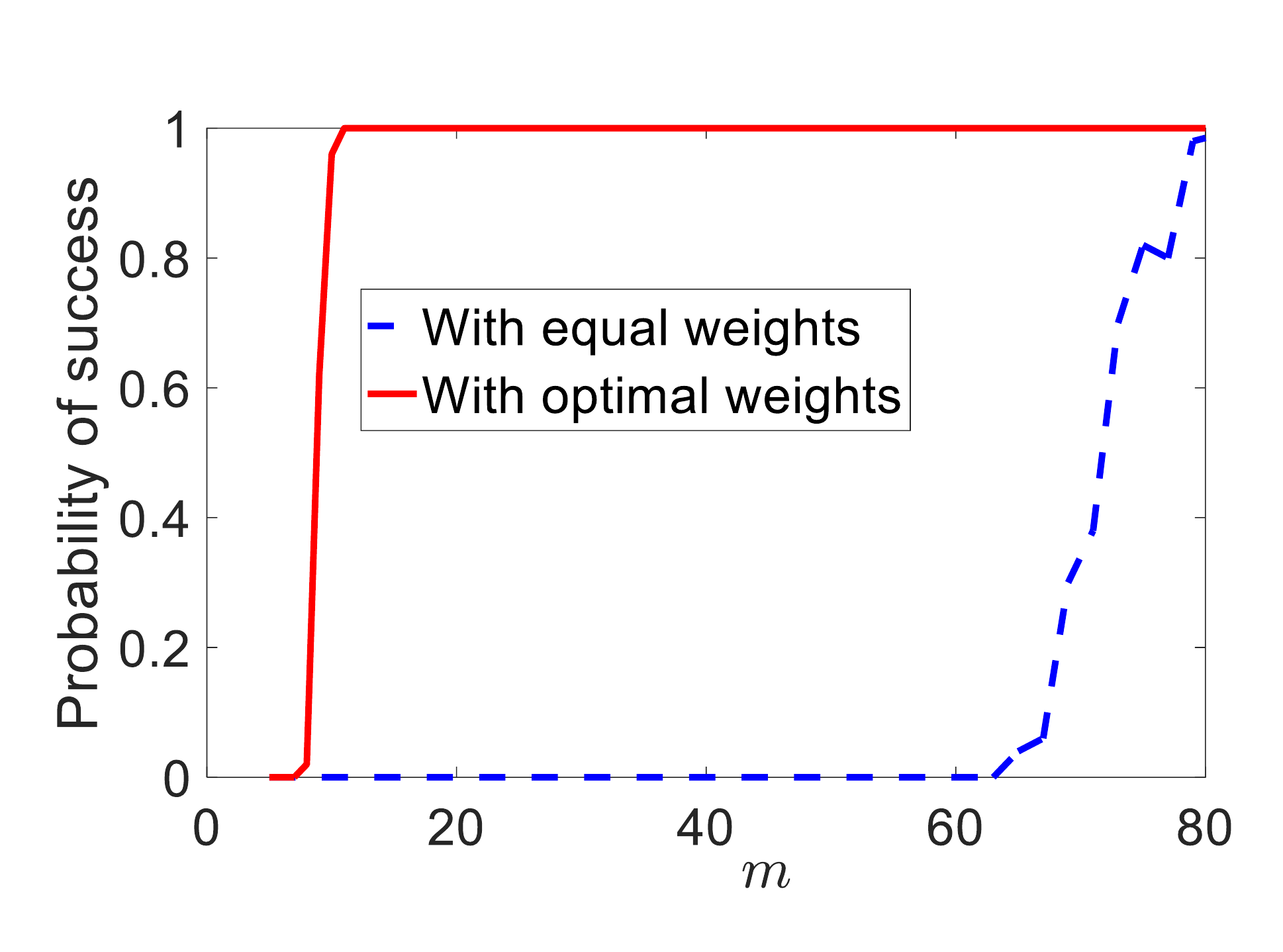}\label{fig.paperfig2}}\quad
		\subfigure[]{\includegraphics[width=2.34in]{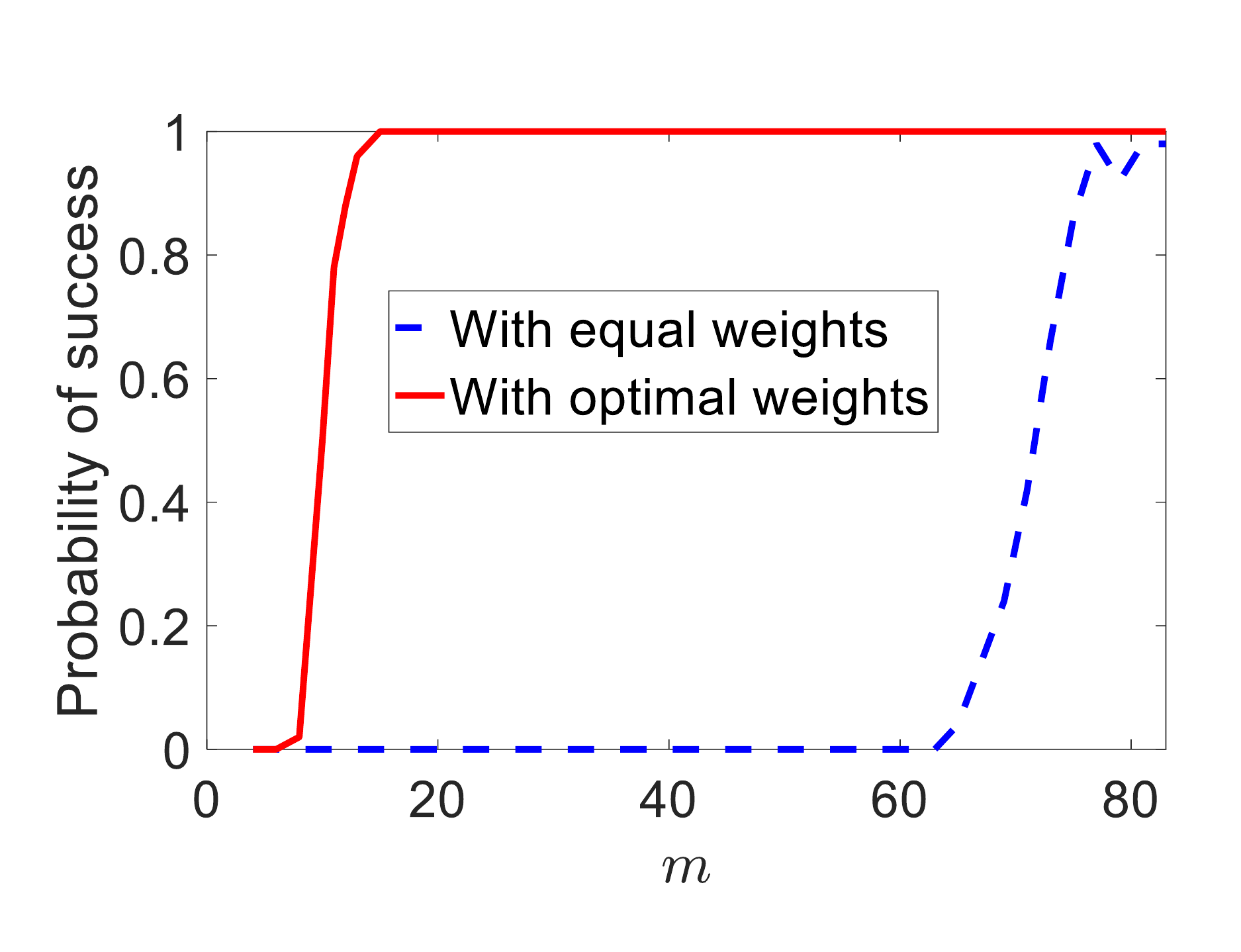}\label{fig.paperfig1}}\quad
		
		\subfigure[]{\includegraphics[width=2.34in]{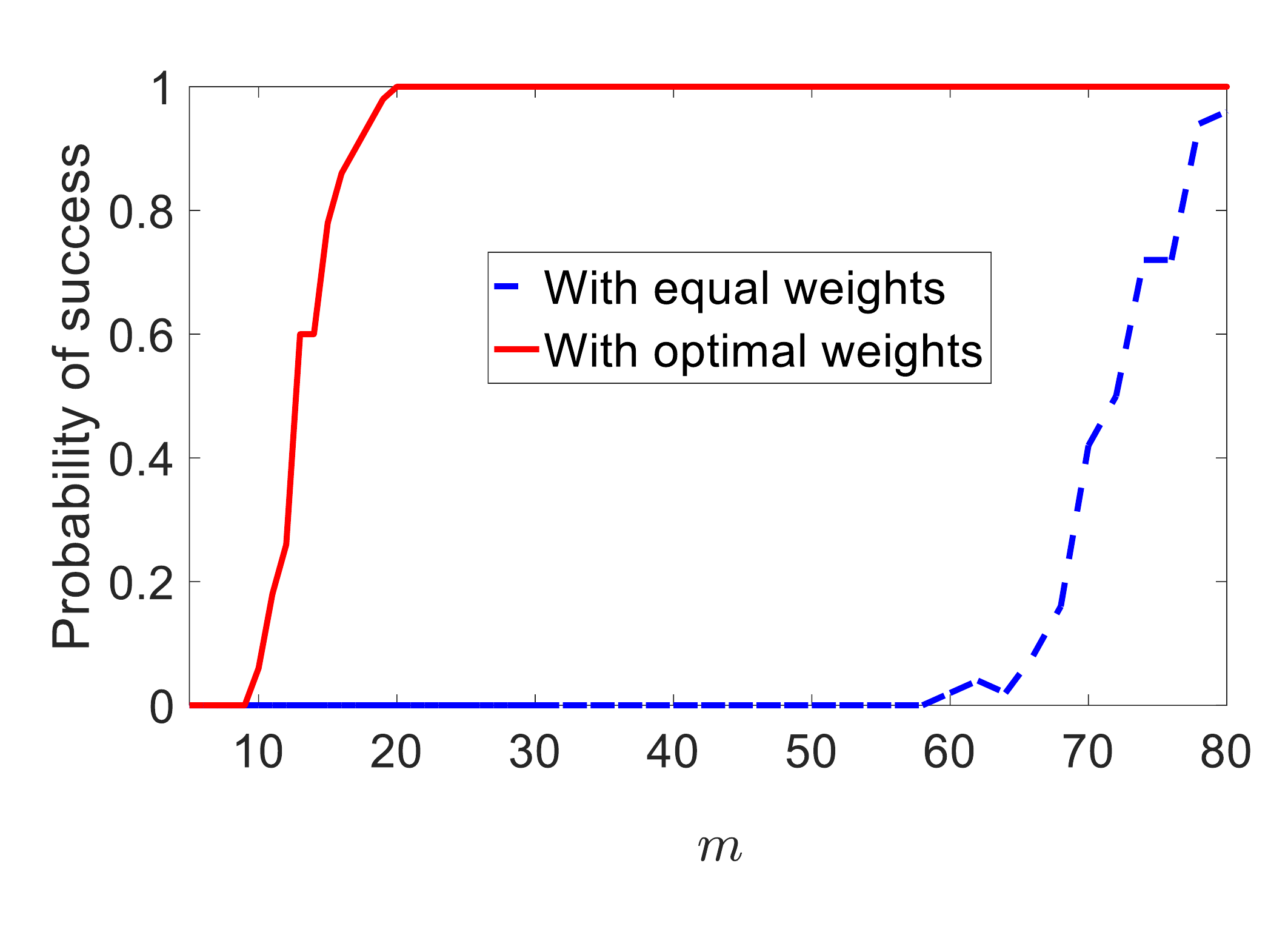}\label{fig.paperfig6}}}
	\mbox{
		\subfigure[]{\includegraphics[width=2.34in]{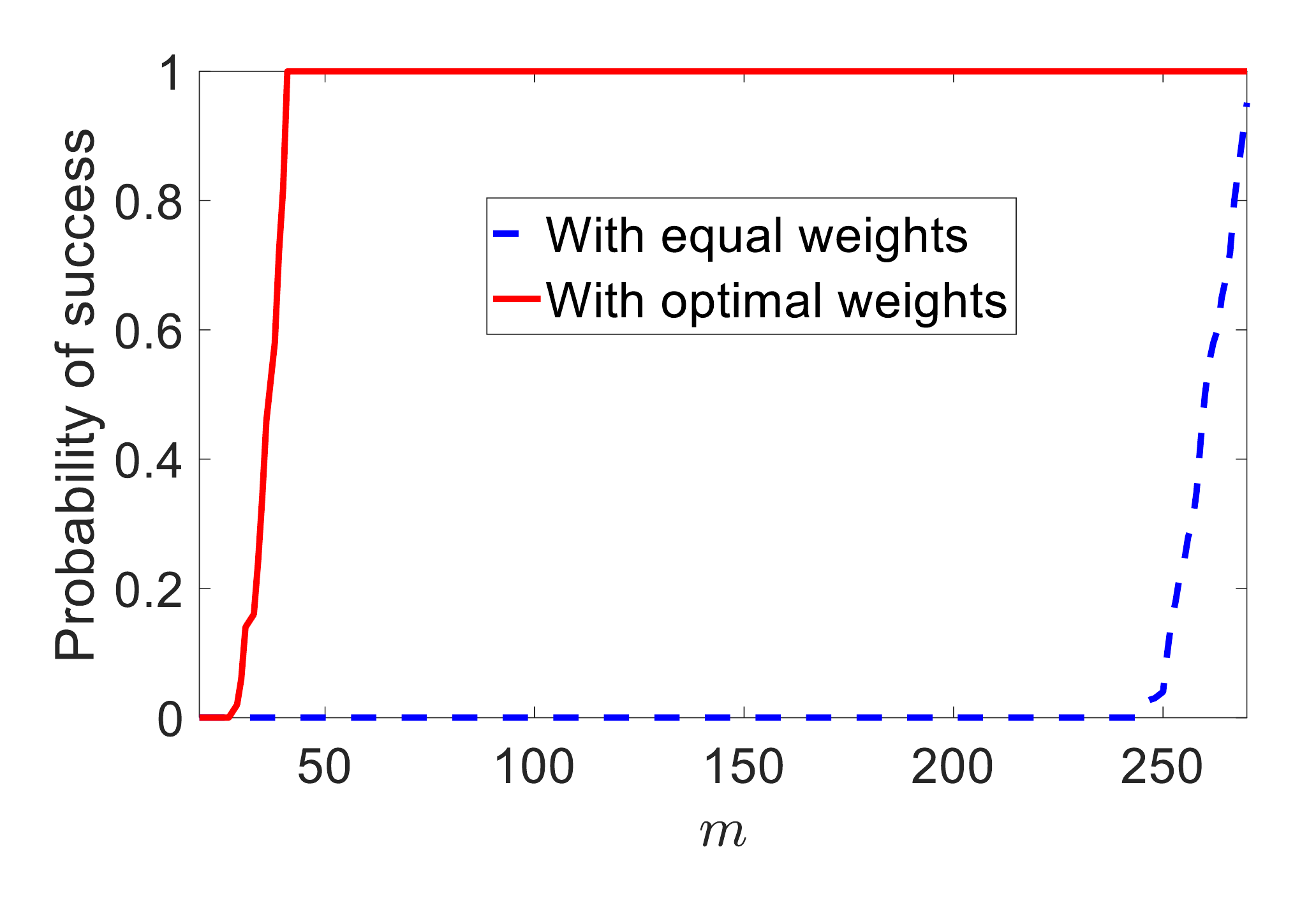}\label{fig.paperfig21}}\quad \subfigure[]{\includegraphics[width=2.34in]{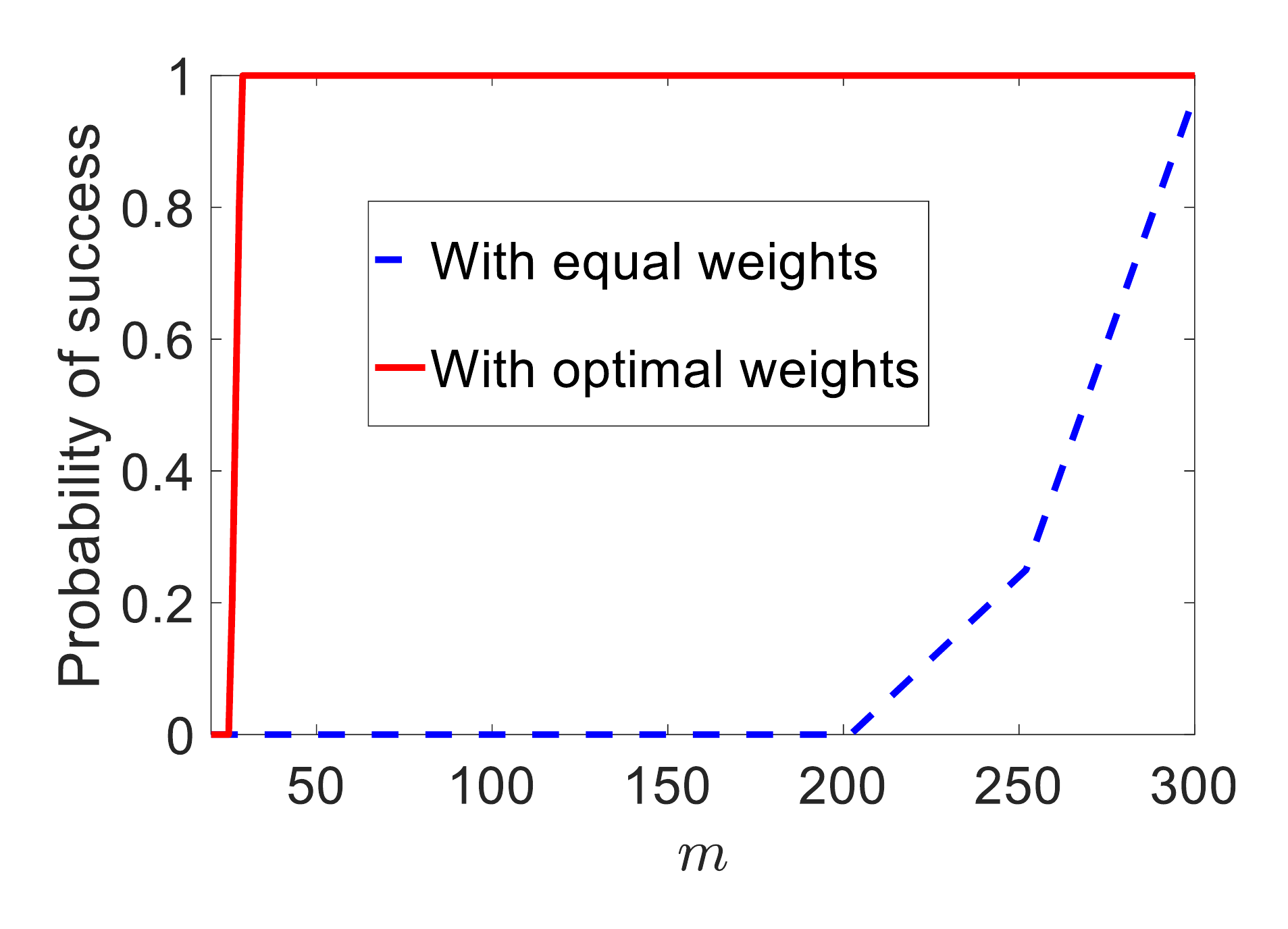}\label{fig.paperfig24}}}
	\caption{The effect of optimal weighting strategy in $\mathsf{P}_{\bm{w},{\rm nuc}}$ when $\widetilde{\bm{\mathcal{U}}}$ and $\widetilde{\bm{\mathcal{V}}}$ have almost the same accuracy and are very close to $\bm{\mathcal{U}}$ and $\bm{\mathcal{V}}$, respectively.  \subref{fig.paperfig2}$r'=r=3$, $n=10$. The principal angles are $\bm{\theta}_u=[0.0196,0.0156,0.005]^T,\bm{\theta}_v=[0.0258,0.0146,0.0098]^T$. The calculated optimal weights are equal to $w_1^*=4.8808\times 10^{-4},w_2^*=0.0907, w_3^*=0.1002, w_4^*=18.6213$. \protect \subref{fig.paperfig1} $r'=r=3$, $n=10$. The principal angles and optimal weights are equal to $\bm{\theta}_u=[0.1858,0.1426,0.0742]^T,\bm{\theta}_v=[0.205,0.1374,0.0878]^T$ $w_1^*=1.1487\times 10^{-4},w_2^*=0.0366, w_3^*=0.0398, w_4^*=12.6870$. \protect \subref{fig.paperfig6} $r'=r=3$, $n=10$. The principal angles are $\bm{\theta}_u=[0.2636,0.1592,0.0281]^T,\bm{\theta}_v=[0.3212,0.1438,0.0470]^T$ and the optimal weights are equal to $w_1^*=0.013,w_2^*=0.4596, w_3^*=0.4917, w_4^*=17.3836$. \protect \subref{fig.paperfig21} $r'=r=5$, $n=20$. The principal angles are $\bm{\theta}_u=[0.3236,0.2660,0.2465,0.2104,0.135]^T,\bm{\theta}_v=[0.2836,0.2667,0.2512,0.1917,0.1703]^T$ and the calculated optimal weights are equal to $w_1^*=0.0008, w_2^*=0.1305, w_3^*=0.1232, w_4^*=19.5313$. \protect \subref{fig.paperfig24} $r'=r=5$, $n=10$. The principal angles are $\bm{\theta}_u=[0.0295,0.024,0.0156,0.0147,0.0108]^T,\bm{\theta}_v=[0.2996,0.2635,0.2346,0.1656,0.1475]^T$ and the calculated optimal weights are equal to $w_1^*=0.0001, w_2^*=0.0357, w_3^*=0.0977, w_4^*=28.6213$.} \label{fig.verygoodprior}
\end{figure*}

\begin{figure*}[t]
	\centering
	\mbox{\subfigure[]{\includegraphics[width=2.34in]{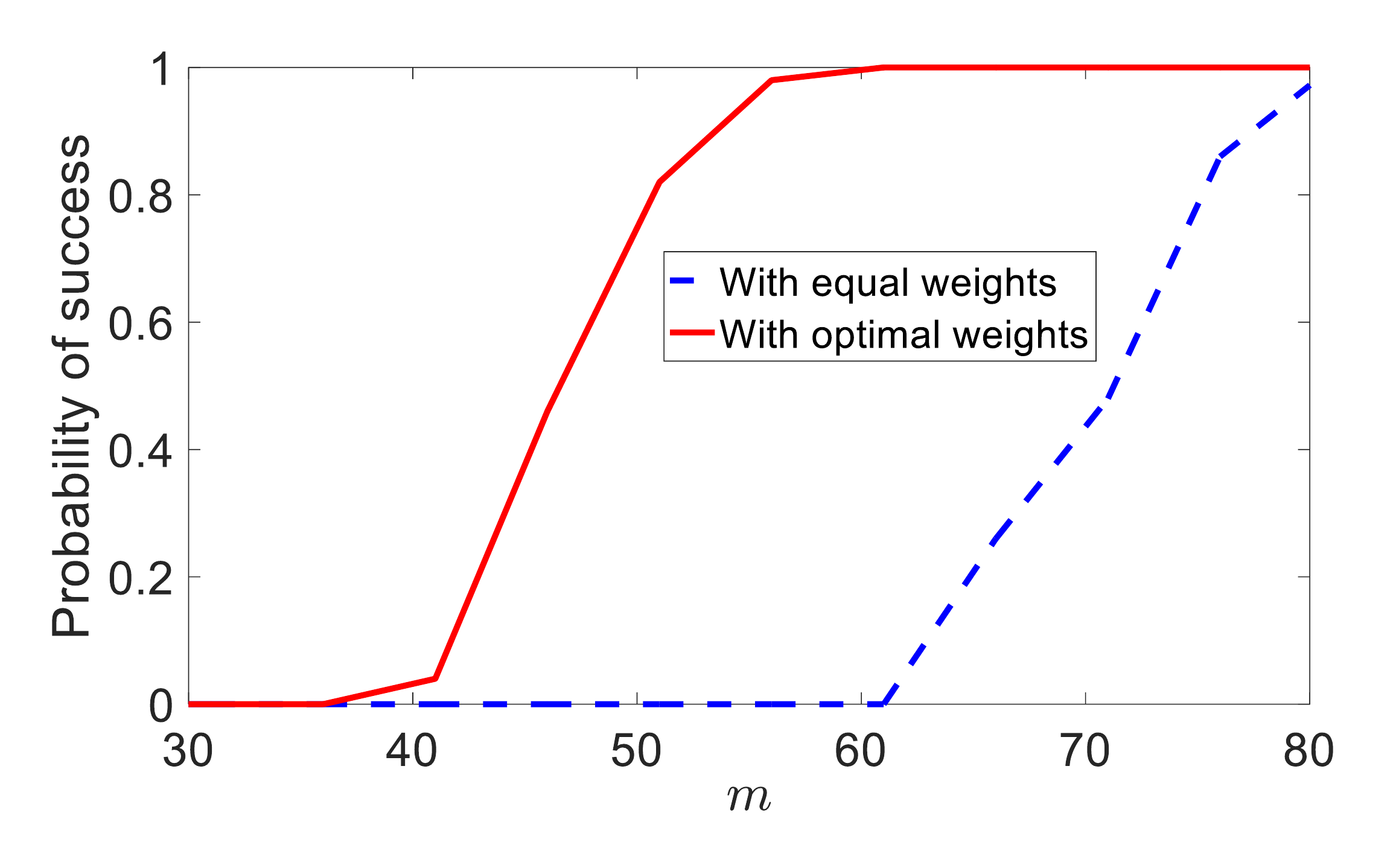}\label{fig.paperfig12}}\quad
		\subfigure[]{\includegraphics[width=2.34in]{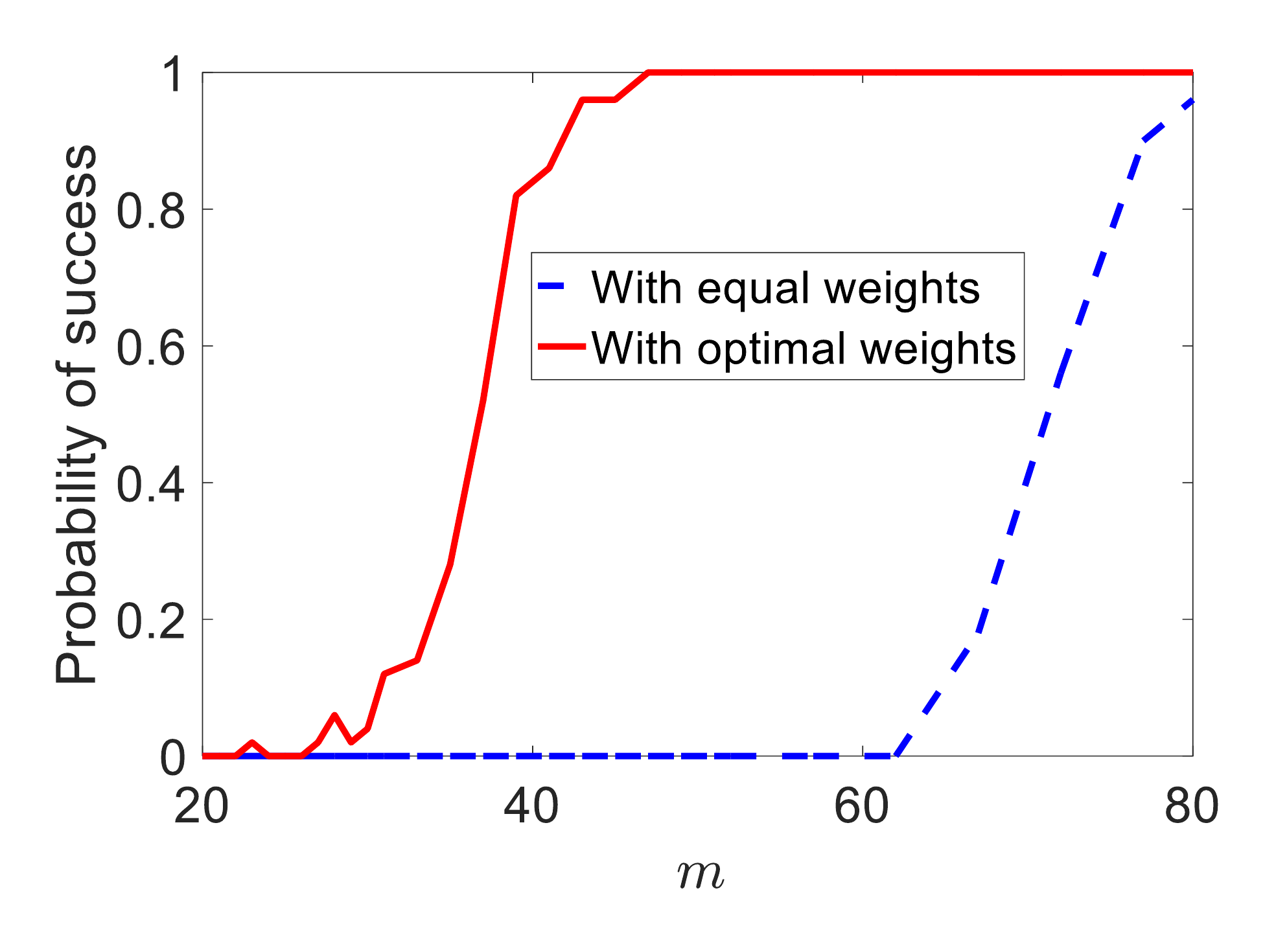}\label{fig.paperfig16}}\quad
		
		\subfigure[]{\includegraphics[width=2.34in]{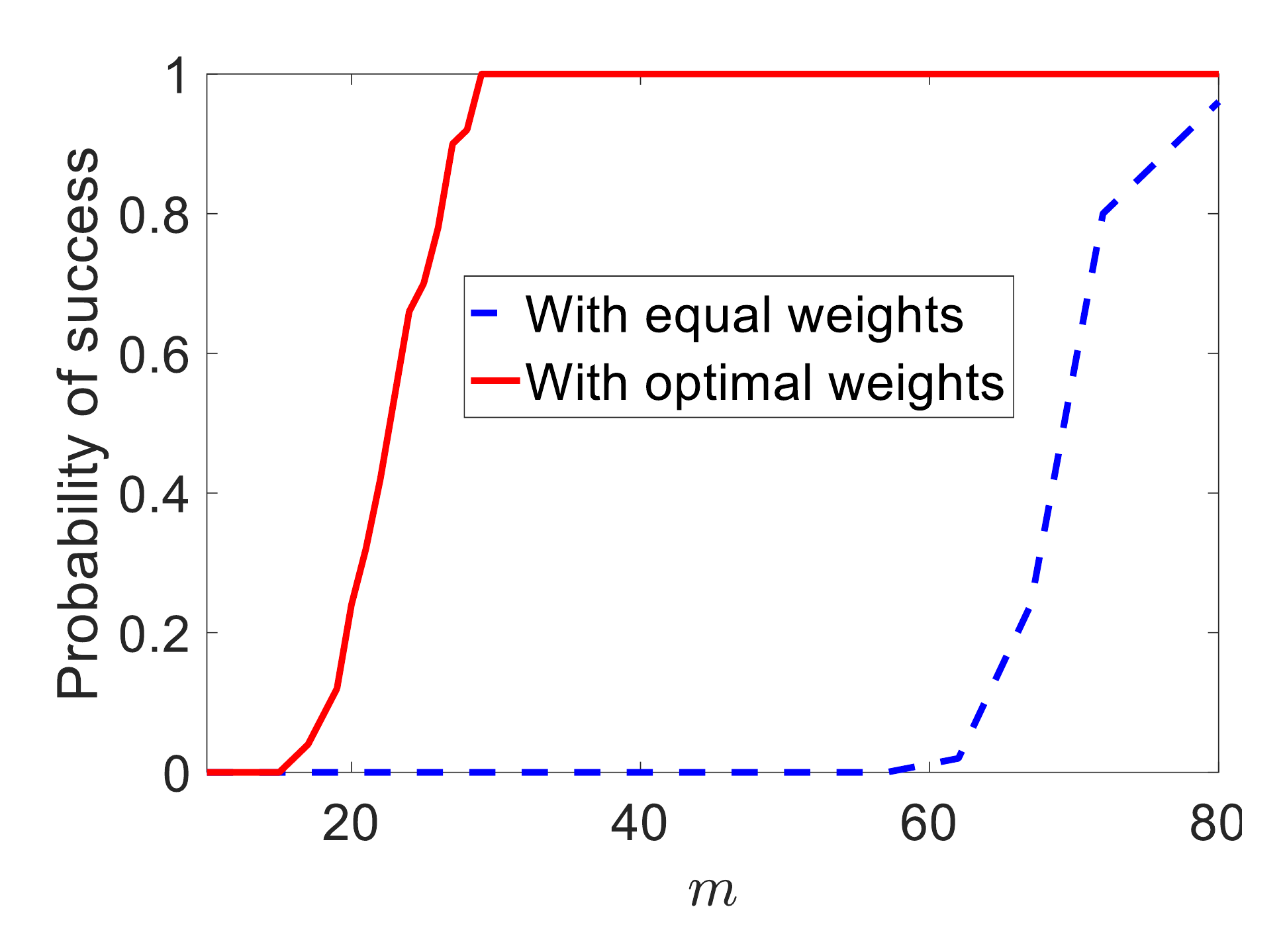}\label{fig.paperfig17}}}
	\caption{The effect of optimal weighting strategy in $\mathsf{P}_{\bm{w},{\rm nuc}}$ when $\widetilde{\bm{\mathcal{U}}}$ and $\widetilde{\bm{\mathcal{V}}}$ have almost the same accuracy and are closer to $\bm{\mathcal{U}}^\perp$ and $\bm{\mathcal{V}}^\perp$ than $\bm{\mathcal{U}}$ and $\bm{\mathcal{V}}$, respectively. \subref{fig.paperfig12} $r'=r=3$, $n=10$. The principal angles and the obtained optimal weights are $\bm{\theta}_u=[89.9832,89.9205,89.8863]^T,\bm{\theta}_v=[89.9845,89.9435,89.9391]^T$ and $w_1^*=91.0229,w_2^*=7.7079, w_3^*=8.0082, w_4^*=0.6781$, respectively. \protect \subref{fig.paperfig16} $r'=7$, $r=3$, $n=10$. The principal angles are $\bm{\theta}_u=[89.8961,89.8351,89.8095]^T,\bm{\theta}_v=[89.8671,89.8480,89.8273]^T$. The optimal weights are $w_1^*=71.003, w_2^*=7.2299, w_3^*=7.2539, w_4^*=0.7386$.  \protect \subref{fig.paperfig17} $r'=7$, $r=3$, $n=10$. The principal angles are $\bm{\theta}_u=[89.9926,89.9872,89.9835]^T,\bm{\theta}_v=[89.9932,89.9892,89.9864]^T$ while  the optimal weights are equal to $w_1^*=109.9399, w_2^*=8.445, w_3^*=8.478, w_4^*=0.6512$.} \label{fig.verygoodfor_perp}
\end{figure*}

In Figure \ref{fig.verygoodprior}, we tested some cases of excellent prior subspace information in which $\widetilde{\bm{\mathcal{U}}}$ and $\widetilde{\bm{\mathcal{V}}}$ are slightly diverged from $\bm{\mathcal{U}}$ and $\bm{\mathcal{V}}$. Also, we set the deviation level of column and row subspaces roughly the same. From Figures \ref{fig.paperfig2}--\ref{fig.paperfig24}, it is observed that the required sample complexity of $\mathsf{P}_{\bm{w},{\rm nuc}}$ reaches the optimal number of measurements i.e. $r^2$. Besides, its sample complexity is far from that in $\mathsf{P}_{{\rm nuc}}$. 
In Figure \ref{fig.verygoodfor_perp}, $\widetilde{\bm{{\mathcal{U}}}}$ and $\widetilde{\bm{{\mathcal{V}}}}$, are close to $\bm{\mathcal{U}}^\perp$ and $\bm{\mathcal{V}}^\perp$, respectively. Figures \ref{fig.paperfig12}--\ref{fig.paperfig17} show that even when $\widetilde{\bm{{\mathcal{U}}}}$ and $\widetilde{\bm{{\mathcal{V}}}}$ are very far from $\bm{\mathcal{U}}$ and $\bm{\mathcal{V}}$, respectively, the reduction of sample complexity is possible. It is worth mentioning that one can also hope to reach the optimal number of measurements when there exists a subspace with dimension $r'=n-r$ that is very close to $\bm{\mathcal{U}}^\perp$. This case can be observed in Figure \ref{fig.paperfig17}.

\begin{figure*}[t]
	\centering
	\mbox{\subfigure[]{\includegraphics[width=2.34in]{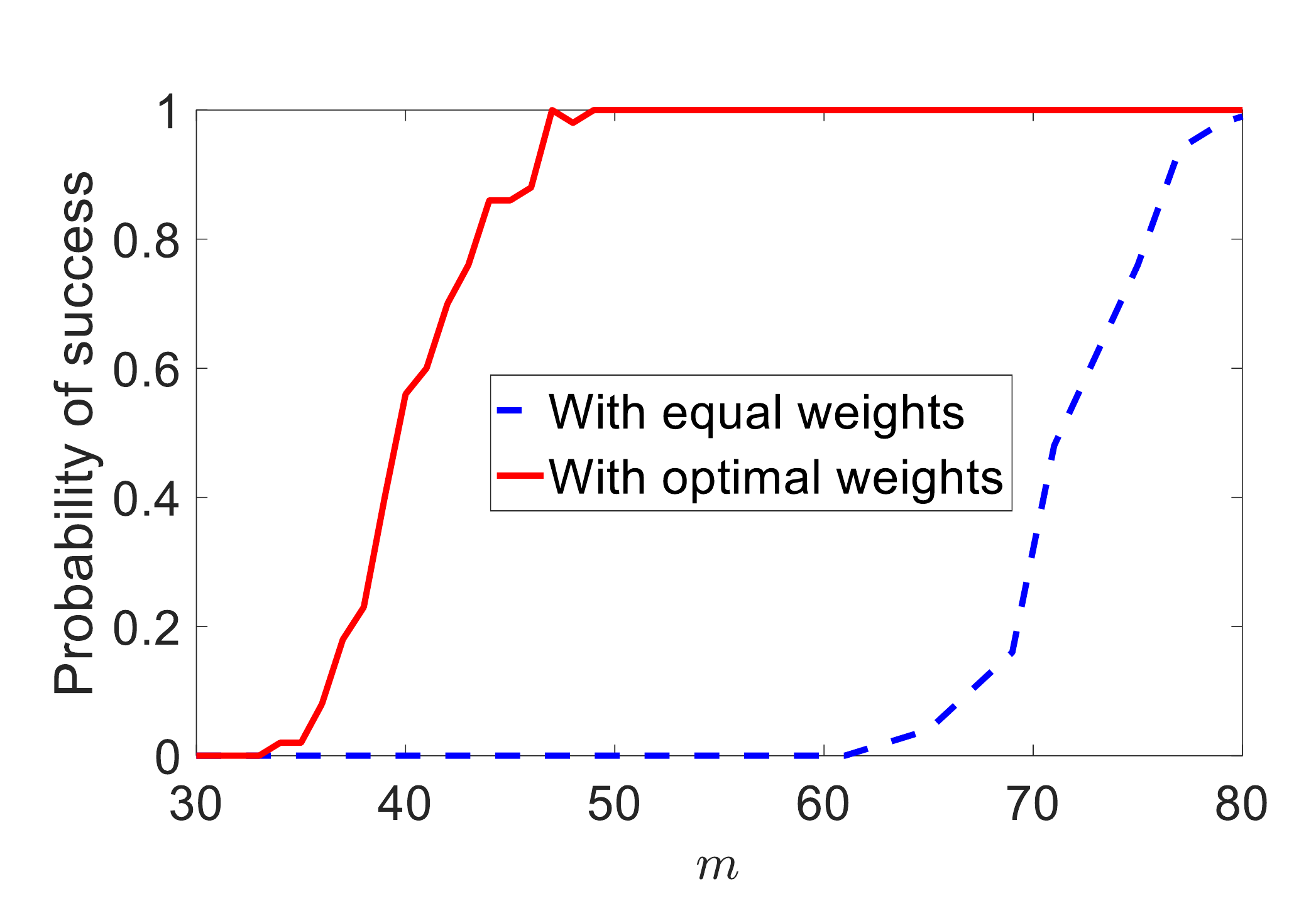}\label{fig.paperfig3}}\quad
		\subfigure[]{\includegraphics[width=2.34in]{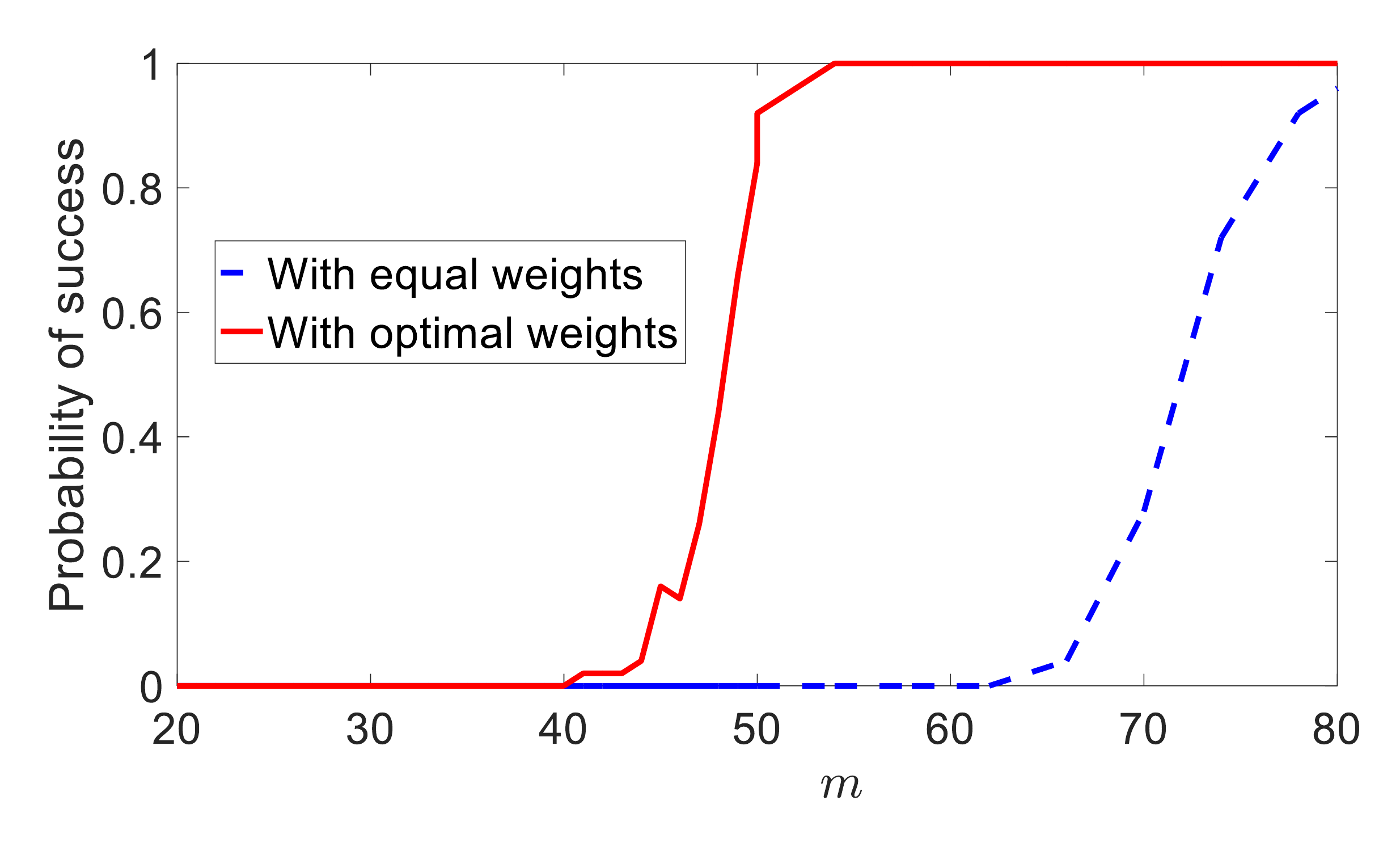}\label{fig.paperfig7}}\quad
		
		\subfigure[]{\includegraphics[width=2.34in]{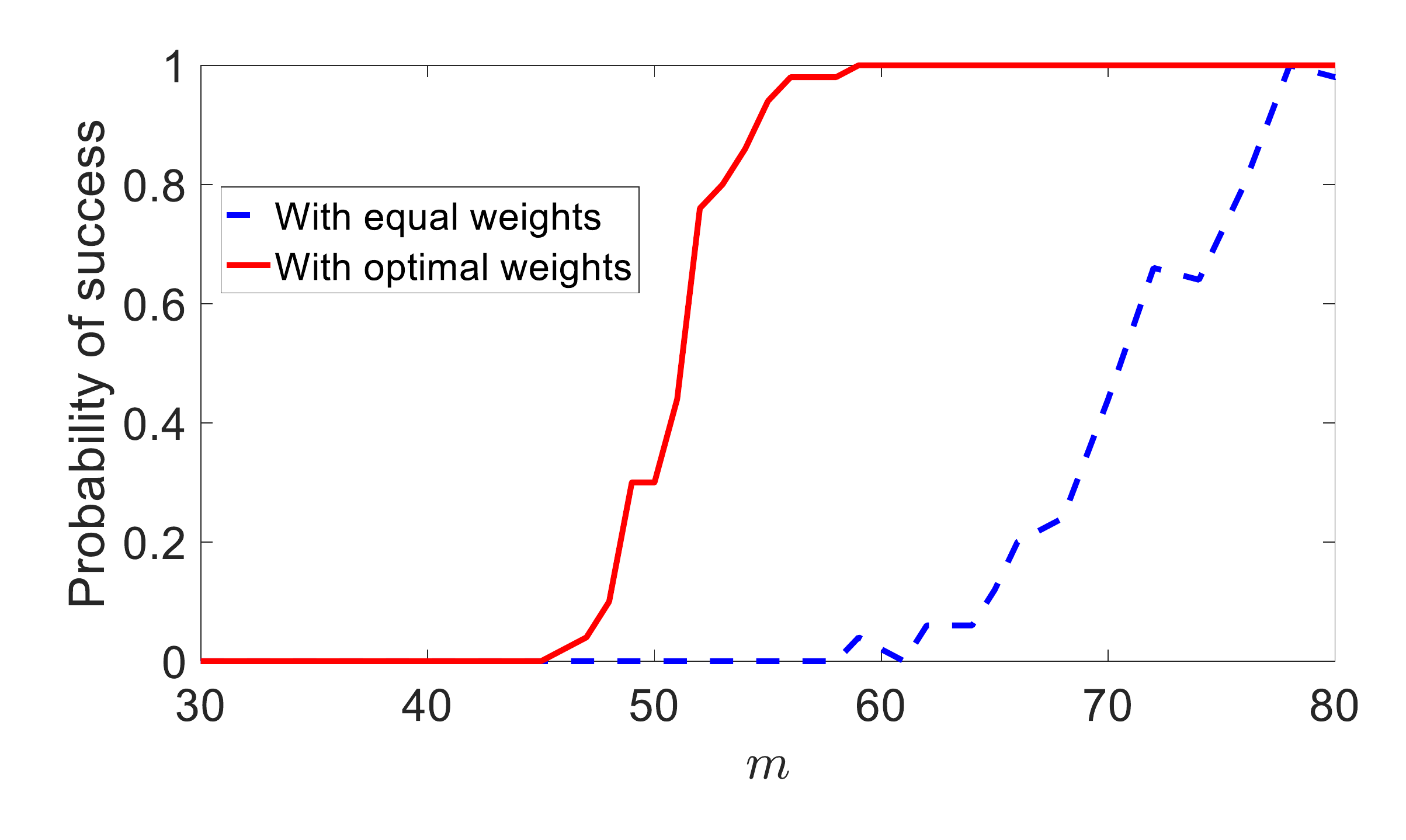}\label{fig.paperfig8}}}
	\mbox{
		\subfigure[]{\includegraphics[width=2.34in]{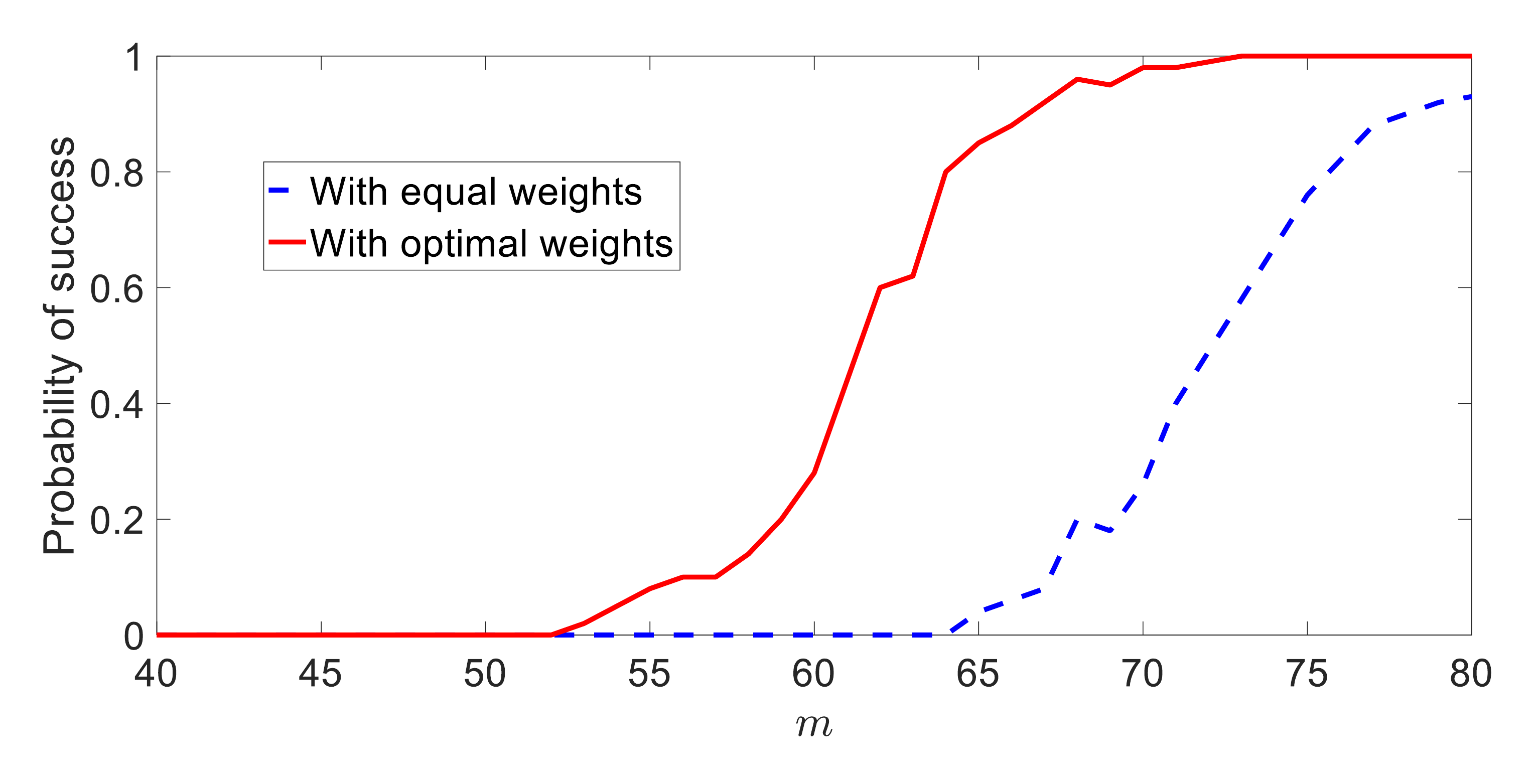}\label{fig.paperfig10}}\quad
		\subfigure[]{\includegraphics[width=2.34in]{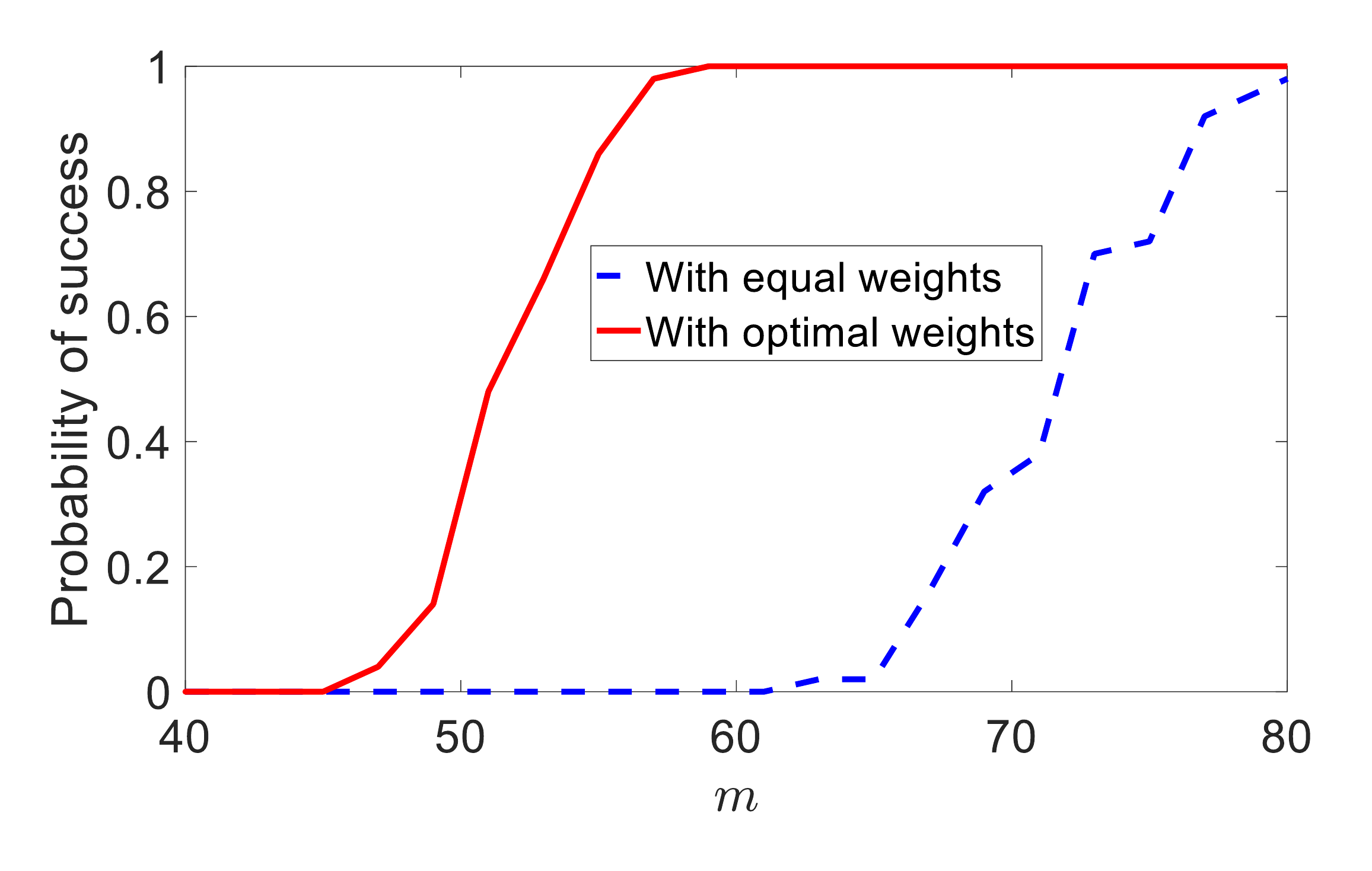}\label{fig.paperfig13}}}
	\caption{The effect of optimal weighting strategy in $\mathsf{P}_{\bm{w},{\rm nuc}}$ when $\widetilde{\bm{\mathcal{U}}}$ and $\widetilde{\bm{\mathcal{V}}}$ are closer to $\bm{\mathcal{U}}$ and $\bm{\mathcal{V}}$ than $\bm{\mathcal{U}}^\perp$ and $\bm{\mathcal{V}}^\perp$, respectively Also, the accuracies of $\widetilde{\bm{\mathcal{U}}}$ and $\widetilde{\bm{\mathcal{V}}}$ are almost equal. \subref{fig.paperfig3} $r'=r=3$, $n=10$. The principal angles are $\bm{\theta}_u=[2.1069,1.5826,0.9226]^T,\bm{\theta}_v=[1.6620,1.0637,0.7858]^T$. The calculated optimal weights are $w_1^*=0.0112,w_2^*=0.3625, w_3^*=0.3257, w_4^*=10.5095$. \protect \subref{fig.paperfig7} $r'=r=3$, $n=10$. The principal angles are $\bm{\theta}_u=[2.7698,1.5071,1.3792]^T,\bm{\theta}_v=[2.4161,1.0778,0.4847]^T$. The calculated optimal weights are $w_1^*=0.0134,w_2^*=0.4132, w_3^*=0.3539, w_4^*=10.9387$. \protect \subref{fig.paperfig8} $r'=r=3$, $n=10$. The principal angles are $\bm{\theta}_u=[21.3380,6.2792,3.5496]^T,\bm{\theta}_v=[16.0732,6.0245,2.5172]^T$. The optimal weights are $w_1^*=0.4832,w_2^*=1.6806, w_3^*=1.6169, w_4^*=5.6237$. \protect \subref{fig.paperfig10} $r'=r=3$, $n=10$. The principal angles are $\bm{\theta}_u=[32.0793,16.3673,11.5502]^T,\bm{\theta}_v=[25.0806,14.4304,6.7770]^T$. The optimal weights are $w_1^*=0.5842,w_2^*=1.6130, w_3^*=1.4967, w_4^*=4.1325$. \protect \subref{fig.paperfig13} $r'=r=3$, $n=10$. The principal angles are $\bm{\theta}_u=[2.0528,1.1229,0.8021]^T,\bm{\theta}_v=[1.4690,0.4807,0.2612]^T$ and the optimal weights are $w_1^*=0.0608,w_2^*=0.9920, w_3^*=0.7907, w_4^*=12.8981$.
	} \label{fig.notgoodnotbad}
\end{figure*}

In Figure \ref{fig.notgoodnotbad}, we test a scenario where the principal angles are not so small but less than $45^\circ$. One can see from Figures \ref{fig.paperfig3}--\ref{fig.paperfig13} that as much as the principal angles get less, more reduction is achievable in the required sample complexity.

\begin{figure*}[t]
	\centering
	\mbox{\subfigure[]{\includegraphics[width=2.34in]{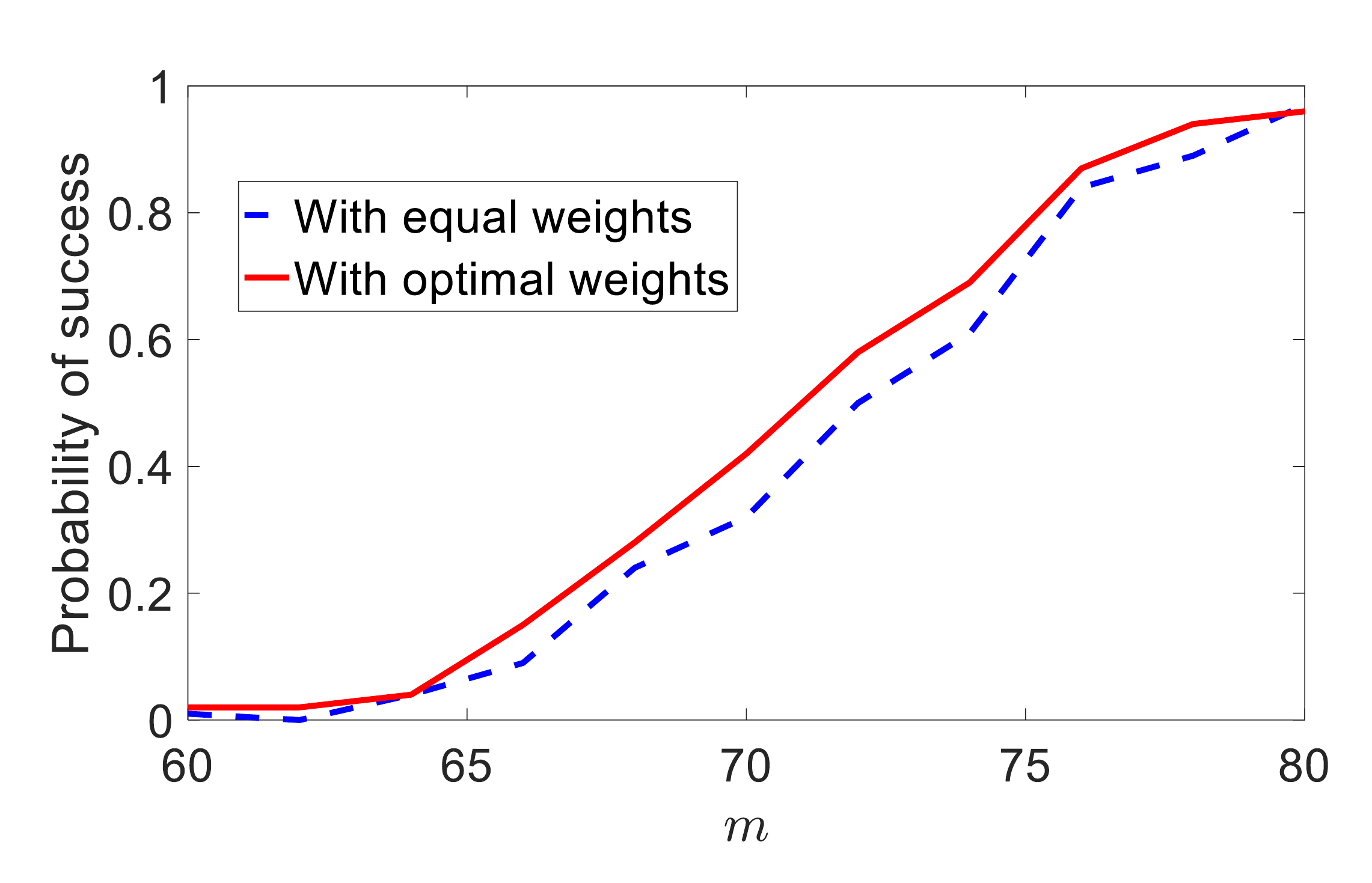}\label{fig.paperfig5}}\quad
		\subfigure[]{\includegraphics[width=2.34in]{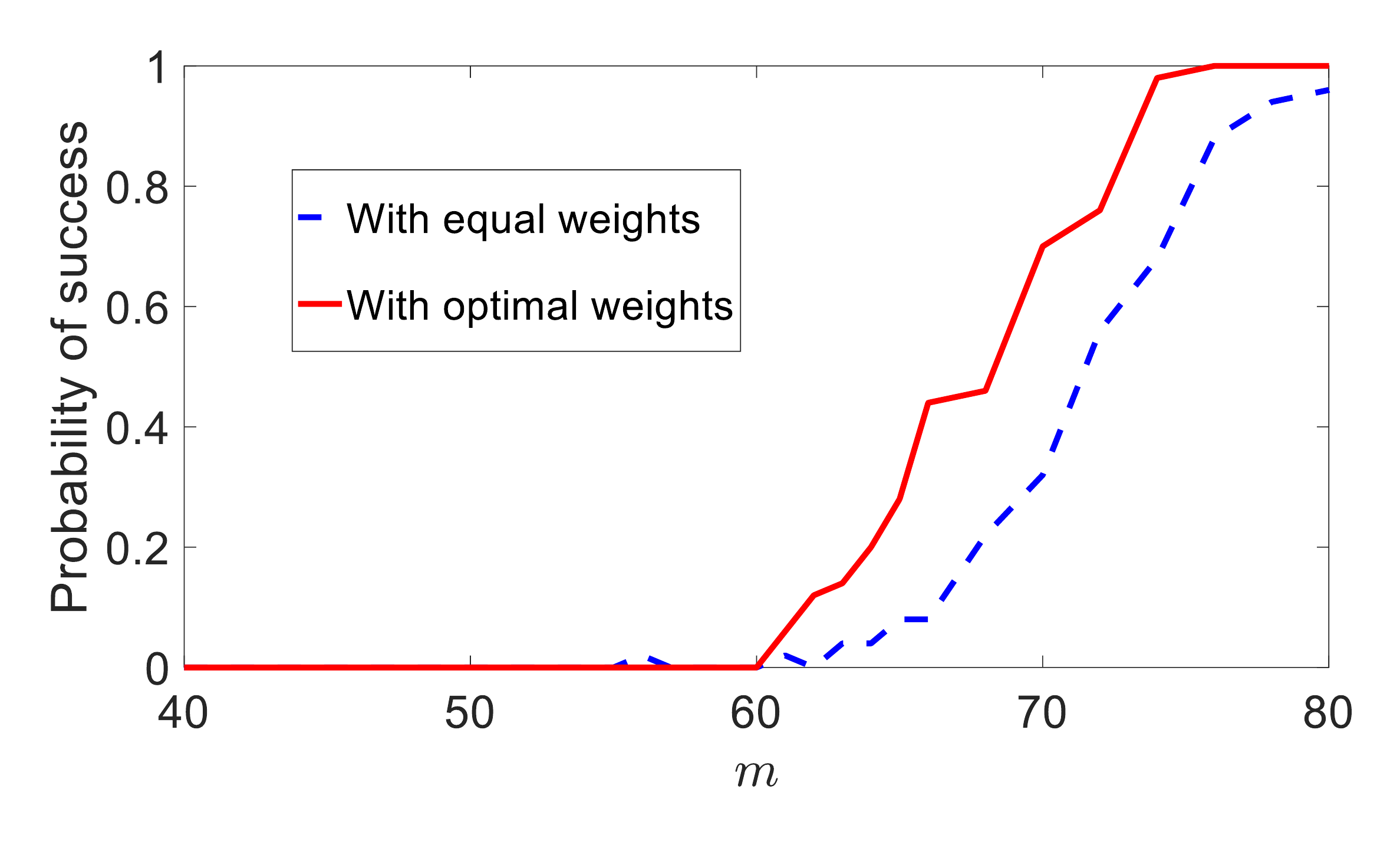}\label{fig.paperfig9}}\quad
		
		\subfigure[]{\includegraphics[width=2.34in]{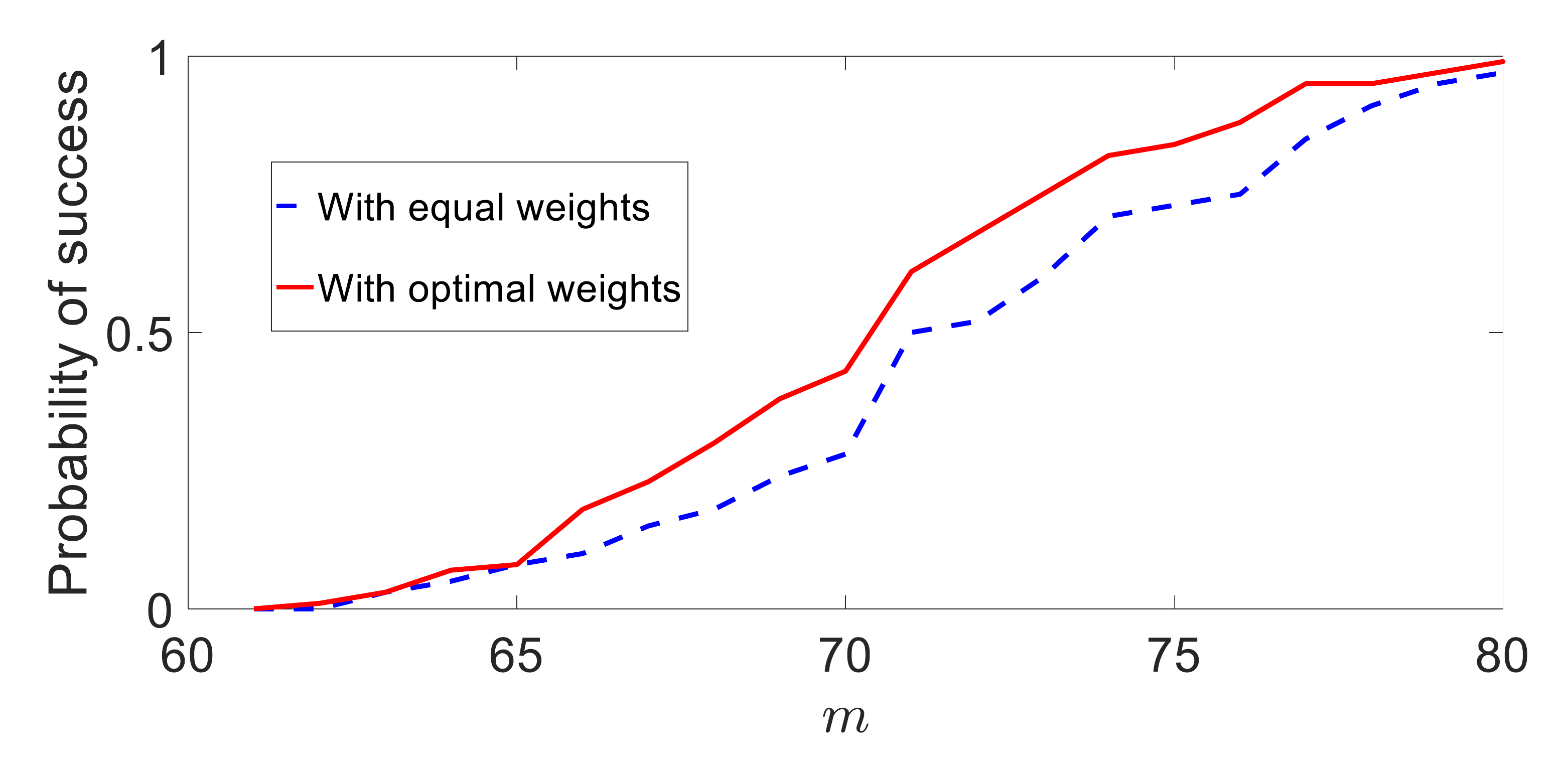}\label{fig.paperfig11}}}
	\mbox{
		\subfigure[]{\includegraphics[width=2.34in]{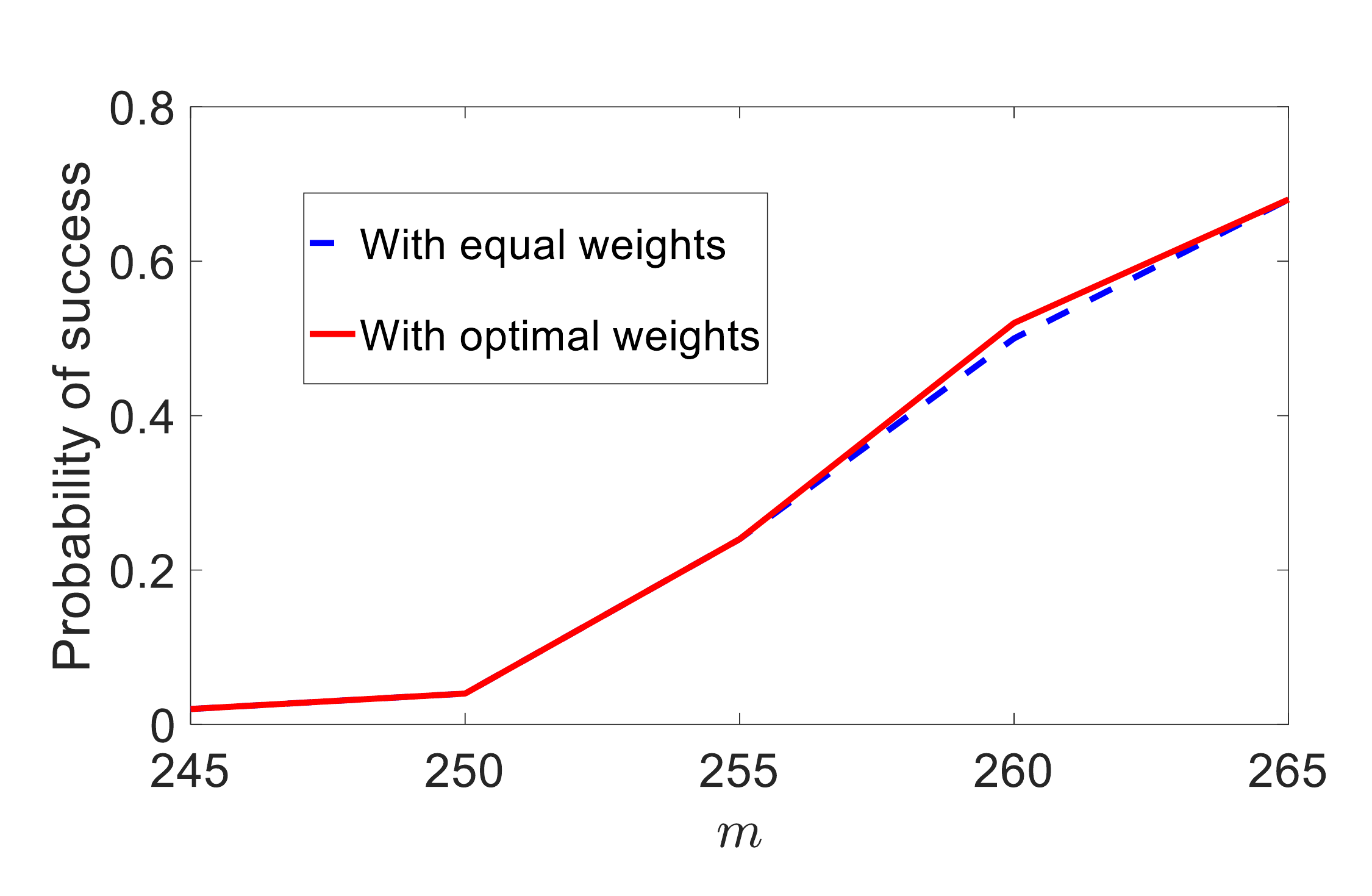}\label{fig.paperfig19}}\quad
	}
	\caption{The effect of optimal weighting strategy in $\mathsf{P}_{\bm{w},{\rm nuc}}$ in case of weak and almost equal accuracies of $\widetilde{\bm{\mathcal{U}}}$ and $\widetilde{\bm{\mathcal{V}}}$.  \subref{fig.paperfig5} $r'=r=3$, $n=10$. The principal angles are $\bm{\theta}_u=[81.7481,54.6846,40.0055]^T,\bm{\theta}_v=[88.8679,79.7605, 60.8808]^T$. The calculated optimal weights are $w_1^*=2.0781,w_2^*=1.3021, w_3^*=2.5499, w_4^*=1.5976$. \protect \subref{fig.paperfig9} $r'=r=3$, $n=10$. The principal angles are $\bm{\theta}_u=[76.6703,15.0108,5.9896]^T,\bm{\theta}_v=[89.6091,12.5393,5.0507]^T$. The corresponding optimal weights are $w_1^*=0.6691,w_2^*=1.4492, w_3^*=1.3059, w_4^*=2.8285$. \protect \subref{fig.paperfig11} $r'=r=3$, $n=10$. The principal angles are $\bm{\theta}_u=[89.4788,72.0998,42.8150]^T,\bm{\theta}_v=[88.5496,84.0146,58.8733]^T$. The optimal weights are $w_1^*=2.6334,w_2^*=1.7578, w_3^*=2.2170, w_4^*=1.4799$. \protect \subref{fig.paperfig19} $r'=r=5$, $n=20$. The principal angles are $\bm{\theta}_u=[74.75,68.0787,65.8337,56.3507,52.5944]^T, \bm{\theta}_v=[89.2984,73.4526,62.7018,55.48,46.3011]^T$. The optimal weights are $w_1^*=2.9837, w_2^*=2.9356, w_3^*=2.9153, w_4^*=2.8683$.} \label{fig.weakprior}
\end{figure*}

In Figure \ref{fig.weakprior}, optimal weighting strategy is investigated when there exists weak prior subspace information about the column and row space of $\bm{X}$. By \textit{weak prior}, we mean a case that $\widetilde{\bm{\mathcal{U}}}$ and $\widetilde{\bm{\mathcal{V}}}$ are almost as close to $\bm{\mathcal{U}}$ and $\bm{\mathcal{V}}$ as they are to $\bm{\mathcal{U}}^\perp$ and $\bm{\mathcal{V}}^\perp$. In these cases, (see Figures \ref{fig.paperfig5}--\ref{fig.paperfig19}) the sample complexity of our algorithm approaches the one in $\mathsf{P}_{\rm nuc}$.

\begin{figure*}[t]
	\centering
	\mbox{\subfigure[]{\includegraphics[width=2.34in]{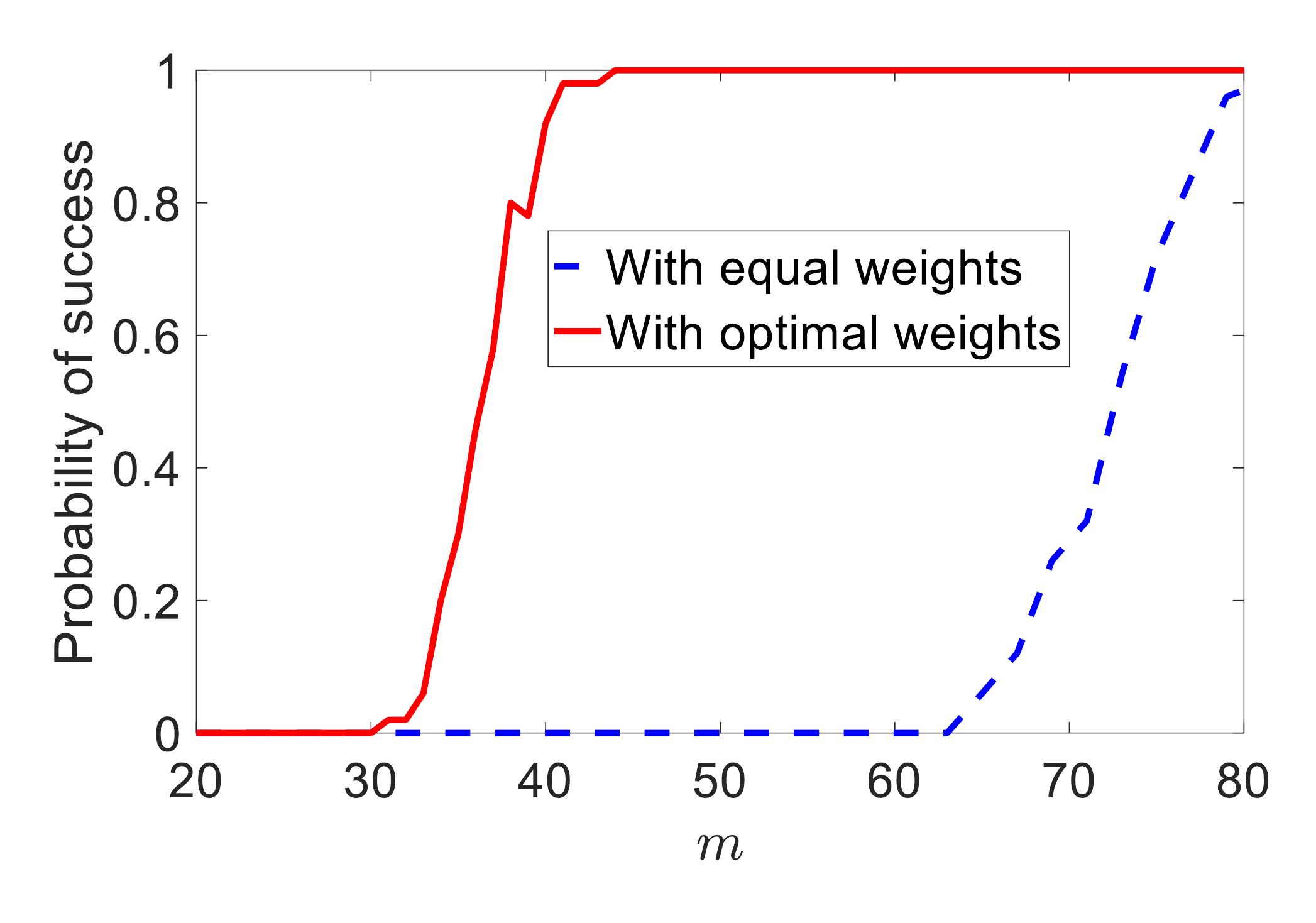}\label{fig.paperfig14}}\quad
		\subfigure[]{\includegraphics[width=2.34in]{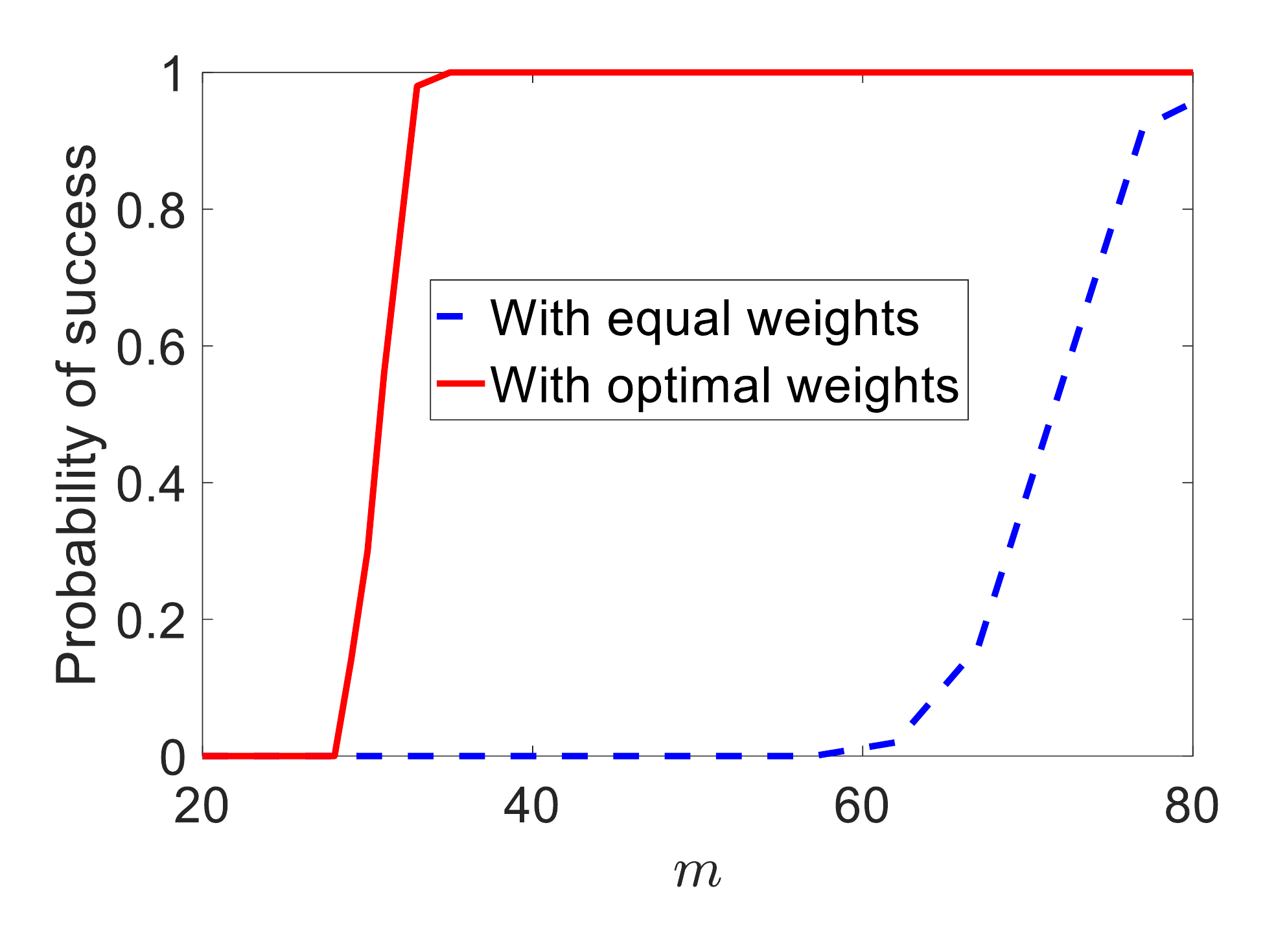}\label{fig.paperfig15}}\quad
		
			\subfigure[]{\includegraphics[width=2.34in]{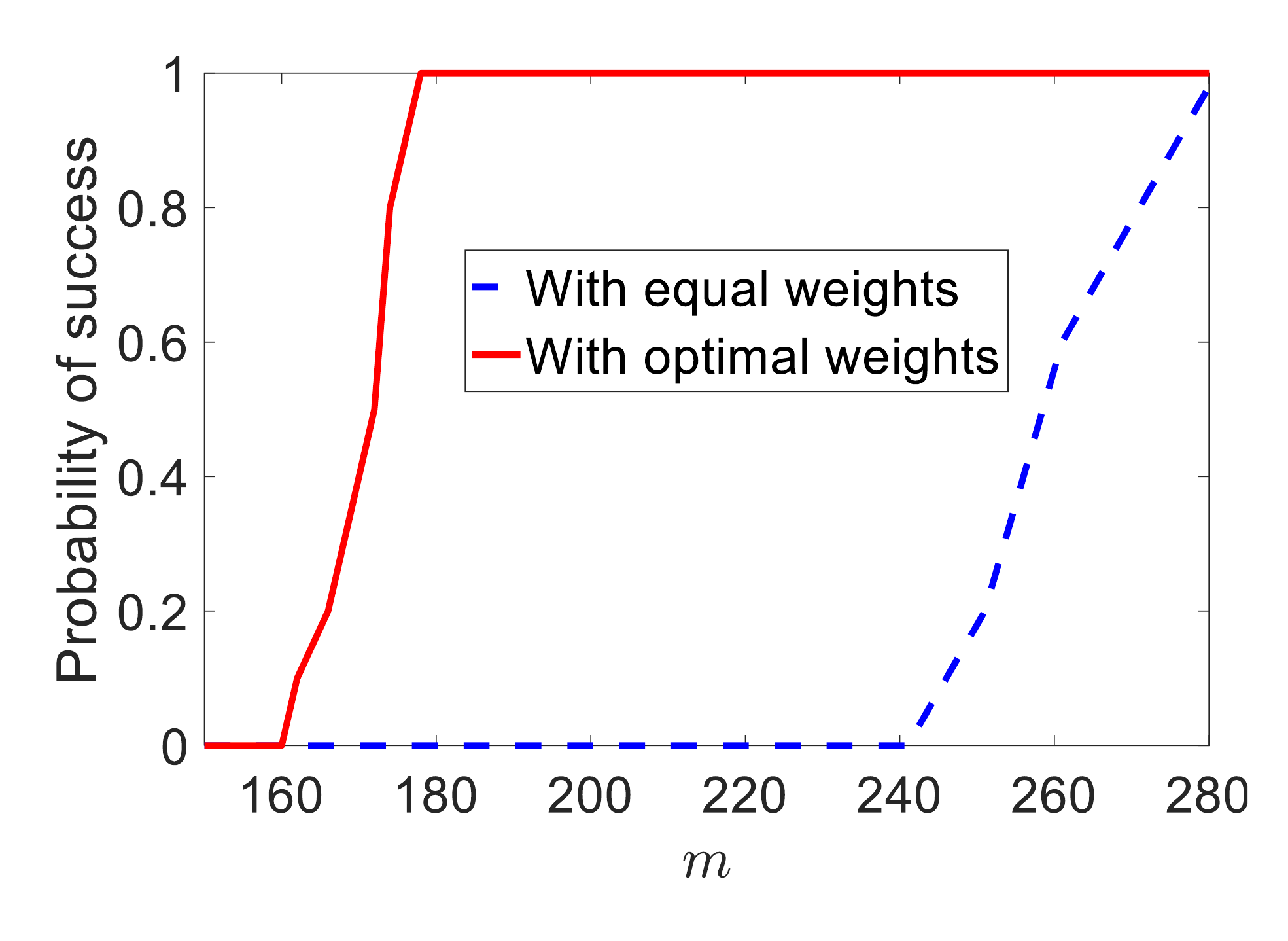}\label{fig.paperfig25}}
	}
	
		\mbox{
			\subfigure[]{\includegraphics[width=2.34in]{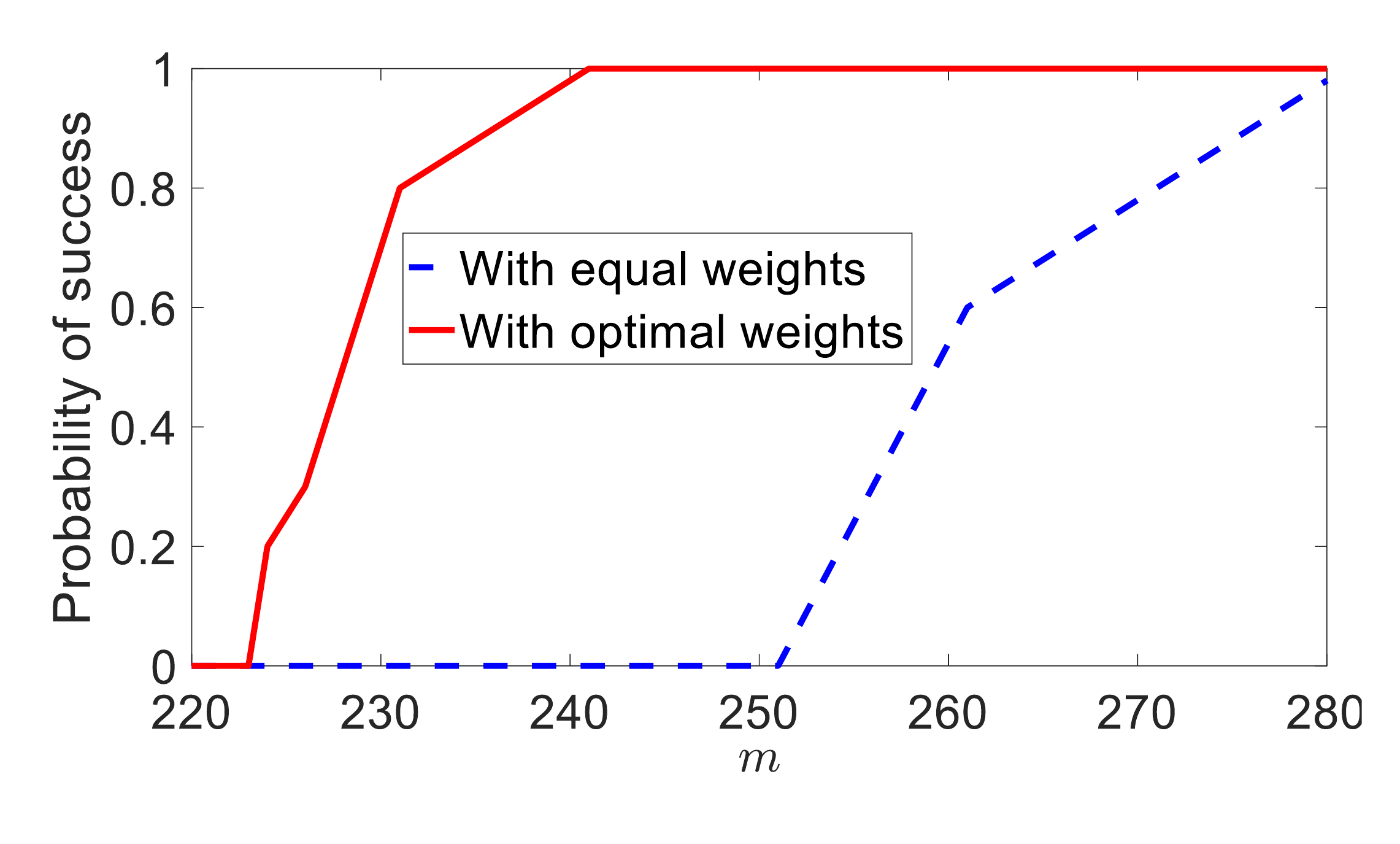}\label{fig.paperfig26}}\quad
		}
	\caption{The effect of optimal weighting strategy in $\mathsf{P}_{\bm{w},{\rm nuc}}$ when the accuracy of $\widetilde{\bm{\mathcal{U}}}$ and $\widetilde{\bm{\mathcal{V}}}$ are different.  \subref{fig.paperfig14} $r'=r=3$, $n=10$. The principal angles are $\bm{\theta}_u=[0.2094,0.1374,0.0668]^T,\bm{\theta}_v=[77.0117,53.6449,47.4287]^T$ and the optimal weights are $w_1^*=0.5910,w_2^*=0.4501, w_3^*=12.1409, w_4^*=9.2470$. \protect \subref{fig.paperfig15} $r'=r=3$, $n=10$. The principal angles are $\bm{\theta}_u=[0.0210, 0.0137,0.0067]^T,\bm{\theta}_v=[19.8050,13.4433,12.8398]^T$. The optimal weights are equal to $w_1^*=0.1078, w_2^*=0.2473, w_3^*=7.5826, w_4^*=17.3987$.
		\protect \subref{fig.paperfig25} $r'=r=5$, $n=20$. The principal angles are $\bm{\theta}_u=[2.5765,2.5291,1.852,1.6211,1.1702]^T,\bm{\theta}_v=[89.9748,89.6046,89.1707,88.5476,87.9445]^T$. The optimal weights are $w_1^*=5.1241,w_2^*=1.4695, w_3^*=28.0858, w_4^*=8.0542$.
 \protect \subref{fig.paperfig26} $r'=r=5$, $n=20$. The principal angles are $\bm{\theta}_u=[27.4179,26.3887,21.2222,16.8778,9.2861]^T,\bm{\theta}_v=[88.5979,83.8165,81.5085,77.0814,73.7583]^T$. The optimal weights are $w_1^*=3.6265, w_2^*=1.7947, w_3^*=7.7617, w_4^*=3.8412$.
	} \label{fig.differentprior}
\end{figure*}

In the last experiment shown in Figure \ref{fig.differentprior}, we consider the case where  accuracies of $\widetilde{\bm{\mathcal{U}}}$ and $\widetilde{\bm{\mathcal{V}}}$ are different. From Figures \ref{fig.paperfig14}--\ref{fig.paperfig26}, it is observed that a huge sample complexity reduction is feasible when either prior column or row subspace information is close to the respective subspaces of the ground-truth matrix.
\section{Related works and Key differences}\label{sec.relatedworks}
In \cite{srebro2010collaborative}, a non-uniform sampling distribution is considered for a Netflix data set and is shown that a properly weighted trace norm  of the form
\begin{align}
\|\bm{X}\|_{\mathrm{tr}}:=\|\mathrm{diag}(\sqrt{\bm{p}})\bm{X}\mathrm{diag}(\sqrt{\bm{q}})\|_{*},
\end{align}
works well where $p(i), i=1,..., n$ and $q(j), j=1,..., n$ are the probability of observing row $i$ and column $j$ of the matrix.

In \cite{angst2011generalized}, a non-uniform sampling scheme is considered in which the authors propose a generalized nuclear norm which penalizes the directions in the vector space of $\bm{X}\in\mathbb{R}^{n_1\times n_2}$ non-uniformly; namely, allocates larger weights to certain directions than others.

In \cite{aravkin2014fast}, the authors heuristically propose the following optimization problem to exploit prior subspace information:
\begin{align}
&\min_{\bm{Z}\in\mathbb{R}^{n_1\times n_2}}\|(\lambda \bm{P}_{\widetilde{\bm{\mathcal{U}}}}+\bm{P}_{\widetilde{\bm{\mathcal{U}}}^\perp})\bm{Z}(\rho \bm{P}_{\widetilde{\bm{\mathcal{V}}}}+\bm{P}_{\widetilde{\bm{\mathcal{V}}}^\perp})\|_*\nonumber\\
&\mathrm{s.t.}~~\|\bm{y}-\mathcal{A}(\bm{Z})\|_F\le \epsilon,
\end{align}
where $\widetilde{\bm{\mathcal{U}}}$ and $\widetilde{\bm{\mathcal{V}}}$ with dimension $r$ are the estimates of column and row subspaces of the rank $r$ ground-truth matrix $\bm{X}\in\mathbb{R}^{n\times n}$, $\bm{y}=\mathcal{A}(\bm{X})+\bm{e}$, and $\epsilon$ is an upper-bound for $\|\bm{e}\|_2$.
However, they did not answer how to explicitly find $\lambda$ and $\rho$.

In \cite{eftekhari2018weighted}, the authors investigated the same objective function as in \cite{aravkin2014fast}. They showed that the isometry constant for $\mathcal{A}(\cdot)$ can be more conservative and thus the required bound for robust recovery can be lowered provided that the prior subspace information is good ($\theta_u(1), \theta_v(1)<45^\circ$). Only in case of $\theta_u(1)=\theta_v(1)=\theta$, they suggest to choose $\lambda=\rho=\sqrt{\sqrt{{\rm tan}^4(\theta)+{\rm tan}^2(\theta)}-{\rm tan}^2(\theta)}$ so as to maximize the RIP bound. There are some key differences between our work and \cite{eftekhari2018weighted} which are listed below:
\begin{itemize}
\item They assume that the subspace estimate and the ground-truth subspace are of the same dimension $r$. This assumption fails to occur in practical scenarios in some certain settings for example in Netflix problem where a higher dimensional subspace estimate is available to the practitioner (see Subsection \ref{section.applications} for more explanations). In our work, we consider a generalized case where high dimensional row and column subspaces are angled from the row and column subspaces of interest.
\item The meaning of \textit{optimal} in that work differs from ours in that their weights maximize the RIP constants while ours minimize the required sample complexity.
\item \cite{eftekhari2018weighted} considers only the effect of the largest principal angle on the performance bounds while in fact all principal angles directly affect the performance bounds.
\item The measurement bound in \cite{eftekhari2018weighted} depends on  $\mathcal{A}(\cdot)$ while our bound is independent of the sampling operator. 
\item  There is a wide range of principal angles ($45^\circ\le\theta_u\le 90^\circ$) for which no improvement is predicted in \cite{eftekhari2018weighted}, inevitably reaching the performance bound of $\mathsf{P}_{\rm nuc}$. The only exception that our algorithm reaches the performance bound of $\mathsf{P}_{\rm nuc}$ is the case $\{\theta_u(i),\theta_v(i)\}_{i=1}^r=45^\circ$. For instance, $\theta_u(i)=\theta_v(i)\approx 90^\circ ~i=1,..., r$ is considered to be a weak prior subspace information in \cite{eftekhari2018weighted}, while it is excellent in our work, leading to a huge sample complexity reduction. Also, when $\theta_u\le 45^\circ$, unlike ours, their bound is not optimal in the sense of sample complexity. Overall, our proposed method acts much better in terms of the required sample complexity.
\end{itemize}
\section{Useful Lemmas}\label{sec.usefullem}
This section provides necessary mathematical tools for Sections \ref{section.mainresult} and \ref{sec.optweights}.
\subsection{Constructing a basis for $\mathbb{R}^{n\times n}$}\label{subsection.basis}
In this section, we find a special basis for $\mathbb{R}^{n\times n}$ that simplifies the sample complexity analysis in $\mathsf{P}_{\bm{w},\mathrm{nuc}}$. The following lemma precisely states this.
\begin{lem}\label{lem.basis}
	Consider a rank $r$ matrix $\bm{X}\in\mathbb{R}^{n\times n}$ with column and row subspaces $\bm{\mathcal{U}}$ and $\bm{\mathcal{V}}$, respectively. Also, assume that we are given the subspaces $\widetilde{\bm{\mathcal{U}}}\subseteq\mathbb{R}^n$ and $\widetilde{\bm{\mathcal{V}}}\subseteq\mathbb{R}^n$, each with dimension $r'\ge r$, that have known principal angles $\bm{\theta}_u\in[0^\circ,90^\circ]^r$ and $\bm{\theta}_v\in[0^\circ,90^\circ]^r$ with $\bm{\mathcal{U}}$ and $\bm{\mathcal{V}}$, respectively. Then, there exist bases $\bm{U}\in\mathbb{R}^{n\times r}$, $\bm{V}\in\mathbb{R}^{n\times r}$, $\widetilde{\bm{U}}\in\mathbb{R}^{n\times r'}$, $\widetilde{\bm{V}}\in\mathbb{R}^{n\times r'}$ and
	\begin{align}
	&\bm{B}_L:=\left[\bm{U}_{n\times r}~ {\bm{U}'_1}_{n\times r}~ {\bm{U}'_2}_{n\times (r'-r)}~ {\bm{U}''}_{n\times (n-r-r')}\right],\nonumber\\
	&\bm{B}_R:=\left[\bm{V}_{n\times r}~ {\bm{V}'_1}_{n\times r}~ {\bm{V}'_2}_{n\times (r'-r)}~ {\bm{V}''}_{n\times (n-r-r')}\right],\nonumber\\
	\end{align}
	such that
	\begin{align}
	&\bm{\mathcal{U}}=\mathrm{span}\left(\bm{U}\right),\nonumber\\
	&\widetilde{\bm{\mathcal{U}}}=\mathrm{span}(\widetilde{\bm{U}}),\nonumber\\
	&\bm{\mathcal{V}}=\mathrm{span}\left(\bm{V}\right),\nonumber\\
	&\widetilde{\bm{\mathcal{V}}}=\mathrm{span}(\widetilde{\bm{V}}).\nonumber\\
	&\widetilde{\bm{U}}=\bm{B}_L\begin{bmatrix}
	\cos(\bm{\theta}_u)&\bm{0}_{r\times (r'-r)}\\
	-\sin(\bm{\theta}_u)&\bm{0}\\
	\bm{0}&-\bm{I}_{r'-r}\\
	\bm{0}&\bm{0}
	\end{bmatrix}\label{eq.Util}\\
	&\widetilde{\bm{V}}=\bm{B}_R\begin{bmatrix}
	\cos(\bm{\theta}_v)&\bm{0}\\
	-\sin(\bm{\theta}_v)&\bm{0}\\
	\bm{0}_{r'-r\times r}&-\bm{I}_{r'-r}\\
	\bm{0}_{(n-r-r')\times r}&\bm{0}
	\end{bmatrix},\label{eq.Vtil}
	\end{align}
	where
	\begin{align}
	&\bm{U}'_1=-\bm{P}_{\bm{\mathcal{U}}^\perp}\widetilde{\bm{U}}\begin{bmatrix}
	\sin^{-1}(\bm{\theta}_u)\\
	\bm{0}_{r'-r\times r}
	\end{bmatrix},\nonumber\\
	&\bm{U}'_2=-\bm{P}_{\bm{\mathcal{U}}^\perp}\widetilde{\bm{U}}\begin{bmatrix}
	\bm{0}\\
	\bm{I}_{r'-r\times r}
	\end{bmatrix},\nonumber\\
	&\bm{V}'_1=-\bm{P}_{\bm{\mathcal{V}}^\perp}\widetilde{\bm{V}}\begin{bmatrix}
	\sin^{-1}(\bm{\theta}_v)\\
	\bm{0}_{r'-r\times r}
	\end{bmatrix},\nonumber\\
	&\bm{V}'_2=-\bm{P}_{\bm{\mathcal{V}}^\perp}\widetilde{\bm{V}}\begin{bmatrix}
	\bm{0}\\
	\bm{I}_{r'-r\times r}
	\end{bmatrix},\nonumber\\
	\end{align}
	and $\cos(\bm{\theta}_u)$ is defined as
	\begin{align}
	&\cos(\bm{\theta}_u):=
	\mathrm{diag}\Big[\cos(\theta_u(r)),\cos(\theta_u(r-1)),..., \cos(\theta_u(1))\Big]\nonumber\\
	&\in\mathbb{R}^{n\times n},\nonumber\\
	&\cos(\bm{\theta}_v):=
	\mathrm{diag}\Big[\cos(\theta_v(r)),\cos(\theta_v(r-1)),..., \cos(\theta_v(1))\Big]\nonumber\\
	&\in\mathbb{R}^{n\times n}.
	\end{align}
\end{lem}
Lemma (\ref{lem.basis}) allows us to find $\mathrm{supp}(h_{\bm{w}}(\bm{X}))$ which is later helpful. Below, we state a lemma that includes this, along with a crucial decomposition of $h_{\bm{w}}(\bm{Z})$ for an arbitrary matrix $\bm{Z}\in\mathbb{R}^{n\times n}$.
\begin{lem}\label{lem.decompos}
	Consider a matrix $\bm{X}\in\mathbb{R}^{n\times n}$ with column and row spaces $\bm{\mathcal{U}}$ and $\bm{\mathcal{V}}$, respectively. Then, $h_{\bm{w}}(\bm{Z})$ in (\ref{eq.hZ}) with the convention $w_4:=\frac{w_2w_3}{w_1}$ for an arbitrary matrix $\bm{Z}\in\mathbb{R}^{n\times n}$ is decomposed as:
	\begin{align}\label{eq.decompos}
	h_{\bm{w}}(\bm{Z})=(\frac{1}{w_3})\bm{B}_L\bm{O}_L\bm{L}\bm{B}_L^{\rm H}\bm{Z}\bm{B}_R\bm{R}^{\rm H}\bm{O}_R^H\bm{B}_R^{\rm H},
	\end{align}
	where, $\bm{B}_L\in\mathbb{R}^{n\times n}$ and $\bm{B}_R\in\mathbb{R}^{n\times n}$ are defined in Lemma \ref{lem.basis}. Also, 
	\begin{align}
	&\bm{O}_L:=\left[\begin{array}{ccc}
	\bigg(w_1\cos^2(\bm{\theta}_u)+w_3\sin^2(\bm{\theta}_u)\bigg)(\bm{C}_L)^{-1}\\
	(w_3-w_1)\sin(\bm{\theta}_u)\cos(\bm{\theta}_u)(\bm{C}_L)^{-1}\\
	\bm{0}\\
	\bm{0}
	\end{array}\right.\nonumber\\
	&\left.\begin{array}{ccc}
	(w_1-w_3)\sin(\bm{\theta}_u)\cos(\bm{\theta}_u)(\bm{C}_L)^{-1}&\bm{0}&\bm{0}\\
	\bigg(w_1\cos^2(\bm{\theta}_u)+w_3\sin^2(\bm{\theta}_u)\bigg)(\bm{C}_L)^{-1}&\bm{0}&\bm{0}\\
	\bm{0}&\bm{I}_{r'-r}&\bm{0}\\
	\bm{0}&\bm{0}&\bm{I}_{n-r-r'}
	\end{array}\right],\label{eq.OL}\\
	&\bm{O}_R:=\left[\begin{matrix}
	\bigg(w_3\cos^2(\bm{\theta}_v)+w_4\sin^2(\bm{\theta}_v)\bigg)(\bm{C}_R)^{-1}\\
	(w_4-w_3)\sin(\bm{\theta}_v)\cos(\bm{\theta}_v)(\bm{C}_R)^{-1}\\
	\bm{0}\\
	\bm{0}
	\end{matrix}\right.\nonumber\\
	&\left.\begin{matrix}
	(w_3-w_4)\sin(\bm{\theta}_v)\cos(\bm{\theta}_v)(\bm{C}_R)^{-1}&\bm{0}&\bm{0}\\
	\bigg(w_3\cos^2(\bm{\theta}_v)+w_4\sin^2(\bm{\theta}_v)\bigg)(\bm{C}_R)^{-1}&\bm{0}&\bm{0}\\
	\bm{0}&\bm{I}_{r'-r}&\bm{0}\\
	\bm{0}&\bm{0}&\bm{I}_{n-r-r'}
	\end{matrix}\right],\label{eq.OR}\\
	&\bm{L}:=\begin{bmatrix}
	\bm{C}_L&\bm{L}_{12}&\bm{0}&\bm{0}\\
	\bm{0}&w_1w_3\bm{C}_L^{-1}&\bm{0}&\bm{0}\\
	\bm{0}&\bm{0}&w_1\bm{I}_{r'-r}&\bm{0}\\
	\bm{0}&\bm{0}&\bm{0}&w_3\bm{I}_{n-r-r'}
	\end{bmatrix},\label{eq.L}\\
	&\bm{R}:=\begin{bmatrix}
	\bm{C}_R&\bm{R}_{12}&\bm{0}&\bm{0}\\
	\bm{0}&w_3w_4\bm{\Delta}_R^{-1}&\bm{0}&\bm{0}\\
	\bm{0}&\bm{0}&w_3\bm{I}_{r'-r}&\bm{0}\\
	\bm{0}&\bm{0}&\bm{0}&w_4\bm{I}_{n-r-r'}
	\end{bmatrix},\label{eq.R}
	\end{align}
	where 
	\begin{align}
	&\bm{C}_L=(w_1^2\cos^2(\bm{\theta}_u)+w_3^2\sin(\bm{\theta}_u))^{\frac{1}{2}},\label{eq.C_L}\\
	&\bm{C}_R=(w_3^2\cos^2(\bm{\theta}_v)+w_4^2\sin(\bm{\theta}_v))^{\frac{1}{2}},\label{eq.C_R}\\
	&\bm{L}_{12}=(w_3^2-w_1^2)\sin(\bm{\theta}_u)\cos(\bm{\theta}_u)\bm{C}_L^{-1},\label{eq.L12}\\
	&\bm{R}_{12}=(w_4^2-w_3^2)\sin(\bm{\theta}_v)\cos(\bm{\theta}_v)\bm{C}_R^{-1},\label{eq.R12}
	\end{align}
	and $\bm{B}_L$, $\bm{B}_R$, $\bm{O}_L$ and $\bm{O}_R$ are orthonormal bases. Also, $\bm{L}$ and $\bm{R}$ are upper-triangular matrices. 
\end{lem}
\begin{lem}\label{lem.svd_of_hX}
	Let $\bm{X}=\bm{U}_{n\times r}\bm{\Sigma}_{r\times r} \bm{V}_{n\times r}^{\rm H}$ be the reduced SVD form of $\bm{X}\in\mathbb{R}^{n\times n}$. Then, the unsorted SVD of $h_{\bm{w}}(\bm{X})$ is obtained as
	\begin{align}\label{eq.h_wX}
	h_{\bm{w}}(\bm{X})=\bm{B}_L\bm{O}_L\begin{bmatrix}
	(\frac{1}{w_3})\bm{C}_L\bm{\Sigma}\bm{C}_R&\bm{0}_{r\times (n-r)}\\
	\bm{0}_{n-r\times r}&\bm{0}_{(n-r)\times (n-r)}
	\end{bmatrix}\bm{O}_R^H\bm{B}_R^{\rm H}.
	\end{align}
\end{lem}

\begin{corl}\label{cor.sgn}
Let $\widehat{T}:={\rm supp}(h_{\bm{w}}(\bm{X}))$ and $T_1:={\rm supp}(\bm{I}_r)$.
Then, ${\rm sgn}(h_{\bm{w}}(\bm{X}))$ and $\mathcal{P}_{\widehat{T}^\perp}(\bm{Z})$ for an arbitrary matrix $\bm{Z}\in\mathbb{R}^{n\times n}$ are obtained by
\begin{align}
&{\rm sgn}(h_{\bm{w}}(\bm{X}))=\bm{B}_L\bm{O}_L\mathcal{P}_{T_1}(\bm{I}_n)\bm{O}_R^H\bm{B}_R^{\rm H},\label{eq.sgn_hX}\\
&\mathcal{P}_{\widehat{T}^\perp}(\bm{Z})=\bm{B}_L\bm{O}_L\mathcal{P}_{T_1^\perp}(\bm{O}_L^{\rm H}\bm{B}_L^{\rm H}\bm{Z}\bm{B}_R\bm{O}_R)\bm{O}_R^H\bm{B}_R^{\rm H}\label{eq.Pthatperp}.
\end{align}
\end{corl}
Proof. See Appendix \ref{proof.cor}
\subsection{Spectral Analysis of Large Random Matrices}
\begin{figure}[]
	\centering
	\hspace*{-.6cm}
	\includegraphics[scale=.38]{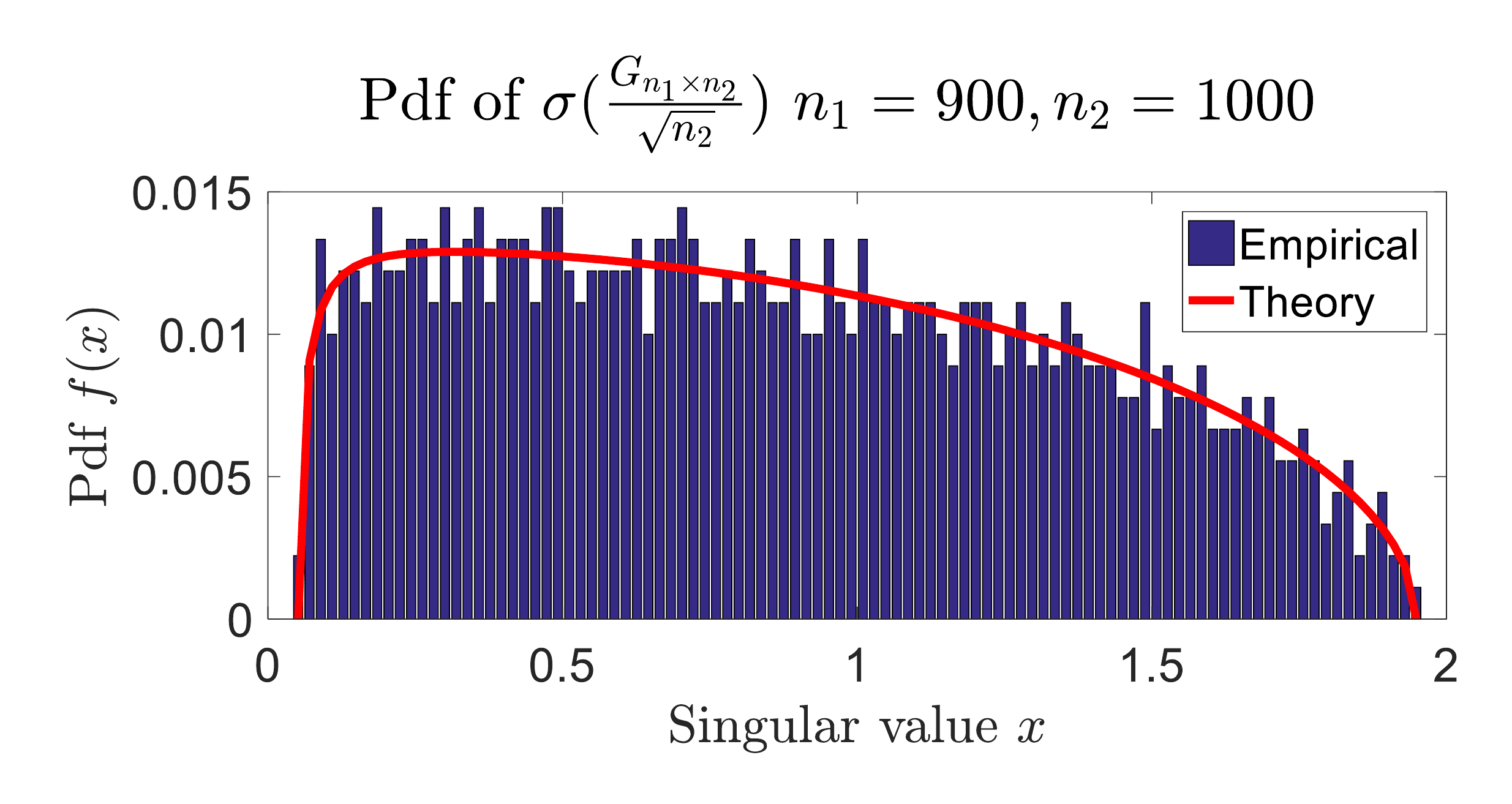}
	\caption{A comparison between probability density estimate of $\sigma(\frac{\bm{G}}{\sqrt{n_2}})$ and the theory obtained in Fact \ref{fact.me}.} 
	\label{fig.marchenko}
\end{figure}

In this part, we aim at specifying the behavior of singular values of large i.i.d. random Gaussian matrices e.g. $\bm{G}\in\mathbb{R}^{n_1\times n_2}$. First, we state a well-known fact that specifies the limiting behavior of eigenvalues of random matrices to the Mar\v{c}enko Pastur law \cite{marvcenko1967distribution} (\cite[Theorem 3.6]{bai2010spectral}). Here, we approximate the distribution of singular values of a random i.i.d. standard normal matrix by a version of Mar\v{c}enko--Pastur Law \cite{marvcenko1967distribution}. The proof uses a change of variable to match the argument for singular values which does not much differ from \cite[Theorem 3.6]{bai2010spectral} and thus we omitted the uninteresting details of this change.
\begin{fact}\label{fact.me}
	Let $\bm{G}\in\mathbb{R}^{n_1\times n_2}$ ($n_1\le n_2$) be a matrix with i.i.d. standard normal distribution, $s:=\frac{n_1}{n_2}$, and $u_b(\cdot), l_b(\cdot)$ be defined as in \eqref{eq.ub-lb}. Then, the probability density function (pdf) of $\sigma(\frac{\bm{G}}{\sqrt{n_2}})$ is given by:
	\begin{align}
	f(u)=\frac{\sqrt{(u_b(s)^2-u^2)(u^2-l_b(s)^2)}}{\pi s u}, ~~\forall u\in[l_b(s), u_b(s)].
	\end{align}	
\end{fact}
One can see from Figure \ref{fig.marchenko} that the empirical density estimate of singular values of a random matrix $\bm{G}$ with Gaussian ensemble (shown with bars) harmonizes with the obtained bound in Fact \ref{fact.me} (shown with dashed line).

In the following lemma, we obtain the limiting behavior of $\mathds{E}\frac{1}{n_1}\sum_{i=1}^{n_1}(\sigma_i(\frac{\bm{G}}{\sqrt{n_2}})-f_i)_{+}^2$ for a non-increasingly ordered vector $\bm{f}$.
\begin{lem}\label{lem.S_vs_Sap}
	Consider a random matrix $\bm{G}\in\mathbb{R}^{n_1\times n_2}$ whose elements are drawn from i.i.d. standard normal distribution. Let $f_i, i=1,\dots, n_1$ be the non-increasingly ordered elements of $\bm{f}\in\mathbb{R}_{+}^n$. Suppose $n_1, n_2\rightarrow \infty$ and $s:=\frac{n_1}{n_2}\in(0,1]$. Then, we have:
	\begin{align}\label{eq.Sap_vs_S}
	&S:=\mathds{E}\frac{1}{n_1}\sum_{i=1}^{n_1}(\sigma_i(\frac{\bm{G}}{\sqrt{n_2}})-f_i)_{+}^2\le\nonumber\\
	&\frac{1}{n_1}\sum_{i=1}^{n_1}\int_{l_b(s)}^{u_b(s)}(u-f_i)_{+}^2\frac{\sqrt{(u_b(s)^2-u^2)(u^2-l_b(s)^2)}}{\pi u s}{\rm{d}}u:=S_{ap},
	\end{align}
	where $u_b(\cdot)$ and $l_b(\cdot)$ are defined in \eqref{eq.ub-lb}.
\end{lem}
Proof. See Appendix \ref{proof.lem.S_vs_Sap}.

\begin{rem}\label{rem.Sap_vs_S}(On the tightness of $S_{ap}$ in \eqref{eq.Sap_vs_S})
In the proof of Proposition \ref{prop.m_hat_nuc_weight}, we face examples of \eqref{eq.Sap_vs_S} for which $n_1 \le n_2$ (usually $n_1$ is chosen to be small while $n_2$ is large) and $\bm{f}$ is chosen as
\begin{align}
\bm{f}=\gamma \bm{\sigma}(\bm{C}_{n_1\times n_2}),
\end{align}
where $\gamma$ is a scaling factor (usually less than one) and $\bm{C}_{n_1\times n_2}$ is a random matrix \footnote{The matrix $\bm{C}$ in the proofs of Proposition \ref{prop.m_hat_nuc_weight} is deterministic and not random. However, the final result shall not be much different.} distributed as $\mathcal{N}(\bm{0},n_2^{-1}\bm{I})$. We test some examples of this flavor in Table \ref{table.Sap_vs_S} and the upper-bound $S_{ap}$ is numerically observed to be tight for the mentioned cases. Noteworthy, if $f_i$s are all equal, the inequality in \eqref{eq.Sap_vs_S} turns into equality.
\end{rem} 
\begin{table}[t]
	\centering
		
		\begin{tabular}{ |c|c|c|c|c|}
			\hline
			$n_1$&$n_2$&$\gamma$ & $S$ & $S_{ap}$\\ 
			\hline
			$10$&$100$&$0.3$ & $0.48$ & $0.487$ \\
			\hline
			$100$&$1000$&$0.5$ & $0.26$ & $0.27$\\ 
			\hline
			$10$&$1000$&$0.9$ & $0.0096$ & $0.01$ \\
			\hline
			$5$&$5$&$0.2$ & $0.69$ & $0.71$ \\
			\hline
		\end{tabular}
	\caption{This table shows the quantities $S$ and $S_{ap}$ in \eqref{eq.Sap_vs_S} for some settings used in our analysis in 
	Proposition \ref{prop.m_hat_nuc_weight}. Notice that $\bm{f}$ \eqref{eq.Sap_vs_S} is obtained using $\bm{f}=\gamma \bm{\sigma}(\bm{C}_{n_1\times n_2})$ where $\bm{C}\sim \mathcal{N}(\bm{0},n_2^{-1}\bm{I})$ and the expectation in $S$ is computed via empirical mean over $5000$ iterations. 
}\label{table.Sap_vs_S}
\end{table}
\section{Conclusion}\label{section.conclusion}
In this work, we presented a new approach for exploiting subspace prior information in matrix sensing. We assumed that two given subspaces form some known angles with the column and row spaces of the ground-truth matrix. We exploited these angles by introducing a new weighted optimization problem and obtained the unique optimal weights that minimize the required number of measurements. The outcome of our work is to use considerably less measurements compared with the regular nuclear norm minimization.
\bibliographystyle{ieeetr}
\bibliography{mypaperbibe}
\appendix
\section{Proof of Main Result and Lemmas}
\subsection{Proof of Lemma \ref{lem.basis}}\label{proof.lem.basis}
\begin{proof}
Assume $\widetilde{\bm{U}}\in\mathbb{R}^{n\times r'}$ and $\widetilde{\bm{U}}^\perp\in\mathbb{R}^{n\times (n-r')}$ be some orthonormal bases for $\widetilde{\bm{\mathcal{U}}}$ and $\widetilde{\bm{\mathcal{U}}}^\perp$, respectively. Also, since $\widetilde{\bm{\mathcal{U}}}$ and $\widetilde{\bm{\mathcal{U}}}^\perp$ are uniquely characterized by their respective projection matrices i.e. $\bm{P}_{\widetilde{\bm{\mathcal{U}}}}\in\mathbb{R}^{n\times n}$ and $\bm{P}_{\widetilde{\bm{\mathcal{U}}}^\perp}\in\mathbb{R}^{n\times n}$, without loss of generality, assume that
\begin{align}
&\bm{U}^{\rm H}\widetilde{\bm{U}}=\begin{bmatrix}
{\rm cos}(\bm{\theta}_u)&\bm{0}_{r\times (r'-r)}
\end{bmatrix}:=\bm{U}^{\rm H}[\widetilde{\bm{U}}_1,\widetilde{\bm{U}}_2],\nonumber\\
&\bm{U}^{\rm H}\widetilde{\bm{U}}^\perp=\begin{bmatrix}
{\rm sin}(\bm{\theta}_u)&\bm{0}_{r\times (n-r'-r)}
\end{bmatrix}:=\bm{U}^{\rm H}[\widetilde{\bm{U}}_1^\perp,\widetilde{\bm{U}}_2^\perp],\nonumber
\end{align} 
where $\widetilde{\bm{U}}_1\in\mathbb{R}^{n\times r}$ $\widetilde{\bm{U}}_2\in\mathbb{R}^{n\times r'-r}$, $\widetilde{\bm{U}}_1^\perp\in\mathbb{R}^{n\times r}$ and $\widetilde{\bm{U}}_2^\perp\in\mathbb{R}^{n\times n-r'-r}$ are orthonormal bases forming the subspaces $\widetilde{\bm{\mathcal{U}}}_1\subseteq \widetilde{\bm{\mathcal{U}}}$,  $\widetilde{\bm{\mathcal{U}}}_2\subseteq \widetilde{\bm{\mathcal{U}}}$, $\widetilde{\bm{\mathcal{U}}}_1^\perp\subseteq \widetilde{\bm{\mathcal{U}}}^\perp$, and $\widetilde{\bm{\mathcal{U}}}_2^\perp\subseteq \widetilde{\bm{\mathcal{U}}}^\perp$, respectively.
(otherwise one could redefine $\widetilde{\bm{U}}$, $\widetilde{\bm{{U}}}^\perp$ and $\bm{U}$ by taking SVD of $\bm{U}^{\rm H}\widetilde{\bm{U}}$ and $\bm{U}^{\rm H}\widetilde{\bm{U}}^\perp$, since rotation in $\widetilde{\bm{U}}$ and $\widetilde{\bm{U}}^\perp$ does not affect $\bm{P}_{\widetilde{\bm{\mathcal{U}}}}$ and $\bm{P}_{\widetilde{\bm{\mathcal{U}}}^\perp}$).

The column space of any matrix in $\mathbb{R}^{n\times n}$ can be decomposed into the spaces $\bm{\mathcal{U}}$, $\bm{\mathcal{U}}^\perp\bigcap \widetilde{\bm{\mathcal{U}}}_1$, $\bm{\mathcal{U}}^\perp\bigcap \widetilde{\bm{\mathcal{U}}}_2$ and $\bm{\mathcal{U}}^\perp\bigcap \widetilde{\bm{\mathcal{U}}}_2^\perp$, where for the last three, we construct an orthonormal basis as follows: 
\begin{align}
&\bm{U}'_1:=-\bm{P}_{\bm{\mathcal{U}}^\perp}\widetilde{\bm{U}}_1\sin^{-1}(\bm{\theta}_u)\in\mathbb{R}^{n\times r},\nonumber\\
&\bm{U}'_2:=-\bm{P}_{\bm{\mathcal{U}}^\perp}\widetilde{\bm{U}}_2\in\mathbb{R}^{n\times r'-r},\nonumber\\
&\bm{U}'':=-\bm{P}_{\bm{\mathcal{U}}^\perp}\widetilde{\bm{U}}_2^\perp\in\mathbb{R}^{n\times n-r-r'};
\end{align}
such that
\begin{align*}
\bm{B}_L:=\left[\bm{U}_{n\times r}~ {\bm{U}'_1}_{n\times r}~ {\bm{U}'_2}_{n\times (r'-r)}~ {\bm{U}''}_{n\times (n-r-r')}\right],
\end{align*}
forms an orthonormal basis for the column span of any matrix in $\mathbb{R}^{n\times n}$. Similar to the above statements, there exist orthonormal bases
\begin{align}
&\bm{V}'_1:=-\bm{P}_{\bm{\mathcal{V}}^\perp}\widetilde{\bm{V}}_1\sin^{-1}(\bm{\theta}_v)\in\mathbb{R}^{n\times r},\nonumber\\
&\bm{V}'_2:=-\bm{P}_{\bm{\mathcal{V}}^\perp}\widetilde{\bm{V}}_2\in\mathbb{R}^{n\times r'-r},\nonumber\\
&\bm{V}'':=-\bm{P}_{\bm{\mathcal{V}}^\perp}\widetilde{\bm{V}}_2^\perp\in\mathbb{R}^{n\times n-r-r'},
\end{align}
such that
\begin{align*}
\bm{B}_R:=\left[\bm{V}_{n\times r}~ {\bm{V}'_1}_{n\times r}~ {\bm{V}'_2}_{n\times (r'-r)}~ {\bm{V}''}_{n\times (n-r-r')}\right],
\end{align*}
forms an orthonormal basis for the row space of any arbitrary matrix in $\mathbb{R}^{n\times n}$. Lastly, it is easy to verify that the matrices $\widetilde{\bm{U}}$ and $\widetilde{\bm{V}}$ can be represented in the bases $\bm{B}_L$ and $\bm{B}_R$ as follows:
\begin{align}
&\widetilde{\bm{U}}=\bm{B}_L\begin{bmatrix}
\cos(\bm{\theta}_u)&\bm{0}_{r\times (r'-r)}\\
-\sin(\bm{\theta}_u)&\bm{0}\\
\bm{0}&-\bm{I}_{r'-r}\\
\bm{0}&\bm{0}
\end{bmatrix}\in\mathbb{R}^{n\times r'},\nonumber\\
&\widetilde{\bm{V}}=\bm{B}_R\begin{bmatrix}
\cos(\bm{\theta}_v)&\bm{0}\\
-\sin(\bm{\theta}_v)&\bm{0}\\
\bm{0}_{r'-r\times r}&-\bm{I}_{r'-r}\\
\bm{0}_{(n-r-r')\times r}&\bm{0}
\end{bmatrix}\in\mathbb{R}^{n\times r'}.\nonumber
\end{align}
\end{proof}	
\subsection{Proof of Lemma \ref{lem.decompos}}\label{proof.lem.decompos}
\begin{proof}
Define 
\begin{align}
w_4:=\frac{w_2w_3}{w_1}.\nonumber
\end{align}
The expression $h_{\bm{w}}(\bm{Z})$ in (\ref{eq.hZ}) can be reformulated as
\begin{align}\label{eq.h_wZ}
h_{\bm{w}}(\bm{Z})=\frac{1}{w_3}\Big(w_1\bm{P}_{\widetilde{\bm{\mathcal{U}}}}+w_3\bm{P}_{\widetilde{\bm{\mathcal{U}}}^\perp}\Big)\bm{Z}\Big(w_3\bm{P}_{\widetilde{\bm{\mathcal{V}}}}+w_4\bm{P}_{\widetilde{\bm{\mathcal{V}}}^\perp}\Big).
\end{align}
We start our derivation by (\ref{eq.Util}) and (\ref{eq.Vtil}) to find $\bm{P}_{\widetilde{\bm{\mathcal{U}}}}$, $\bm{P}_{\widetilde{\bm{\mathcal{V}}}}$ which are the essential components of $h_{\bm{w}}(\bm{Z})$. By (\ref{eq.Util}) and (\ref{eq.Vtil}), it is simply holds that
\begin{align}
&\bm{P}_{\widetilde{\bm{\mathcal{U}}}}=\widetilde{\bm{U}}\widetilde{\bm{U}}^H=\bm{B}_L\nonumber\\
&\begin{bmatrix}
{\rm cos}^2(\bm{\theta}_u)&-{\rm sin}(\bm{\theta}_u){\rm cos}(\bm{\theta}_u)&\bm{0}_{r\times (r'-r)}&\bm{0}\\
-{\rm sin}(\bm{\theta}_u){\rm cos}(\bm{\theta}_u)&{\rm sin}^2(\bm{\theta}_u)&\bm{0}&\bm{0}\\
\bm{0}&\bm{0}&\bm{I}_{r'-r}&\bm{0}\\
\bm{0}_{(n-r-r')\times r}&\bm{0}&\bm{0}&\bm{0}
\end{bmatrix}\bm{B}_L^{\rm H}.
\end{align}
Also, we have:
\begin{align}
&\bm{P}_{\widetilde{\bm{\mathcal{U}}}^\perp}=\bm{I}-\bm{P}_{\widetilde{\bm{\mathcal{U}}}}=\bm{B}_L\nonumber\\
&\begin{bmatrix}
{\rm sin}^2(\bm{\theta}_u)&{\rm sin}(\bm{\theta}_u){\rm cos}(\bm{\theta}_u)&\bm{0}_{r\times (r'-r)}&\bm{0}\\
{\rm sin}(\bm{\theta}_u){\rm cos}(\bm{\theta}_u)&{\rm cos}^2(\bm{\theta}_u)&\bm{0}&\bm{0}\\
\bm{0}&\bm{0}&\bm{0}_{r'-r}&\bm{0}\\
\bm{0}_{(n-r-r')\times r}&\bm{0}&\bm{0}&\bm{I}_{n-r-r'}
\end{bmatrix}\nonumber\\
&\bm{B}_L^{\rm H}.
\end{align}
It also follows that
\begin{align}\label{eq.Qu}
&w_1\bm{P}_{\widetilde{\bm{\mathcal{U}}}}+w_3\bm{P}_{\widetilde{\bm{\mathcal{U}}}^\perp}=\bm{B}_L\nonumber\\
&\left[\begin{matrix}
w_1{\rm cos}^2(\bm{\theta}_u)+w_3{\rm sin}^2(\bm{\theta}_u)&(w_3-w_1){\rm sin}(\bm{\theta}_u){\rm cos}(\bm{\theta}_u)\\
(w_3-w_1){\rm sin}(\bm{\theta}_u){\rm cos}(\bm{\theta}_u)&w_3{\rm cos}^2(\bm{\theta}_u)+w_1{\rm sin}^2(\bm{\theta}_u)\\
\bm{0}&\bm{0}\\
\bm{0}_{(n-r-r')\times r}&\bm{0}
\end{matrix}\right.\nonumber\\
&\left.\begin{matrix}
\bm{0}_{r\times (r'-r)}&\bm{0}\\
\bm{0}&\bm{0}\\
\bm{0}_{r'-r}&\bm{0}\\
\bm{0}&\bm{I}_{n-r-r'}
\end{matrix}\right]
\bm{B}_L^{\rm H}.
\end{align}
We next simplify (\ref{eq.Qu}) by applying a QR decomposition to the matrix in the bracket. Namely,
\begin{align}
&\left[\begin{matrix}
w_1{\rm cos}^2(\bm{\theta}_u)+w_3{\rm sin}^2(\bm{\theta}_u)&(w_3-w_1){\rm sin}(\bm{\theta}_u){\rm cos}(\bm{\theta}_u)\\
(w_3-w_1){\rm sin}(\bm{\theta}_u){\rm cos}(\bm{\theta}_u)&w_3{\rm cos}^2(\bm{\theta}_u)+w_1{\rm sin}^2(\bm{\theta}_u)\\
\bm{0}&\bm{0}\\
\bm{0}_{(n-r-r')\times r}&\bm{0}
\end{matrix}\right.\nonumber\\
&\left.\begin{matrix}
\bm{0}_{r\times (r'-r)}&\bm{0}\\
\bm{0}&\bm{0}\\
\bm{0}_{r'-r}&\bm{0}\\
\bm{0}&\bm{I}_{n-r-r'}
\end{matrix}\right]=\nonumber\\
&\left[\begin{array}{ccc}
\bigg(w_1\cos^2(\bm{\theta}_u)+w_3\sin^2(\bm{\theta}_u)\bigg)(\bm{C}_L)^{-1}\\
(w_3-w_1)\sin(\bm{\theta}_u)\cos(\bm{\theta}_u)(\bm{C}_L)^{-1}\\
\bm{0}\\
\bm{0}
\end{array}\right.\nonumber\\
&\left.\begin{array}{ccc}
(w_1-w_3)\sin(\bm{\theta}_u)\cos(\bm{\theta}_u)(\bm{C}_L)^{-1}&\bm{0}&\bm{0}\\
\bigg(w_1\cos^2(\bm{\theta}_u)+w_3\sin^2(\bm{\theta}_u)\bigg)(\bm{C}_L)^{-1}&\bm{0}&\bm{0}\\
\bm{0}&\bm{I}_{r'-r}&\bm{0}\\
\bm{0}&\bm{0}&\bm{I}_{n-r-r'}
\end{array}\right]\nonumber\\
&\cdot\begin{bmatrix}
\bm{C}_L&\bm{L}_{12}&\bm{0}&\bm{0}\\
\bm{0}&w_1w_3\bm{C}_L^{-1}&\bm{0}&\bm{0}\\
\bm{0}&\bm{0}&w_1\bm{I}_{r'-r}&\bm{0}\\
\bm{0}&\bm{0}&\bm{0}&w_3\bm{I}_{n-r-r'}
\end{bmatrix}
\nonumber\\
&:=\bm{O}_L\bm{L},
\end{align} 
where $\bm{O}_L$ is an orthonormal basis and $\bm{L}$ is an upper-triangular matrix. We rewrite (\ref{eq.Qu}) as
\begin{align}
&w_1\bm{P}_{\widetilde{\bm{\mathcal{U}}}}+w_3\bm{P}_{\widetilde{\bm{\mathcal{U}}}^\perp}=\bm{B}_L\bm{O}_L\bm{L}\bm{B}_L^{\rm H}=\bm{B}_L\bm{L}^{\rm H}\bm{O}_L^{\rm H}\bm{B}_L^{\rm H},
\end{align}	
where the last equality is since $\bm{O}_L\bm{L}=\bm{L}^{\rm H}\bm{O}_L^{\rm H}$. With a similar approach on the row space of $\bm{X}$, one may write
\begin{align}
&w_3\bm{P}_{\widetilde{\bm{\mathcal{V}}}}+w_4\bm{P}_{\widetilde{\bm{\mathcal{V}}}^\perp}=\bm{B}_R\bm{O}_R\bm{R}\bm{B}_R^{\rm H}=\bm{B}_R\bm{R}^{\rm H}\bm{O}_R^H\bm{B}_R^{\rm H}
\end{align}
where 
\begin{align}
&\bm{O}_R:=\left[\begin{matrix}
\bigg(w_3\cos^2(\bm{\theta}_v)+w_4\sin^2(\bm{\theta}_v)\bigg)(\bm{C}_R)^{-1}\\
(w_4-w_3)\sin(\bm{\theta}_v)\cos(\bm{\theta}_v)(\bm{C}_R)^{-1}\\
\bm{0}\\
\bm{0}
\end{matrix}\right.\nonumber\\
&\left.\begin{matrix}
(w_3-w_4)\sin(\bm{\theta}_v)\cos(\bm{\theta}_v)(\bm{C}_R)^{-1}&\bm{0}&\bm{0}\\
\bigg(w_3\cos^2(\bm{\theta}_v)+w_4\sin^2(\bm{\theta}_v)\bigg)(\bm{C}_R)^{-1}&\bm{0}&\bm{0}\\
\bm{0}&\bm{I}_{r'-r}&\bm{0}\\
\bm{0}&\bm{0}&\bm{I}_{n-r-r'}
\end{matrix}\right],
\end{align}
is an orthonormal basis of $\mathbb{R}^n$ and
\begin{align}
&\bm{R}:=\begin{bmatrix}
\bm{C}_R&\bm{R}_{12}&\bm{0}&\bm{0}\\
\bm{0}&w_3w_4\bm{\Delta}_R^{-1}&\bm{0}&\bm{0}\\
\bm{0}&\bm{0}&w_3\bm{I}_{r'-r}&\bm{0}\\
\bm{0}&\bm{0}&\bm{0}&w_4\bm{I}_{n-r-r'}
\end{bmatrix},
\end{align}
is an triangular matrix. Lastly, $h_{\bm{w}}(\bm{Z})$ for an arbitrary $\bm{Z}\in\mathbb{R}^{n\times n}$ (the relation (\ref{eq.h_wZ})) may be written as (\ref{eq.decompos}).
\end{proof}
\subsection{Proof of Lemma \ref{lem.svd_of_hX}}\label{proof.lem.svd_of_hX}
\begin{proof}
Since 
\begin{align}
\bm{B}_L^{\rm H}\bm{X}\bm{B}_R=\begin{bmatrix}
\bm{\Sigma}_{r\times r}&\bm{0}_{r\times (n-r)}\\
\bm{0}_{(n-r)\times r}&\bm{0}_{(n-r)\times(n-r)}
\end{bmatrix},
\end{align}
it follows that
\begin{align}
\bm{L}\bm{B}_L^{\rm H}\bm{X}\bm{B}_R\bm{R}^{\rm H}=\begin{bmatrix}
\bm{C}_L\bm{\Sigma}\bm{C}_R&\bm{0}_{r\times (n-r)}\\
\bm{0}&\bm{0}_{(n-r)\times(n-r)}
\end{bmatrix}.
\end{align} 
As $\bm{C}_L\bm{\Sigma}\bm{C}_R$ is a diagonal matrix, one may deduce from (\ref{eq.decompos}) that
\begin{align}
h_{\bm{w}}(\bm{X})=\bm{B}_L\bm{O}_L\begin{bmatrix}
(\frac{1}{w_3})\bm{C}_L\bm{\Sigma}\bm{C}_R&\bm{0}_{r\times (n-r)}\\
\bm{0}_{n-r\times r}&\bm{0}_{(n-r)\times (n-r)}
\end{bmatrix}\bm{O}_R^H\bm{B}_R^{\rm H},
\end{align}
provides an unsorted SVD form for $h_{\bm{w}}(\bm{X})$. 
\end{proof}
\subsection{Proof of Proposition \ref{prop.mhat_nuc}}\label{proof.prop.mhat_nuc}
\begin{proof}
Before proving the result, we define some notations which are required in our analysis.	
	\begin{align}
	&T:=\mathrm{supp}(\bm{X}),\nonumber\\
	&T_1:=\Bigg\{\bm{Z}\in\mathbb{R}^{n\times n},~\bm{Z}=
	\begin{bmatrix}
	{\bm{Z}_{11}}_{r\times r}&{\bm{Z}_{12}}_{r\times n-r}\\
	{\bm{Z}_{21}}_{n-r\times r}&{\bm{0}}_{n-r\times n-r}\\
	\end{bmatrix} 
	\Bigg\},\nonumber\\
	&T_1^\perp:=\Bigg\{\bm{Z}\in\mathbb{R}^{n\times n},~\bm{Z}=
	\begin{bmatrix}
	{\bm{0}}_{r\times r}&\bm{0}_{r\times n-r}\\
	{\bm{0}}_{n-r\times r}&{\bm{Z}_{22}}_{n-r\times n-r}\\
	\end{bmatrix} 
	\Bigg\},\nonumber\\
	&T_{11}:=\Bigg\{\bm{Z}\in\mathbb{R}^{n\times n},~\bm{Z}=\nonumber\\
	&
	\begin{bmatrix}
	{\bm{Z}_{11}}_{r\times r}&{\bm{0}}_{r\times r}&{\bm{0}}_{r\times r'-r}&{\bm{0}}_{r\times n-r-r'}\\
	{\bm{0}}_{r\times r}&{\bm{0}}_{r\times r}&{\bm{0}}_{r\times r'-r}&{\bm{0}}_{r\times n-r-r'}\\
	{\bm{0}}_{r'-r\times r}&{\bm{0}}_{r'-r\times r}&{\bm{0}}_{r'-r\times r'-r}&{\bm{0}}_{r'-r\times n-r-r'}\\
	{\bm{0}}_{n-r-r'\times r}&{\bm{0}}_{n-r-r'\times r}&{\bm{0}}_{n-r-r'\times r'-r}&{\bm{0}}_{n-r-r'\times n-r-r'}\\
	\end{bmatrix}\label{eq.T11}\\
	&
	\Bigg\}.\nonumber
	\end{align}
	$\{T_{ij}\}_{i,j=1}^4$ are defined in the same way as $T_{11}$. We begin the proof by bounding the statistical dimension as follows:
	\begin{align}\label{eq.mhat_nuc}
	&\delta(\mathcal{D}(\|\cdot\|_*,\bm{X}))\stackrel{(\RN{1})}{\le}\inf_{t\ge 0}\mathds{E}\inf_{\bm{Z}\in\partial\|\cdot\|_*(\bm{X})}\|\bm{G}-t\bm{Z}\|_F^2,
	\end{align}
	where $(\RN{1})$ follows from the fact that the infimum of an affine function is concave and Jensen's inequality. Next, we proceed by showing that 
	\begin{align}
	&\mathds{E}\inf_{\bm{Z}\in\partial\|\cdot\|_*(\bm{X})}\|\bm{G}-t\bm{Z}\|_F^2\stackrel{(\RN{2})}{=}\mathds{E}\|\mathcal{P}_{T}(\bm{G})-t\mathrm{sgn}(\bm{X})\|_F^2+\nonumber\\
	&\mathds{E}\inf_{\|\mathcal{P}_{T^\perp}(\bm{Z})\|_{2\rightarrow 2}\le 1}\|\mathcal{P}_{T^\perp}(\bm{G})-t\mathcal{P}_{T^\perp}(\bm{Z})\|_F^2\Bigg\},
	\end{align}
where in $(\RN{2})$, we decomposed the term in the Frobenius norm into $T$ and $T^\perp$ and used the relation
\begin{align}
\partial\|\cdot\|_*(\bm{X}):=\Big\{\mathrm{sgn}(X)+\mathcal{P}_{T^\perp}(\bm{Z})~:~\|\mathcal{P}_{T^\perp}(\bm{Z})\|_{2\rightarrow 2}\le 1\Big\}.
\end{align}	
By using the definitions in Lemma \ref{lem.basis}, it is straightforward to check that
\begin{align}
&\mathcal{P}_{T}(\bm{A})=\bm{B}_L\mathcal{P}_{T_1}(\bm{B}_L^{\rm H}\bm{A}\bm{B}_R)\bm{B}_R^{\rm H},\nonumber\\
&\mathcal{P}_{T^\perp}(\bm{A})=\bm{B}_L\mathcal{P}_{T_1^\perp}(\bm{B}_L^{\rm H}\bm{A}\bm{B}_R)\bm{B}_R^{\rm H},\nonumber
\end{align}
for an arbitrary matrix $\bm{A}\in\mathbb{R}^{n\times n}$. Hence,
	\begin{align}
	&\mathds{E}\inf_{\bm{Z}\in\partial\|\cdot\|_*(\bm{X})}\|\bm{G}-t\bm{Z}\|_F^2\stackrel{(\RN{3})}{=}\nonumber\\
	&\mathds{E}\|\bm{B}_L\mathcal{P}_{T_1}(\bm{B}_L^{\rm H}\bm{G}\bm{B}_R)\bm{B}_R^{\rm H}-t\bm{B}_L\begin{bmatrix}
	\bm{I}_r&\bm{0}\\
	\bm{0}&\bm{0}
	\end{bmatrix}\bm{B}_R^{\rm H}\|_F^2+\nonumber\\
	&\mathds{E}\inf_{\|\mathcal{P}_{T_1^\perp}(\bm{B}_L^{\rm H}\bm{Z}\bm{B}_R)\|_{2\rightarrow 2}\le 1}\|\bm{B}_L\mathcal{P}_{T_1^\perp}(\bm{B}_L^{\rm H}\bm{G}\bm{B}_R)\bm{B}_R^{\rm H}\nonumber\\
	&-t\bm{B}_L\mathcal{P}_{T_1^\perp}(\bm{B}_L^{\rm H}\bm{Z}\bm{B}_R)\bm{B}_R^{\rm H}\|_F^2,
	\end{align}
	 where in $(\RN{3})$, we also used the rotational invariance of spectral norm. By rewriting and simplifying the above expression, we reach
	\begin{align}
	&\inf_{t\ge 0}\mathds{E}\inf_{\bm{Z}\in\partial\|\cdot\|_*(\bm{X})}\|\bm{G}-t\bm{Z}\|_F^2\stackrel{(\RN{4})}{=}\nonumber\\
	&\inf_{t\ge 0}\Bigg\{\mathds{E}\|\begin{bmatrix}
	\bm{G}_{11}-\bm{I}_r&{\bm{G}_{12}}_{r\times n-r}\\
	{\bm{G}_{21}}_{n-r \times r}&\bm{0}
	\end{bmatrix}\|_F^2+\mathds{E}\inf_{\|\mathcal{P}_{T_1^\perp}(\bm{Z})\|_{2\rightarrow 2}\le 1}\nonumber\\
	&\|\mathcal{P}_{T_1^\perp}(\bm{G})-t\mathcal{P}_{T_1^\perp}(\bm{Z})\|_F^2\Bigg\},
	\end{align}
	where in $(\RN{4})$, as $\bm{B}_L$ and $\bm{B}_R$ have orthonormal columns, the entries of $\bm{B}_L^{\rm H}\bm{G}\bm{B}_R$ are i.i.d. standard Gaussian which, without loss of generality, we denote by $\bm{G}$ again. For simplicity, we replaced $\bm{B}_L^{\rm H}\bm{Z}\bm{B}_R$ by $\bm{Z}$. We also used the rotational invariance of Frobenius and spectral norms. By further simplifying, we reach
	\begin{align}
	&\Bigg\{\mathds{E}\|\begin{bmatrix}
	\bm{G}_{11}-\bm{I}_r&{\bm{G}_{12}}_{r\times n-r}\\
	{\bm{G}_{21}}_{n-r \times r}&\bm{0}
	\end{bmatrix}\|_F^2+\mathds{E}\inf_{\|\mathcal{P}_{T_1^\perp}(\bm{Z})\|_{2\rightarrow 2}\le 1}\nonumber\\
	&\|\mathcal{P}_{T_1^\perp}(\bm{G})-t\mathcal{P}_{T_1^\perp}(\bm{Z})\|_F^2\Bigg\},\stackrel{(\RN{5})}{=}\Bigg\{3r^2+t^2r+\nonumber\\
	&\mathds{E}\inf_{\|\mathcal{P}_{T_1^\perp}(\bm{Z})\|_{2\rightarrow 2}\le 1}\|\mathcal{P}_{T_1^\perp}(\bm{G})-t\mathcal{P}_{T_1^\perp}(\bm{Z})\|_F^2\Bigg\},
	\end{align}
	In $(\RN{5})$, we only used the fact that the entries of $\bm{G}_{11}$, $\bm{G}_{12}$ and $\bm{G}_{13}$ have i.i.d. standard normal distribution. Borrowing the notations of \eqref{eq.T11}, one may write
	\begin{align}
	&\mathds{E}\inf_{\bm{Z}\in\partial\|\cdot\|_*(\bm{X})}\|\bm{G}-t\bm{Z}\|_F^2\nonumber\\
	&\stackrel{(\RN{6})}{\le}\Bigg\{3r^2+t^2r+\mathds{E}\inf_{\|\bm{Z}_{22}\|_{2\rightarrow 2}\le \alpha_{22}}\|\bm{G}_{22}-t\bm{Z}_{22}\|_F^2+\nonumber\\
	&\mathds{E}\inf_{\|\bm{Z}_{23}\|_{2\rightarrow 2}\le \alpha_{23}}\|\bm{G}_{23}-t\bm{Z}_{23}\|_F^2+\mathds{E}\inf_{\|\bm{Z}_{24}\|_{2\rightarrow 2}\le \alpha_{24}}\nonumber\\
	&\|\bm{G}_{24}-t\bm{Z}_{24}\|_F^2+\mathds{E}\inf_{\|\bm{Z}_{32}\|_{2\rightarrow 2}\le \alpha_{32}}\|\bm{G}_{32}-t\bm{Z}_{32}\|_F^2+\mathds{E}\nonumber\\
	&\inf_{\|\bm{Z}_{33}\|_{2\rightarrow 2}\le \alpha_{33}}\|\bm{G}_{33}-t\bm{Z}_{33}\|_F^2+\mathds{E}\inf_{\|\bm{Z}_{34}\|_{2\rightarrow 2}\le \alpha_{34}}\nonumber\\
	&\|\bm{G}_{34}-t\bm{Z}_{34}\|_F^2+\mathds{E}\inf_{\|\bm{Z}_{42}\|_{2\rightarrow 2}\le \alpha_{42}}\|\bm{G}_{42}-t\bm{Z}_{42}\|_F^2+\nonumber\\
	&\mathds{E}\inf_{\|\bm{Z}_{43}\|_{2\rightarrow 2}\le \alpha_{43}}\|\bm{G}_{43}-t\bm{Z}_{43}\|_F^2+\nonumber\\
	&\mathds{E}\inf_{\|\bm{Z}_{44}\|_{2\rightarrow 2}\le \alpha_{44}}\|\bm{G}_{44}-t\bm{Z}_{44}\|_F^2\Bigg\}.
	\end{align}
	In $(\RN{6})$, we decomposed the space $T_1^\perp$ into the spaces $T_{ij}$ for $i,j=\{2,3,4\}$. Also, we used the relation
	\begin{align}
	\|\mathcal{P}_{T_1^\perp}(\bm{Z})\|_{2\rightarrow 2}\le \sum_{i,j=2}^{4}\|\mathcal{P}_{T_{ij}}(\bm{Z})\|_{2\rightarrow 2}\le \sum_{i,j=2}^4\alpha_{ij}\le 1,
	\end{align}
	where the first inequality is due to the triangle inequality of spectral norm. The second is due to the definition of $\{\alpha_{ij}\}_{i,j=2}^4$ in \eqref{eq.alphadef}. In fact,
	\begin{align}\label{eq.fact1}
	&\{\bm{Z}:~\|\mathcal{P}_{T_1^\perp}(\bm{Z})\|_{2\rightarrow 2}\le 1\}\supseteq\nonumber\\
	&\{\bm{Z}:~\|\mathcal{P}_{T_{ij}}(\bm{Z})\|_{2\rightarrow 2}\le \alpha_{ij}~~ \forall i,j\in\{2,3,4\}\}.
	\end{align}
	We further use Hoffman-Wielandt Theorem \cite[Corollary 7.3.5]{horn2013matrix} to reach
	\begin{align}\label{eq.re1}
	&\mathds{E}\inf_{\bm{Z}\in\partial\|\cdot\|_*(\bm{X})}\|\bm{G}-t\bm{Z}\|_F^2\le\nonumber\\
	&3r^2+t^2r+\mathds{E}\sum_{i=1}^{r}\inf_{\sigma_i(\bm{Z}_{22})\le \alpha_{22}}(\sigma_i(\bm{G}_{22})-t\sigma_i(\bm{Z}_{22}))^2\nonumber\\
	&+\mathds{E}\sum_{i=1}^{r\wedge(r'-r)}\inf_{\sigma_i(\bm{Z}_{23})\le 1}(\sigma_i(\bm{G}_{23})-t\sigma_i(\bm{Z}_{23}))^2+\nonumber\\
	&\mathds{E}\sum_{i=1}^{r\wedge(n-r-r')}\inf_{\sigma_i(\bm{Z}_{24})\le 1}(\sigma_i(\bm{G}_{24})-t\sigma_i(\bm{Z}_{24}))^2\nonumber\\
	&+\mathds{E}\sum_{i=1}^{r\wedge(r'-r)}\inf_{\sigma_i(\bm{Z}_{32})\le 1}(\sigma_i(\bm{G}_{32})-t\sigma_i(\bm{Z}_{32}))^2+\mathds{E}\sum_{i=1}^{r'-r}\nonumber\\
	&\inf_{\sigma_i(\bm{Z}_{33})\le 1}(\sigma_i(\bm{G}_{33})-t\sigma_i(\bm{Z}_{33}))^2+\mathds{E}\sum_{i=1}^{(r'-r)\wedge(n-r-r')}\nonumber\\
	&\inf_{\sigma_i(\bm{Z}_{34})\le 1}(\sigma_i(\bm{G}_{34})-t\sigma_i(\bm{Z}_{34}))^2+\mathds{E}\sum_{i=1}^{r\wedge(n-r-r')}\inf_{\sigma_i(\bm{Z}_{42})\le 1}\nonumber\\
	&(\sigma_i(\bm{G}_{42})-t\sigma_i(\bm{Z}_{42}))^2+\mathds{E}\sum_{i=1}^{(r'-r)\wedge(n-r-r')}\inf_{\sigma_i(\bm{Z}_{43})\le 1}\nonumber\\
	&(\sigma_i(\bm{G}_{43})-t\sigma_i(\bm{Z}_{43}))^2+\mathds{E}\sum_{i=1}^{n-r-r'}\inf_{\sigma_i(\bm{Z}_{44})\le 1}\nonumber\\
	&(\sigma_i(\bm{G}_{44})-t\sigma_i(\bm{Z}_{44}))^2.
	\end{align}
	The minimizations in the above expression have closed form relations. Indeed, its is not hard to check that
	\begin{align}\label{eq.shrinkage}
	\inf_{|z|\le a}(g-z)^2=(|g|-a)_{+}^2,
	\end{align}
	for arbitrary scalar $g$ and positive $a$. We incorporate this fact into \eqref{eq.re1} to get
	\begin{align}
	&\mathds{E}\inf_{\bm{Z}\in\partial\|\cdot\|_*(\bm{X})}\|\bm{G}-t\bm{Z}\|_F^2\le 3r^2+t^2r+\nonumber\\
	&\mathds{E}\sum_{i=1}^{r}(\sigma_i(\bm{G}_{22})-t\alpha_{22})_{+}^2+\mathds{E}\sum_{i=1}^{r\wedge(r'-r)}(\sigma_i(\bm{G}_{23})-t\alpha_{23})_{+}^2\nonumber\\
	&+\mathds{E}\sum_{i=1}^{r\wedge(n-r-r')}(\sigma_i(\bm{G}_{24})-t\alpha_{24})_{+}^2\nonumber\\
	&+\mathds{E}\sum_{i=1}^{r\wedge(r'-r)}(\sigma_i(\bm{G}_{32})-t\alpha_{32})_{+}^2+\mathds{E}\sum_{i=1}^{r'-r}(\sigma_i(\bm{G}_{33})-t\alpha_{33})_{+}^2\nonumber\\
	&+\mathds{E}\sum_{i=1}^{(r'-r)\wedge(n-r-r')}(\sigma_i(\bm{G}_{34})-t\alpha_{34})_{+}^2+\mathds{E}\sum_{i=1}^{r\wedge(n-r-r')}\nonumber\\
	&(\sigma_i(\bm{G}_{42})-t\alpha_{42})_{+}^2+\mathds{E}\sum_{i=1}^{(r'-r)\wedge(n-r-r')}(\sigma_i(\bm{G}_{43})-t\alpha_{43})_{+}^2\nonumber\\
	&+\mathds{E}\sum_{i=1}^{n-r-r'}(\sigma_i(\bm{G}_{44})-t\alpha_{44})_{+}^2.
	\end{align}
	Lastly, by invoking Lemma \ref{lem.S_vs_Sap} and Remark \ref{rem.Sap_vs_S}, we obtain
	\begin{align}
	&\mathds{E}\inf_{\bm{Z}\in\partial\|\cdot\|_*(\bm{X})}\|\bm{G}-t\bm{Z}\|_F^2\le 3r^2+t^2r+\nonumber\\
	&r^2\phi(\tfrac{t\alpha_{22}}{\sqrt{r}},1)+2r(r'-r)\phi(\tfrac{t\alpha_{22}}{\sqrt{r\vee (r'-r)}},1)+2r(n-r-r')\nonumber\\
	&\phi(\tfrac{t\alpha_{24}}{\sqrt{r\vee (n-r-r')}},1)+2(r'-r)(n-r-r')\nonumber\\
	&\phi(\tfrac{t\alpha_{34}}{\sqrt{(r'-r)\vee (n-r-r')}},1)+(r'-r)^2\phi(\tfrac{t\alpha_{33}}{\sqrt{r'-r}},1)\nonumber\\
	&+(n-r-r')^2\phi(\tfrac{t\alpha_{44}}{\sqrt{n-r-r'}},1).	
	\end{align}	
\end{proof}
\subsection{Proof of Proposition \ref{prop.m_hat_nuc_weight}}\label{proof.prop.m_hat_nuc_weight}
\begin{proof}
Before proving the result, we define some notations:
\begin{align}
&\widehat{T}:=\mathrm{supp}(h_{\bm{w}}(\bm{X})),\nonumber\\
&\bm{w}:=[w_1, w_2, w_3]^T,\nonumber\\
&w_4:=\frac{w_2w_3}{w_1}.
\end{align}	
We begin with the definition of $\Psi_t(\bm{w},\bm{\theta}_u,\bm{\theta}_v)$ which is used in $\widehat{m}_{\bm{w},\bm{\theta}_u, \bm{\theta}_v}$:
\begin{align}\label{eq.psi1}
&\Psi_t(\bm{w},\bm{\theta}_u,\bm{\theta}_v)=\mathds{E}\inf_{\bm{Z}\in\partial\|h_{\bm{w}}(\cdot)\|_*(\bm{X})}\|\bm{G}-t\bm{Z}\|_F^2\stackrel{(\RN{1})}{=}\nonumber\\
&\mathds{E}\inf_{\bm{Z}\in\partial\|\cdot\|_*(h_{\bm{w}}(\bm{X}))}\|\bm{G}-th_{\bm{w}}^*(\bm{Z})\|_F^2,
\end{align}
where in $(\RN{1})$, we used the chain rule lemma of subdifferential \cite[Theorem 23.9]{rockafellar2015convex}. By using the facts that $h_{\bm{w}}$ is a self-adjoint function i.e. $h_{\bm{w}}^*=h_{\bm{w}}$ and also the decomposition in \eqref{eq.decompos}, we rewrite the above expression as
\begin{align}\label{eq.relw1}
&\mathds{E}\inf_{\bm{Z}\in\partial\|\cdot\|_*(h_{\bm{w}}(\bm{X}))}\|\bm{G}-th_{\bm{w}}^*(\bm{Z})\|_F^2=\mathds{E}\inf_{\bm{Z}\in\partial\|\cdot\|_*(h_{\bm{w}}(\bm{X}))}\|\bm{G}\nonumber\\
&-\frac{t}{w_3}\bm{B}_L\bm{O}_L\bm{L\bm{B}_L^{\rm H}\bm{Z}\bm{B}_R\bm{R}^{\rm H}\bm{O}_R^H\bm{B}_R^{\rm H}}\|_F^2.
\end{align}
The set $\partial\|\cdot\|_*(h_{\bm{w}}(\bm{X}))$ is defined as
\begin{align}\label{eq.partial_nuc_hw}
&\partial\|\cdot\|_*(h_{\bm{w}}(\bm{X})):=\Big\{\mathrm{sgn}(h_{\bm{w}}(\bm{X}))+\mathcal{P}_{\widehat{T}^\perp}(\bm{W})~~:\nonumber\\
&~\|\mathcal{P}_{\widehat{T}^\perp}(\bm{W})\|_{2\rightarrow 2}\le1\Big\}.
\end{align}
By incorporating \eqref{eq.partial_nuc_hw} into \eqref{eq.relw1}, we have that
\begin{align}
&\mathds{E}\inf_{\bm{Z}\in\partial\|\cdot\|_*(h_{\bm{w}}(\bm{X}))}\|\bm{G}-th_{\bm{w}}^*(\bm{Z})\|_F^2\mathds{E}\inf_{\|\mathcal{P}_{\widehat{T}^\perp}(\bm{Z})\|_{2\rightarrow 2}\le 1}\nonumber\\
&\|\bm{G}-\frac{t}{w_3}\bm{B}_L\bm{O}_L\bm{L}\bm{B}_L^{\rm H}\mathrm{sgn}(h_{\bm{w}}(\bm{X}))\bm{B}_R\bm{R}^{\rm H}\bm{O}_R^H\bm{B}_R^{\rm H}\nonumber\\
&-\frac{t}{w_3}\bm{B}_L\bm{O}_L\bm{L}\bm{B}_L^{\rm H}\mathcal{P}_{\widehat{T}^\perp}(\bm{Z})\bm{B}_R\bm{R}^{\rm H}\bm{O}_R^H\bm{B}_R^{\rm H}\|_F^2.
\end{align}
We proceed by writing
\begin{align}\label{eq.relw2}
&\mathds{E}\inf_{\|\mathcal{P}_{\widehat{T}^\perp}(\bm{Z})\|_{2\rightarrow 2}\le 1}\|\bm{G}-\frac{t}{w_3}\bm{B}_L\bm{O}_L\bm{L}\bm{B}_L^{\rm H}\mathrm{sgn}(h_{\bm{w}}(\bm{X}))\nonumber\\
&\bm{B}_R\bm{R}^{\rm H}\bm{O}_R^H\bm{B}_R^{\rm H}-\frac{t}{w_3}\bm{B}_L\bm{O}_L\bm{L}\bm{B}_L^{\rm H}\mathcal{P}_{\widehat{T}^\perp}(\bm{Z})\bm{B}_R\bm{R}^{\rm H}\bm{O}_R^H\bm{B}_R^{\rm H}\|_F^2\nonumber\\
&\stackrel{(\RN{2})}{=}\mathds{E}\inf_{\|\mathcal{P}_{\widehat{T}^\perp}(\bm{Z})\|_{2\rightarrow 2}\le 1}\|\bm{G}-\frac{t}{w_3}\bm{L}^{\rm H}\bm{O}_L^{\rm H}\bm{B}_L^{\rm H}\mathrm{sgn}(h_{\bm{w}}(\bm{X}))\bm{B}_R\bm{O}_R\nonumber\\
&\bm{R}-\frac{t}{w_3}\bm{L}^{\rm H}\bm{O}_L^{\rm H}\bm{B}_L^{\rm H}\mathcal{P}_{\widehat{T}^\perp}(\bm{Z})\bm{B}_R\bm{O}_R\bm{R}\|_F^2,
\end{align}
where $(\RN{2})$ is since $\bm{B}_L,\bm{O}_L,\bm{B}_R,\bm{O}_R$ are orthonormal bases and Frobenius norm has rotational invariance property.  Also, we used the fact that $\bm{O}_L^{\rm H}\bm{B}_L^{\rm H}\bm{G}\bm{B}_R\bm{O}_R$ has the same distribution as $\bm{G}$. So, for simplicity, we replace $\bm{G}$ instead of $\bm{O}_L^{\rm H}\bm{B}_L^{\rm H}\bm{G}\bm{B}_R\bm{O}_R$. By using \eqref{eq.sgn_hX} and \eqref{eq.Pthatperp} in Lemma \ref{lem.svd_of_hX} and replacing $\bm{O}_L^{\rm H}\bm{B}_L^{\rm H}\bm{Z}\bm{B}_R\bm{O}_R$ by $\bm{Z}$, \eqref{eq.relw2} can be further simplified:
\begin{align}\label{eq.relw3}
&\mathds{E}\inf_{\|\mathcal{P}_{\widehat{T}^\perp}(\bm{Z})\|_{2\rightarrow 2}\le 1}\|\bm{G}-\frac{t}{w_3}\bm{L}^{\rm H}\bm{O}_L^{\rm H}\bm{B}_L^{\rm H}\mathrm{sgn}(h_{\bm{w}}(\bm{X}))\bm{B}_R\bm{O}_R\bm{R}-\nonumber\\
&\frac{t}{w_3}\bm{L}^{\rm H}\bm{O}_L^{\rm H}\bm{B}_L^{\rm H}\mathcal{P}_{\widehat{T}^\perp}(\bm{Z})\bm{B}_R\bm{O}_R\bm{R}\|_F^2=\mathds{E}\inf_{\|\mathcal{P}_{T_1^\perp}(\bm{Z})\|_{2\rightarrow 2}\le 1}\nonumber\\
&\Big\|\bm{G}-\frac{t}{w_3}\bm{L}^{\rm H}\mathcal{P}_{T_1}(\bm{I}_n)\bm{R}-\frac{t}{w_3}\bm{L}^{\rm H}\mathcal{P}_{T_1^\perp}(\bm{Z})\bm{R}\Big\|_F^2.
\end{align}
In the last expression of \eqref{eq.relw3}, we decompose the matrices inside the Frobenius norm into the disjoint sets $\{T_{ij}\}_{i,j=1}^4$ as follows:
\begin{align}
&\frac{t}{w_3}\bm{L}^{\rm H}\mathcal{P}_{T_1}(\bm{I}_n)\bm{R}=\nonumber\\
&\begin{bmatrix}
\frac{t}{w_3}\bm{C}_L\bm{C}_R&\frac{t}{w_3}\bm{C}_L\bm{R}_{12}&\bm{0}_{r\times (n-2r)}\\
\frac{t}{w_3}\bm{L}_{12}\bm{C}_R&\frac{t}{w_3}\bm{L}_{12}\bm{R}_{12}&\bm{0}_{r\times (n-2r)}\\
\bm{0}_{r\times r}&\bm{0}_{r\times r}&\bm{0}_{r\times (n-2r)}
\end{bmatrix},\label{eq.support_part}\\
&\frac{t}{w_3}\bm{L}^{\rm H}\mathcal{P}_{T_1^\perp}(\bm{Z})\bm{R}=\nonumber\\
&\left[\begin{matrix}
\bm{0}_{r\times r}&\bm{0}_{r\times r}\\
\bm{0}_{r\times r}&\frac{t}{w_3}w_1w_3w_3w_4\bm{C}_L^{-1}\bm{Z}_{22}\bm{C}_R^{-1}\\
\bm{0}_{(r'-r)\times r}&\frac{t}{w_3}w_1w_3w_4\bm{Z}_{32}\bm{C}_R^{-1}\\
\bm{0}_{(n-r-r')\times r}&\frac{t}{w_3}w_3w_3w_4\bm{Z}_{42}\bm{C}_R^{-1}
\end{matrix}\right.\nonumber\\
&\hspace{30pt}\left.\begin{matrix}
\bm{0}_{r\times (r'-r)}&\bm{0}_{r\times (n-r-r')}\\
\frac{t}{w_3}w_1w_3w_3\bm{C}_L^{-1}\bm{Z}_{23}&\frac{t}{w_3}w_1w_3w_4\bm{C}_L^{-1}\bm{Z}_{24}\\
\frac{t}{w_3}w_1w_3\bm{Z}_{33}&\frac{t}{w_3}w_1w_4\bm{Z}_{34}\\
\frac{t}{w_3}w_3w_3\bm{Z}_{43}&\frac{t}{w_3}w_3w_4\bm{Z}_{44}
\end{matrix}\right],\label{eq.offsupport_part}
\end{align}
where $\bm{C}_L$ and $\bm{C}_R$ are defined as
\begin{align}
&\bm{C}_L=\Big((tw_1)^2{\cos}^2(\bm{\theta}_u)+(tw_3)^2{\sin}^2(\bm{\theta}_u)\Big)^{\frac{1}{2}},\label{eq.CL1}\\
&\bm{C}_R=\Big((tw_3)^2{\cos}^2(\bm{\theta}_v)+(tw_4)^2{\sin}^2(\bm{\theta}_v)\Big)^{\frac{1}{2}},\label{eq.CR1}
\end{align} 
in this part\footnote{The definitions of $\bm{C}_L$ and $\bm{C}_R$ in \eqref{eq.CL1} and \eqref{eq.CR1} slightly differ from those in \eqref{eq.C_L} and \eqref{eq.C_R} as the weights are accompanied with $t$.}. Moreover, it is straightforward to check that \eqref{eq.support_part} and \eqref{eq.offsupport_part} can be more simplified and rewritten as:
\begin{align}\label{eq.relw4}
&\frac{t}{w_3}\bm{L}^{\rm H}\mathcal{P}_{T_1}(\bm{I}_n)\bm{R}=\begin{bmatrix}
\bm{E}_{11}&\bm{E}_{12}&\bm{0}_{r\times (n-2r)}\\
\bm{E}_{21}&\bm{E}_{22}&\bm{0}_{r\times (n-2r)}\\
\bm{0}_{r\times r}&\bm{0}_{r\times r}&\bm{0}_{r\times (n-2r)}
\end{bmatrix},\\
&\frac{t}{w_3}\bm{L}^{\rm H}\mathcal{P}_{T_1^\perp}(\bm{Z})\bm{R}=\nonumber\\
&\left[\begin{matrix}
\bm{0}_{r\times r}&\bm{0}_{r\times r}\\
\bm{0}_{r\times r}&(tw_1)(tw_3)(tw_4)\bm{C}_L^{-1}\bm{Z}_{22}\bm{C}_R^{-1}\\
\bm{0}_{(r'-r)\times r}&(tw_2)(tw_3)\bm{Z}_{32}\bm{C}_R^{-1}\\
\bm{0}_{(n-r-r')\times r}&(tw_3)(tw_4)\bm{Z}_{42}\bm{C}_R^{-1}
\end{matrix}\right.\nonumber\\
&\hspace{70pt}\left.\begin{matrix}
\bm{0}_{r\times (r'-r)}&\bm{0}_{r\times (n-r-r')}\\
(tw_1)(tw_3)\bm{C}_L^{-1}\bm{Z}_{23}&(tw_1)(tw_4)\bm{C}_L^{-1}\bm{Z}_{24}\\
tw_1\bm{Z}_{33}&tw_2\bm{Z}_{34}\\
tw_3\bm{Z}_{43}&tw_4\bm{Z}_{44}
\end{matrix}\right],\nonumber
\end{align}
where
\begin{align}
&\bm{E}_{11}=\Bigg((tw_1)^2{\cos}^2(\bm{\theta}_u){\cos}^2(\bm{\theta}_v)+(tw_2)^2{\cos}^2(\bm{\theta}_u){\sin}^2(\bm{\theta}_v),\nonumber\\
&+(tw_3)^2{\sin}^2(\bm{\theta}_u){\cos}^2(\bm{\theta}_v)+(tw_4)^2{\sin}^2(\bm{\theta}_u){\sin}^2(\bm{\theta}_v)\Bigg)^{\frac{1}{2}},\nonumber\\
&\bm{E}_{12}=(\frac{tw_4}{tw_3}-1)(tw_1+tw_2)\Big((tw_1)^2{\cos}^2(\bm{\theta}_u)+(tw_3)^2\nonumber\\
&{\sin}^2(\bm{\theta}_u)\Big)^{\frac{1}{2}}\Big((tw_1)^2{\cos}^2(\bm{\theta}_v)+(tw_2)^2{\sin}^2(\bm{\theta}_v)\Big)^{-\frac{1}{2}} {\sin}(\bm{\theta}_v)\nonumber\\
&{\cos}(\bm{\theta}_v),\nonumber\\
&\bm{E}_{21}=(\frac{tw_3}{tw_1}-1)(tw_1+tw_3)\Big((tw_1)^2{\cos}^2(\bm{\theta}_v)+\nonumber\\
&(tw_2)^2{\sin}^2(\bm{\theta}_v)\Big)^{\frac{1}{2}}\Big((tw_1)^2{\cos}^2(\bm{\theta}_u)+(tw_3)^2{\sin}^2(\bm{\theta}_u)\Big)^{-\frac{1}{2}} \nonumber\\
&{\sin}(\bm{\theta}_u){\cos}(\bm{\theta}_u),\nonumber\\
&\bm{E}_{22}=\Big((tw_4)^2-(tw_3)^2-(tw_2)^2+(tw_1)^2\Big)\nonumber\\
&\Bigg((tw_1)^2{\cos}^2(\bm{\theta}_u){\cos}^2(\bm{\theta}_v)+(tw_2)^2{\cos}^2(\bm{\theta}_u){\sin}^2(\bm{\theta}_v)+\nonumber\\
&(tw_3)^2{\sin}^2(\bm{\theta}_u){\cos}^2(\bm{\theta}_v)+(tw_4)^2{\sin}^2(\bm{\theta}_u){\sin}^2(\bm{\theta}_v)\Bigg)^{-\frac{1}{2}}
\nonumber\\
&{\sin}(\bm{\theta}_u){\cos}(\bm{\theta}_u){\sin}(\bm{\theta}_v){\cos}(\bm{\theta}_v).\nonumber
\end{align}
Incorporate \eqref{eq.relw4} into \eqref{eq.relw3} to reach
\begin{align}\label{eq.relw5}
&\mathds{E}\inf_{\|\mathcal{P}_{T_1^\perp}(\bm{Z})\|_{2\rightarrow 2}\le 1}\Big\|\bm{G}-\frac{t}{w_3}\bm{L}^{\rm H}\mathcal{P}_{T_1}(\bm{I}_n)\bm{R}\nonumber\\
&-\frac{t}{w_3}\bm{L}^{\rm H}\mathcal{P}_{T_1^\perp}(\bm{Z})\bm{R}\Big\|_F^2\stackrel{(\RN{6})}{=}\mathds{E}\|\mathcal{P}_{T_{11}}(\bm{G})-\bm{E}_{11}\|_F^2+\nonumber\\
&\mathds{E}\|\mathcal{P}_{T_{12}}(\bm{G})-\bm{E}_{12}\|_F^2+\mathds{E}\|\mathcal{P}_{T_{21}}(\bm{G})-\bm{E}_{21}\|_F^2+\nonumber\\
&\mathds{E}\inf_{\|\mathcal{P}_{T_1^\perp}(\bm{Z})\|_{2\rightarrow 2}\le 1}\Bigg\{\|\mathcal{P}_{T_{22}}(\bm{G})-\bm{E}_{22}-(tw_1)(tw_3)(tw_4)\bm{C}_L^{-1}\nonumber\\
&\mathcal{P}_{T_{22}}(\bm{Z})\bm{C}_R^{-1}\|_F^2+\|\mathcal{P}_{T_{23}}(\bm{G})-(tw_1)(tw_3)\bm{C}_L^{-1}\mathcal{P}_{T_{23}}(\bm{Z})\|_F^2+\nonumber\\
&\|\mathcal{P}_{T_{24}}(\bm{G})-(tw_1)(tw_4)\bm{C}_L^{-1}\mathcal{P}_{T_{24}}(\bm{Z})\|_F^2+\|\mathcal{P}_{T_{32}}(\bm{G})-(tw_2)\nonumber\\
&(tw_3)\mathcal{P}_{T_{32}}(\bm{Z})\bm{C}_R^{-1}\|_F^2+\|\mathcal{P}_{T_{33}}(\bm{G})-(tw_1)\mathcal{P}_{T_{33}}(\bm{Z})\|_F^2+\nonumber\\
&\|\mathcal{P}_{T_{34}}(\bm{G})-(tw_2)\mathcal{P}_{T_{34}}(\bm{Z})\|_F^2+\|\mathcal{P}_{T_{42}}(\bm{G})-(tw_3)(tw_4)\nonumber\\
&\mathcal{P}_{T_{42}}(\bm{Z})\bm{C}_R^{-1}\|_F^2+\|\mathcal{P}_{T_{43}}(\bm{G})-(tw_3)\mathcal{P}_{T_{43}}(\bm{Z})\|_F^2+\|\mathcal{P}_{T_{44}}(\bm{G})\nonumber\\
&-(tw_4)\mathcal{P}_{T_{44}}(\bm{Z})\|_F^2\Bigg\}.
\end{align}
Since the entries of ${\bm{G}_{11}}_{r\times r}, {\bm{G}_{12}}_{r\times r}$ and ${\bm{G}_{21}}_{r\times r}$ are i.i.d. standard normal variables, we have $\mathds{E}\|\bm{G}_{11}\|_F^2=\mathds{E}\|\bm{G}_{12}\|_F^2=\mathds{E}\|\bm{G}_{21}\|_F^2=r^2$. Combining \eqref{eq.relw5} with the fact \eqref{eq.fact1}, we reach\footnote{Notice that using the fact \eqref{eq.fact1} leads to an asymptotically equal expression due to the explanations in Remark \ref{rem.ame} and Figure \ref{fig.error_vs_rprim}.}:
\begin{align}\label{eq.psi2}
&\Psi_t(\bm{w},\bm{\theta}_u,\bm{\theta}_v)=r^2+\|\bm{E}_{11}\|_F^2+r^2+\|\bm{E}_{12}\|_F^2+r^2+\|\bm{E}_{21}\|_F^2\nonumber\\
&+\mathds{E}\inf_{\|\mathcal{P}_{T_{22}}(\bm{Z})\|_{2\rightarrow 2}\le \alpha_{22}}\|\mathcal{P}_{T_{22}}(\bm{G})-\bm{E}_{22}-(tw_1)(tw_3)(tw_4)\nonumber\\
&\bm{C}_L^{-1}\mathcal{P}_{T_{22}}(\bm{Z})\bm{C}_R^{-1}\|_F^2+\mathds{E}\inf_{\|\mathcal{P}_{T_{23}}(\bm{Z})\|_{2\rightarrow 2}\le \alpha_{23}}\|\mathcal{P}_{T_{23}}(\bm{G})-(tw_1)\nonumber\\
&(tw_3)\bm{C}_L^{-1}\mathcal{P}_{T_{23}}(\bm{Z})\|_F^2+\mathds{E}\inf_{\|\mathcal{P}_{T_{24}}(\bm{Z})\|_{2\rightarrow 2}\le \alpha_{24}}\|\mathcal{P}_{T_{24}}(\bm{G})-(tw_1)\nonumber\\
&(tw_4)\bm{C}_L^{-1}\mathcal{P}_{T_{24}}(\bm{Z})\|_F^2+\mathds{E}\inf_{\|\mathcal{P}_{T_{32}}(\bm{Z})\|_{2\rightarrow 2}\le \alpha_{32}}\|\mathcal{P}_{T_{32}}(\bm{G})-(tw_2)\nonumber\\
&(tw_3)\mathcal{P}_{T_{32}}(\bm{Z})\bm{C}_R^{-1}\|_F^2+\mathds{E}\inf_{\|\mathcal{P}_{T_{33}}(\bm{Z})\|_{2\rightarrow 2}\le \alpha_{33}}\|\mathcal{P}_{T_{33}}(\bm{G})-(tw_1)\nonumber\\
&\mathcal{P}_{T_{33}}(\bm{Z})\|_F^2+\mathds{E}\inf_{\|\mathcal{P}_{T_{34}}(\bm{Z})\|_{2\rightarrow 2}\le \alpha_{34}}\|\mathcal{P}_{T_{34}}(\bm{G})-(tw_2)\mathcal{P}_{T_{34}}(\bm{Z})\|_F^2\nonumber\\
&+\mathds{E}\inf_{\|\mathcal{P}_{T_{42}}(\bm{Z})\|_{2\rightarrow 2}\le \alpha_{42}}\|\mathcal{P}_{T_{42}}(\bm{G})-(tw_3)(tw_4)\mathcal{P}_{T_{42}}(\bm{Z})\bm{C}_R^{-1}\|_F^2\nonumber\\
&+\mathds{E}\inf_{\|\mathcal{P}_{T_{43}}(\bm{Z})\|_{2\rightarrow 2}\le \alpha_{43}}\|\mathcal{P}_{T_{43}}(\bm{G})-(tw_3)\mathcal{P}_{T_{43}}(\bm{Z})\|_F^2+\nonumber\\
&\mathds{E}\inf_{\|\mathcal{P}_{T_{44}}(\bm{Z})\|_{2\rightarrow 2}\le \alpha_{44}}\|\mathcal{P}_{T_{44}}(\bm{G})-(tw_4)\mathcal{P}_{T_{44}}(\bm{Z})\|_F^2.\nonumber\\
\end{align}
We solve the minimizations in \eqref{eq.psi2}, one by one: First for the minimization in the second line of \eqref{eq.psi2}, we have that 
\begin{align}\label{eq.relw6}
&\inf_{\|\mathcal{P}_{T_{22}}(\bm{Z})\|_{2\rightarrow 2}\le \alpha_{22}}\|\mathcal{P}_{T_{22}}(\bm{G})-\bm{E}_{22}-(tw_1)(tw_3)(tw_4)\nonumber\\
&\bm{C}_L^{-1}\mathcal{P}_{T_{22}}(\bm{Z})\bm{C}_R^{-1}\|_F^2\stackrel{(\RN{1})}{=}
\sum_{i=1}^{r}\inf_{\sigma_i(\bm{Z}_{22})\le \alpha_{22}}\Bigg(\sigma_i(\bm{G}_{22})-\nonumber\\
&\sigma_i\bigg(\bm{E}_{22}+(tw_1)(tw_3)(tw_4)\bm{C}_L^{-1}\bm{Z}_{22}\bm{C}_R^{-1}\bigg)\Bigg)^2\stackrel{(\RN{2})}{=}\nonumber\\
&\sum_{i=1}^{r}
\inf_{\sigma_i(\bm{Z}_{22})\le \alpha_{22}}\Bigg(\sigma_i(\bm{G}_{22})-\sigma_i(\bm{E}_{22})-\nonumber\\
&(tw_1)(tw_3)(tw_4)\sigma_i(\bm{C}_L^{-1})\sigma_i(\bm{Z}_{22})\sigma_i(\bm{C}_R^{-1})\Bigg)^2,
\end{align} 
where in the above equations, the first equality $(\RN{1})$ is due to Hoffman--Wielandt Theorem \cite[Corollary 7.3.5]{horn1990matrix}. The equality $(\RN{2})$ in \eqref{eq.relw6} is because of the relations
\begin{align}\label{eq.rel111}
&\sum_{i=1}^{r}\Bigg(\sigma_i(\bm{G}_{22})-\sigma_i\bigg(\bm{E}_{22}+(tw_1)(tw_3)(tw_4)\bm{C}_L^{-1}\bm{Z}_{22}\bm{C}_R^{-1}\bigg)\Bigg)\nonumber\\
&\ge \sum_{i=1}^{r}
\Bigg(\sigma_i(\bm{G}_{22})-\sigma_i(\bm{E}_{22})-\nonumber\\
&(tw_1)(tw_3)(tw_4)\sigma_i(\bm{C}_L^{-1})\sigma_i(\bm{Z}_{22})\sigma_i(\bm{C}_R^{-1})\Bigg),
\end{align}
\cite[Lemma 3.3.8]{horn1990matrix}, and the fact that $f(x)=x^2$ is an increasing convex function on $[0,\infty]$. In (\ref{eq.rel111}), we benefited from \cite[Theorem 3.3.14 a]{horn1990matrix} and \cite[Problem 3]{horn1990matrix}. In the third line of \eqref{eq.psi2}, we use
\begin{align}\label{eq.relw7}
&\inf_{\|\mathcal{P}_{T_{23}}(\bm{Z})\|_{2\rightarrow 2}\le \alpha_{23}}\|\mathcal{P}_{T_{23}}(\bm{G})-(tw_1)(tw_3)\bm{C}_L^{-1}\mathcal{P}_{T_{23}}(\bm{Z})\|_F^2=\nonumber\\
&\sum_{i=1}^{r\wedge(r'-r)}\inf_{\sigma_i(\bm{Z}_{23})\le 1}\bigg(\sigma_i(\bm{G}_{23})-(tw_1)(tw_3)\sigma_i(\bm{C}_L^{-1}\bm{Z}_{23})\bigg)^2=\nonumber\\
&\sum_{i=1}^{r\wedge(r'-r)}\inf_{\sigma_i(\bm{Z}_{23})\le 1}\bigg(\sigma_i(\bm{G}_{23})-(tw_1)(tw_3)\sigma_i(\bm{C}_L^{-1})\sigma_i(\bm{Z}_{23})\bigg)^2,
\end{align}
where the first equality comes from Hoffman--Wielandt Theorem \cite[Corollary 7.3.5]{horn1990matrix}. The second is the result of \cite[Lemma 3.3.8]{horn1990matrix}, convexity besides monotonicity of $f(x)=x^2$ in the interval $[0,\infty]$ and
\begin{align}
&\sum_{i=1}^{r\wedge(r'-r)}\bigg(\sigma_i(\bm{G}_{23})-(tw_1)(tw_3)\sigma_i(\bm{C}_L^{-1}\bm{Z}_{23})\bigg)\nonumber\\\ge
&\sum_{i=1}^{r\wedge(r'-r)}(\sigma_i(\bm{G}_{23})-(tw_1)(tw_3)\sigma_i(\bm{C}_L^{-1})\sigma_i(\bm{Z}_{23})),
\end{align}
which follows from \cite[Theorem 3.3.14 a]{horn1990matrix}. Other minimizations in \eqref{eq.psi2} can be solved using similar strategies. After simplifying $\|\bm{E}_{11}\|_F^2, \|\bm{E}_{12}\|_F^2, \|\bm{E}_{21}\|_F^2$, $\|\bm{E}_{22}\|_F^2$ in \eqref{eq.psi2} and employing
\eqref{eq.relw6}, \eqref{eq.relw7}, we may rewrite \eqref{eq.psi2} as

\begin{align}\label{eq.psi3}
&\Psi_t(\bm{w},\bm{\theta}_u,\bm{\theta}_v)=3r^2+w_1^2\sum_{i=1}^{r}\cos^2(\theta_u(i))\cos^2(\theta_v(i))+w_2^2\nonumber\\
&\sum_{i=1}^{r}\cos^2(\theta_u(i))\sin^2(\theta_v(i))+w_3^2\sum_{i=1}^{r}\sin^2(\theta_u(i))\cos^2(\theta_v(i))\nonumber\\
&+w_4^2\sum_{i=1}^{r}\sin^2(\theta_u(i))\sin^2(\theta_v(i))+(\tfrac{w_4}{w_3}-1)^2(w_1+w_2)^2\nonumber\\
&\sum_{i=1}^{r}\Bigg\{\tfrac{w_1^2{\cos}^2(\bm{\theta}_u(i))+w_3^2{\sin}^2(\bm{\theta}_u(i))}{w_1^2{\cos}^2(\bm{\theta}_v(i))+w_2^2{\sin}^2(\bm{\theta}_v(i))}\sin^2(\theta_v(i))\cos^2(\theta_v(i))\Bigg\}+\nonumber\\
&(\tfrac{w_3}{w_1}-1)^2(w_1+w_3)^2\sum_{i=1}^{r}\Bigg\{\tfrac{w_1^2{\cos}^2(\bm{\theta}_v(i))+w_2^2{\sin}^2(\bm{\theta}_v(i))}{w_1^2{\cos}^2(\bm{\theta}_u(i))+w_3^2{\sin}^2(\bm{\theta}_u(i))}\nonumber\\
&\sin^2(\theta_u(i))\cos^2(\theta_u(i))\Bigg\}+\mathds{E}\sum_{i=1}^{r}\inf_{\sigma_i(\bm{Z}_{22})\le \alpha_{22}}\Big(\sigma_i(\bm{G}_{22})-\nonumber\\
&\sigma_i(\bm{E}_{22})-(tw_1)(tw_3)(tw_4)\sigma_i(\bm{C}_L^{-1})\sigma_i(\bm{Z}_{22})\sigma_i(\bm{C}_R^{-1})\Big)^2+\mathds{E}\nonumber\\
&\sum_{i=1}^{r\wedge(r'-r)}\inf_{\sigma_i(\bm{Z}_{23})\le 1}\Big(\sigma_i(\bm{G}_{23})-(tw_1)(tw_3)\sigma_i(\bm{C}_L^{-1})\sigma_i(\bm{Z}_{23})\Big)^2\nonumber\\
&+\mathds{E}\sum_{i=1}^{r\wedge(n-r-r')}\inf_{\sigma_i(\bm{Z}_{24})\le 1}\Big(\sigma_i(\bm{G}_{24})-(tw_1)(tw_4)\sigma_i(\bm{C}_L^{-1})\nonumber\\
&\sigma_i(\bm{Z}_{24})\Big)^2+\mathds{E}\sum_{i=1}^{r\wedge(r'-r)}\inf_{\sigma_i(\bm{Z}_{32})\le 1}\Big(\sigma_i(\bm{G}_{32})-(tw_2)(tw_3)\nonumber\\
&\sigma_i(\bm{Z}_{32})\sigma_i(\bm{C}_R^{-1})\Big)^2+\mathds{E}\sum_{i=1}^{r'-r}\inf_{\sigma_i(\bm{Z}_{33})\le 1}\Big(\sigma_i(\bm{G}_{33})-(tw_1)\nonumber\\
&\sigma_i(\bm{Z}_{33})\Big)^2+\mathds{E}\sum_{i=1}^{(r'-r)\wedge(n-r-r')}\inf_{\sigma_i(\bm{Z}_{34})\le 1}\Big(\sigma_i(\bm{G}_{34})-(tw_2)\nonumber\\
&\sigma_i(\bm{Z}_{34})\Big)^2+\mathds{E}\sum_{i=1}^{r\wedge(n-r-r')}\inf_{\sigma_i(\bm{Z}_{42})\le 1}\Big(\sigma_i(\bm{G}_{42})-(tw_3)(tw_4)\nonumber\\
&\sigma_i(\bm{Z}_{42})\sigma_i(\bm{C}_R^{-1})\Big)^2+\mathds{E}\sum_{i=1}^{(r'-r)\wedge(n-r-r')}\inf_{\sigma_i(\bm{Z}_{43})\le 1}\Big(\sigma_i(\bm{G}_{43})-\nonumber\\
&(tw_3)\sigma_i(\bm{Z}_{43})\Big)^2+\mathds{E}\sum_{i=1}^{n-r-r'}\inf_{\sigma_i(\bm{Z}_{44})\le 1}\Big(\sigma_i(\bm{G}_{44})-(tw_4)\nonumber\\
&\sigma_i(\bm{Z}_{44})\Big)^2.
\end{align}
By benefiting the relation \eqref{eq.shrinkage} for the minimizations in \eqref{eq.psi3}, it is straightforward to conclude
\begin{align}\label{eq.psi4}
&\Psi_t(\bm{w},\bm{\theta}_u,\bm{\theta}_v)=3r^2+(tw_1)^2\sum_{i=1}^{r}\cos^2(\theta_u(i))\cos^2(\theta_v(i))+\nonumber\\
&(tw_2)^2\sum_{i=1}^{r}\cos^2(\theta_u(i))\sin^2(\theta_v(i))+(tw_3)^2\sum_{i=1}^{r}\sin^2(\theta_u(i))\nonumber\\
&\cos^2(\theta_v(i))+(tw_4)^2\sum_{i=1}^{r}\sin^2(\theta_u(i))\sin^2(\theta_v(i))+(\frac{w_4}{w_3}-1)^2\nonumber\\
&(tw_1+tw_2)^2\sum_{i=1}^{r}\Bigg\{\frac{w_1^2{\cos}^2(\bm{\theta}_u(i))+w_3^2{\sin}^2(\bm{\theta}_u(i))}{w_1^2{\cos}^2(\bm{\theta}_v(i))+w_2^2{\sin}^2(\bm{\theta}_v(i))}\nonumber\\
&\sin^2(\theta_v(i))\cos^2(\theta_v(i))\Bigg\}+(\frac{w_3}{w_1}-1)^2(tw_1+tw_3)^2\nonumber\\
&\sum_{i=1}^{r}\Bigg\{\frac{w_1^2{\cos}^2(\bm{\theta}_v(i))+w_2^2{\sin}^2(\bm{\theta}_v(i))}{w_1^2{\cos}^2(\bm{\theta}_u(i))+w_3^2{\sin}^2(\bm{\theta}_u(i))}\sin^2(\theta_u(i))\cos^2(\theta_u(i))\Bigg\}\nonumber\\
&+\mathds{E}\sum_{i=1}^{r}\Big(\sigma_i(\bm{G}_{22})-\sigma_i(\bm{E}_{22})-(tw_1)(tw_3)(tw_4)\sigma_i(\bm{C}_L^{-1})\nonumber\\
&\sigma_i(\bm{C}_R^{-1})\alpha_{22}\Big)_{+}^2+\mathds{E}\sum_{i=1}^{r\wedge(r'-r)}\Big(\sigma_i(\bm{G}_{23})-(tw_1)(tw_3)\sigma_i(\bm{C}_L^{-1})\nonumber\\
&\alpha_{23}\Big)_{+}^2+\mathds{E}\sum_{i=1}^{r\wedge(n-r-r')}\Big(\sigma_i(\bm{G}_{24})-(tw_1)(tw_4)\sigma_i(\bm{C}_L^{-1})\alpha_{24}\Big)_{+}^2	\nonumber\\
&+\mathds{E}\sum_{i=1}^{r\wedge(r'-r)}\Big(\sigma_i(\bm{G}_{32})-(tw_2)(tw_3)\sigma_i(\bm{C}_R^{-1})\alpha_{32}\Big)_{+}^2\nonumber\\
&+\mathds{E}\sum_{i=1}^{r'-r}\Big(\sigma_i(\bm{G}_{33})-(tw_1)\alpha_{33}\Big)_{+}^2+\mathds{E}\sum_{i=1}^{(r'-r)\wedge(n-r-r')}\nonumber\\
&\Big(\sigma_i(\bm{G}_{34})-tw_2\alpha_{34}\Big)_{+}^2+\mathds{E}\sum_{i=1}^{r\wedge(n-r-r')}\Big(\sigma_i(\bm{G}_{42})-(tw_3)\nonumber\\
&(tw_4)\sigma_i(\bm{C}_R^{-1})\alpha_{42}\Big)_{+}^2+\mathds{E}\sum_{i=1}^{(r'-r)\wedge(n-r-r')}\Big(\sigma_i(\bm{G}_{43})\nonumber\\
&-tw_3\alpha_{43}\Big)_{+}^2+\mathds{E}\sum_{i=1}^{n-r-r'}\Big(\sigma_i(\bm{G}_{44})-tw_4\alpha_{44}\Big)_{+}^2.
\end{align}
Lastly, we invoke Lemma \ref{lem.S_vs_Sap} which helps to obtain the limiting value of expectations in (\ref{eq.psi4}):
\begin{align}\label{eq.psi5}
&\Psi_t(\bm{w},\bm{\theta}_u,\bm{\theta}_v)=3r^2+(tw_1)^2\sum_{i=1}^{r}\cos^2(\theta_u(i))\cos^2(\theta_v(i))+\nonumber\\
&(tw_2)^2\sum_{i=1}^{r}\cos^2(\theta_u(i))\sin^2(\theta_v(i))+(tw_3)^2\sum_{i=1}^{r}\sin^2(\theta_u(i))\nonumber\\
&\cos^2(\theta_v(i))+(tw_4)^2\sum_{i=1}^{r}\sin^2(\theta_u(i))\sin^2(\theta_v(i))+\nonumber\\
&(\tfrac{w_4}{w_3}-1)^2(tw_1+tw_2)^2\sum_{i=1}^{r}\Bigg\{\tfrac{w_1^2{\cos}^2(\bm{\theta}_u(i))+w_3^2{\sin}^2(\bm{\theta}_u(i))}{w_1^2{\cos}^2(\bm{\theta}_v(i))+w_2^2{\sin}^2(\bm{\theta}_v(i))}\nonumber\\
&\sin^2(\theta_v(i))\cos^2(\theta_v(i))\Bigg\}+(\tfrac{w_3}{w_1}-1)^2(tw_1+tw_3)^2\sum_{i=1}^{r}\nonumber\\
&\Bigg\{\tfrac{w_1^2{\cos}^2(\bm{\theta}_v(i))+w_2^2{\sin}^2(\bm{\theta}_v(i))}{w_1^2{\cos}^2(\bm{\theta}_u(i))+w_3^2{\sin}^2(\bm{\theta}_u(i))}\sin^2(\theta_u(i))\cos^2(\theta_u(i))\Bigg\}+\nonumber\\
&\sum_{i=1}^{r}r\phi(\tfrac{\sigma_i(\bm{E}_{22})+(tw_1)(tw_3)(tw_4)\sigma_i(\bm{C}_L^{-1})\sigma_i(\bm{C}_R^{-1})\alpha_{22}}{\sqrt{r}},1)+\nonumber\\
&r\vee (r'-r)\sum_{i=1}^{r\wedge (r'-r)}\phi(\tfrac{(tw_1)(tw_3)\sigma_i(\bm{C}_L^{-1})\alpha_{23}}{\sqrt{r\vee (r'-r)}}, s_1)+\nonumber\\
&r\vee(n-r-r')\sum_{i=1}^{r\wedge (n-r'-r)}\phi(\tfrac{(tw_1)(tw_4)\sigma_i(\bm{C}_L^{-1})\alpha_{24}}{\sqrt{r\vee (n-r-r')}},s_2)+\nonumber\\
&r\vee(r'-r)\sum_{i=1}^{r\wedge (r'-r)}\phi(\tfrac{(tw_2)(tw_3)\sigma_i(\bm{C}_R^{-1})\alpha_{32}}{\sqrt{r\vee (r'-r)}},s_1)+(r'-r)^2\nonumber\\
&\phi(\tfrac{(tw_1)\alpha_{33}}{\sqrt{(r'-r)}},1)+(r'-r)\vee (n-r-r')\nonumber\\
&\sum_{i=1}^{(r'-r)\wedge (n-r-r')}\phi(\tfrac{tw_2\alpha_{34}}{\sqrt{(r'-r)\vee (n-r-r')}}, s_3)+r\vee(n-r-r')\nonumber\\
&\sum_{i=1}^{r \wedge n-r-r'}\phi(\tfrac{(tw_3)(tw_4)\sigma_i(\bm{C}_R^{-1})\alpha _{42}}{\sqrt{r\wedge (n-r-r')}}, s_2)+(r'-r)\vee(n-r-r')\nonumber\\
&\sum_{i=1}^{(r'-r)\wedge(n-r-r')}\phi(\tfrac{tw_3\alpha_{43}}{\sqrt{(r'-r)\vee (n-r-r')}}, s_3)+(n-r-r')^2\nonumber\\
&\phi(\tfrac{tw_4\alpha_{44}}{\sqrt{n-r-r'}}, 1).
\end{align}
\end{proof}	
\subsection{Proof of Lemma  \ref{lem.errorbound}}\label{proof.errorbound}
\begin{proof}
First, define $w_4:=\frac{w_2w_3}{w_1}$. To control the error term, we benefit from 
\begin{align}\label{eq.errorame}
\mathrm{Error}:=\frac{2\sup_{\bm{S}\in\partial\| h_{\bm{w}}(\cdot)\|_*(\bm{X})}\|\bm{S}\|_F}{\frac{h_{\bm{w}}(\bm{X})}{\|\bm{X}\|_F}}.
\end{align}
For any $\bm{S}\in\partial \|h_{\bm{w}}(\cdot)\|_*(\bm{X})$, there exists $\bm{Z}\in\partial\|\cdot\|_*(h_{\bm{w}}(\bm{X}))$ such that:
\begin{align}\label{eq.numer}
&\|\bm{S}\|_F= \|h_{\bm{w}}(\bm{Z})\|_F\stackrel{(\RN{1})}{=}\nonumber\\
&\|\frac{1}{w_3}\bm{B}_L\bm{O}_L\bm{L}\bm{B}_L^{\rm H}\bm{Z}\bm{B}_R\bm{R}^{\rm H}\bm{O}_R^H\bm{B}_R^{\rm H}\|_F\stackrel{(\RN{2})}{=}\nonumber\\
&\|\frac{1}{w_3}\bm{L}\bm{B}_L^{\rm H}\bm{Z}\bm{B}_R\bm{R}^{\rm H}\|_F\stackrel{(\RN{3})}{\le}\frac{1}{w_3}\|\bm{L}\|_{2\rightarrow 2}\|\bm{B}_L^{\rm H}\bm{Z}\bm{B}_R\|_F\|\bm{R}\|_{2\rightarrow 2}\nonumber\\
&\stackrel{(\RN{4})}{\le}\frac{\sqrt{n}}{w_3}\|\bm{L}\|_{2\rightarrow 2}\|\bm{Z}\|_{2\rightarrow 2}\|\bm{R}\|_{2\rightarrow 2}\nonumber\\
&\stackrel{(\RN{5})}{\le}
\frac{\sqrt{n}}{w_3}\|\bm{L}\|_{2\rightarrow 2}\|\bm{R}\|_{2\rightarrow 2}\stackrel{(\RN{6})}{=}
\sqrt{n}\max\{w_1, w_2, w_3, w_4\}\stackrel{(\RN{7})}{\le}\nonumber\\
&\sqrt{n}\sqrt{w_1^2+w_2^2+w_3^2+w_4^2}.
\end{align}
In (\ref{eq.numer}), $(\RN{1})$ follows from a chain rule lemma in subdifferential \cite[Chapter 4]{bertsekas2014convex} namely $\partial\| h_{\bm{w}}(\cdot)\|_*(\bm{X})=h_{\bm{w}}^*(\partial\|\cdot\|_*(h_{\bm{w}}(\bm{X})))$ and Lemma \ref{lem.basis}. In $(\RN{2})$, the rotational invariance of Frobenius norm is used. In $(\RN{3})$, we used the relation $\|\bm{A}\bm{B}\bm{C}\|_F\le \|\bm{A}\|_{2\rightarrow 2}\|\bm{B}\|_F\|\bm{C}\|_{2\rightarrow 2}$ for any conforming matrices $\bm{A}, \bm{B}, \bm{C}$. $(\RN{4})$ is the result of $\|\bm{A}\|_F\le \sqrt{n}\|\bm{A}\|_{2\rightarrow 2}$ and rotational invariance of spectral norm. $(\RN{5})$ follows from (\ref{eq.partial_nuc_hw}) and that $\|\bm{Z}\|_{2\rightarrow2}\le 1$ for any $\bm{Z}\in\partial \|\cdot\|_*(h_{\bm{w}}(\bm{X}))$. In $(\RN{6})$, since the singular values of the matrix
\begin{align}
\begin{bmatrix}
\bm{C}_L&\bm{L}_{12}\\
\bm{0}&w_1w_3\bm{C}_L^{-1}
\end{bmatrix}
\end{align}
which is a submatrix of $\bm{L}$ in \ref{lem.basis}, are $[w_1,w_3]^T$\cite[Equation 6]{singularofL},
it holds that the singular values of $\bm{L}$ are
\begin{align}\label{eq.sigma_L}
\bm{\sigma}(\bm{L})=[w_1\vee w_3,w_1\vee w_3,w_1\wedge w_3,w_1\wedge w_3]^T. 
\end{align}
Also, the same result holds for the singular values of $\bm{R}$. Namely,
\begin{align}\label{eq.sigma_R}
\bm{\sigma}(\bm{R})=[w_3\vee w_4,w_3\vee w_4,w_3\wedge w_4,w_3\wedge w_4]^T. 
\end{align}
 Hence, $\frac{1}{w_3}\|\bm{L}\|_{2\rightarrow 2}\|\bm{R}\|_{2\rightarrow 2}=(\frac{1}{w_3})(w_1\vee w_3)(w_3\vee w_4)=\max\{w_1, w_2, w_3, w_4\}$. Lastly, $(\RN{7})$ is the result of the inequality $\|\cdot\|_\infty\le \|\cdot\|_2$.
Moreover, for the denominator of (\ref{eq.errorame}), it holds that
\begin{align}\label{eq.denum}
&\frac{\|h_{\bm{w}}(\bm{X})\|_*}{\|\bm{X}\|_F}\stackrel{(\RN{1})}{=}\frac{\langle \mathrm{sgn}(h_{\bm{w}}(\bm{X})), h_{\bm{w}}(\bm{X}) \rangle_F}{\|\bm{X}\|_F}\stackrel{(\RN{2})}{=}\nonumber\\
&\frac{\Big\langle h_{\bm{w}}^*\Big(\bm{B}_L\bm{O}_L\begin{bmatrix}
	\bm{I}_{r\times r} &\bm{0}\\
	\bm{0}&\bm{0}
	\end{bmatrix}\bm{O}_R^H\bm{B}_R^{\rm H}\Big), \bm{X}\Big\rangle_F}{\|\bm{X}\|_F}\stackrel{(\RN{3})}{=}\nonumber\\
&\frac{\frac{1}{w_3}\Big\langle \bm{B}_L\bm{L}^{\rm H}\begin{bmatrix}
	\bm{I}_{r\times r} &\bm{0}\\
	\bm{0}&\bm{0}
	\end{bmatrix}\bm{R}\bm{B}_R^{\rm H}, \bm{X}\Big\rangle_F}{\|\bm{X}\|_F}\stackrel{(\RN{4})}{=}\nonumber\\
&\frac{\frac{1}{w_3}\Big\langle \bm{B}_L\begin{bmatrix}
	\bm{C}_L\bm{C}_R &\bm{0}\\
	\bm{0}&\bm{0}
	\end{bmatrix}\bm{B}_R^{\rm H}, \bm{X}\Big\rangle_F}{\|\bm{X}\|_F}\stackrel{(\RN{5})}{\le}\frac{1}{w_3}\|\bm{C}_L\bm{C}_R\|_F\stackrel{(\RN{6})}{=}\nonumber\\
&\sqrt{w_1^2\beta_1+w_2^2\beta_2+w_3^2\beta_3+w_4^2\beta_4},
\end{align}
where
\begin{align}
&\beta_1=\sum_{i=1}^{r}\cos^2(\theta_u(i))\cos^2(\theta_v(i)),\nonumber\\
&\beta_2=\sum_{i=1}^{r}\cos^2(\theta_u(i))\sin^2(\theta_v(i)),\nonumber\\
&\beta_3=\sum_{i=1}^{r}\sin^2(\theta_u(i))\cos^2(\theta_v(i)),\nonumber\\
&\beta_4=\sum_{i=1}^{r}\sin^2(\theta_u(i))\cos^2(\theta_v(i)).
\end{align}
In $(\RN{1})$, we used the definition of nuclear norm. In $(\RN{2})$, $\mathrm{sgn}(h_{\bm{w}}(\bm{X}))$ is obtained from \ref{eq.sgn_hX}. Also, we used the fact that $h_{\bm{w}}=h_{\bm{w}}^*$. In $(\RN{3})$, we used \ref{eq.decompos} and the facts $\bm{O}_L\bm{L}=\bm{L}^{\rm H}\bm{O}_L^{\rm H}$ and $\bm{O}_R\bm{R}=\bm{R}^{\rm H}\bm{O}_R^{\rm H}$. $(\RN{4})$ is the consequence of the fact that
\begin{align}
\bm{X}=\bm{B}_L\begin{bmatrix}
\bm{\Sigma}&\bm{0}_{r\times n-r}\\
\bm{0}_{r\times r}&\bm{0}_{(n-r)\times (n-r)}
\end{bmatrix}\bm{B}_R^{\rm H},
\end{align}
and only $\mathcal{P}_{T_{11}}(\bm{L}^{\rm H}\bm{R})=\bm{C}_L\bm{C}_R$ contributes to the Frobenius inner product. $(\RN{5})$ follows from Cauchy Schwartz inequality and the rotational invariance of Frobenius norm. By considering (\ref{eq.D(f,x)}), (\ref{eq.decompos}), the fact that 
\begin{align}
&\partial \|h_{\bm{w}}(\cdot)\|_*(\bm{X})=h_{\bm{w}}^*(\partial \|\cdot\|_*(h_{\bm{w}}(\bm{X})))=\nonumber\\
&(\frac{1}{w_3})\bm{B}_L\bm{O}_L\bm{L}\bm{B}_L^{\rm H}\partial \|\cdot\|_*(h_{\bm{w}}(\bm{X}))\bm{B}_R\bm{R}^{\rm H}\bm{O}_R^{\rm H}\bm{B}_R^{\rm H},
\end{align}
and \eqref{eq.sgn_hX}, $\mathcal{D}(\|h_{\bm{w}}(\cdot)\|_*,\bm{X})$ does not depend on the singular values of $\bm{X}$, i.e. $\bm{\Sigma}$. So, a matrix
\begin{align}
\bm{Z}=\bm{U}_{n\times r}\bm{C}_L\bm{C}_R\bm{V}_{n\times r}^{\rm H}
\end{align}
can be chosen to have equality in $(\RN{5})$. Lastly, $(\RN{6})$ follows from the definitions (\ref{eq.C_L}) and (\ref{eq.C_R}) and some simplifications.

 Therefore, by (\ref{eq.numer} and (\ref{eq.denum})), the error bound reads:
\begin{align}
\frac{2\sqrt{n}\sqrt{w_1^2+w_2^2+w_3^2+w_4^2}}{\sqrt{w_1^2\beta_1+w_2^2\beta_2+w_3^2\beta_3+w_4^2\beta_4}}\le 2\sqrt{\frac{n}{\min_\{\beta_1,\beta_2,\beta_3,\beta_4\}}}.
\end{align}
Since $\cos(\bm{\theta}_u)$ and $\sin(\bm{\theta}_u)$ are arranged in decreasing and increasing order, respectively, it holds that
\begin{align}
&\min\{\beta_1,\beta_2,\beta_3,\beta_4\}\ge\nonumber\\
& r \min\{\cos(\theta_u(r)),\sin(\theta_u(1)) \}\min\{\cos(\theta_v(r)),\sin(\theta_v(1)) \},
\end{align}
and hence the result concludes.
\end{proof}
\subsection{Proof of Lemma \ref{lem.objective_convex}}\label{proof.lem.objective_convex}
\begin{proof}
	\textit{Continuity in bounded points}. For continuity, it must be shown that sufficiently small changes in $\bm{v}$ result in arbitrary small changes in $J(\bm{v})$. Let $\bm{v}_1,\bm{v}_2\in\mathbb{R}_{+}^3$. By definition of $J_{\bm{G}}$, it holds that
	\begin{align}
	&J_{\bm{G}}(\bm{v}_1)-J_{\bm{G}}(\bm{v}_2)=\nonumber\\
	&\|\bm{G}-\mathcal{P}_{h_{\bm{v}_1}(\mathcal{C})}(\bm{G})\|_F^2-\|\bm{G}-\mathcal{P}_{h_{\bm{v}_2}(\mathcal{C})}(\bm{G})\|_F^2=\nonumber\\
	&2\langle \bm{G},\mathcal{P}_{h_{\bm{v}_2}(\mathcal{C})}(\bm{G})-\mathcal{P}_{h_{\bm{v}_1}(\mathcal{C})}(\bm{G})\rangle_F+\Big(\|\mathcal{P}_{h_{\bm{v}_1}(\mathcal{C})}(\bm{G})\|_F-\nonumber\\
	&\|\mathcal{P}_{h_{\bm{v}_2}(\mathcal{C})}(\bm{G})\|_F\Big)\Big(\|\mathcal{P}_{h_{\bm{v}_1}(\mathcal{C})}(\bm{G})\|_F+\|\mathcal{P}_{h_{\bm{v}_2}(\mathcal{C})}(\bm{G})\|_F\Big).
	\end{align}
	Since
	\begin{align}
	\|\mathcal{P}_{h_{\bm{v}}(\mathcal{C})}(\bm{G})\|_F\le \sup_{\bm{Z}\in\mathcal{C}}\|h_{\bm{v}}(\bm{Z})\|_F\le \sqrt{n}\max\bigg\{\|\bm{v}\|_{\infty},\frac{v(2)v(3)}{v(1)}\bigg\},
	\end{align}
	  (due to \ref{eq.numer})) and
	\begin{align}
	&\|\mathcal{P}_{h_{\bm{v}_1}(\mathcal{C})}(\bm{G})\|_F-\|\mathcal{P}_{h_{\bm{v}_2}(\mathcal{C})}(\bm{G})\|_F\le\nonumber\\
	&\sup_{\bm{Z}\in\mathcal{C}}\Bigg(\|h_{\bm{v}_1}(\bm{Z})\|_F-\|h_{\bm{v}_2}(\bm{Z})\|_F\Bigg)\le\nonumber\\
	& \sup_{\bm{Z}\in\mathcal{C}}\|h_{\bm{v}_1}(\bm{Z})-h_{\bm{v}_2}(\bm{Z})\|_F=\sup_{\bm{Z}\in\mathcal{C}}\|h_{\bm{v}_1-\bm{v}_2}(\bm{Z})\|_F\le\nonumber\\
	& \sqrt{n}\max\bigg\{\|\bm{v}_1-\bm{v}_2\|_{\infty},\frac{(v_1(2)-v_2(2))(v_1(3)-v_2(3))}{(v_1(1)-v_2(1))}\bigg\},
	\end{align}
	(see (\ref{eq.numer})), we have
	\begin{align}
	&|J_{\bm{G}}(\bm{v}_1)-J_{\bm{G}}(\bm{v}_2)|\le\nonumber\\
	&\Bigg(2\|\bm{G}\|_F\sqrt{n}+n\Big(\max\bigg\{\|\bm{v}_1\|_{\infty},\frac{v_1(2)v_1(3)}{v_1(1)}\bigg\}+\nonumber\\
	&\max\bigg\{\|\bm{v}_2\|_{\infty},\frac{v_2(2)v_2(3)}{v_2(1)}\bigg\}\Big)\Bigg)\nonumber\\
	&\max\Big\{\|\bm{v}_1-\bm{v}_2\|_{\infty},\frac{(v_1(2)-v_2(2))(v_1(3)-v_2(3))}{(v_1(1)-v_2(1))}\Big\}.
	\end{align}
	As a consequence, we obtain
	\begin{align}
	|J_{\bm{G}}(\bm{v}_1)-J_{\bm{G}}(\bm{v}_2)|\rightarrow 0~~ \text{as}~~\bm{v}_1\rightarrow \bm{v}_2.
	\end{align}
	Since $\|\bm{v}\|_{\infty}$ is bounded, continuity holds.
	
	\textit{Convexity}. Let $\bm{v}_1,\bm{v}_2\in\mathbb{R}_{+}^3$ and $\theta\in[0,1]$. Then,
	\begin{align}
	&\forall \epsilon,\widetilde{\epsilon}>0, \exists \bm{Z},\widetilde{\bm{Z}}\in\mathcal{C}~\text{such that}\nonumber\\
	&\|\bm{G}-h_{\bm{v}_1}(\bm{Z})\|_F\le \mathrm{dist}(\bm{G},h_{\bm{v}_1}(\mathcal{C}))+\epsilon,\nonumber\\
	&\|\bm{G}-h_{\bm{v}_2}(\bm{Z})\|_F\le \mathrm{dist}(\bm{G},h_{\bm{v}_2}(\mathcal{C}))+\widetilde{\epsilon}.\nonumber
	\end{align}
	Since otherwise, we have:
	\begin{align}
	&\forall \bm{Z},\widetilde{\bm{Z}}\in\mathcal{C}:\nonumber\\
	&\|\bm{G}-h_{\bm{v}_1}(\bm{Z})\|_F> \mathrm{dist}(\bm{G},h_{\bm{v}_1}(\mathcal{C}))+\epsilon,\nonumber\\
	&\|\bm{G}-h_{\bm{v}_2}(\bm{Z})\|_F> \mathrm{dist}(\bm{G},h_{\bm{v}_2}(\mathcal{C}))+\widetilde{\epsilon}.\nonumber
	\end{align}
	By taking the infimum over $\bm{Z},\widetilde{\bm{Z}}\in\mathcal{C}$, we reach a contradiction. Below, we proceed to prove the convexity of $\mathrm{dist}(\bm{G},h_{\bm{v}}(\mathcal{C}))$:
	\begin{align}\label{eq.distconvexity}
	&\mathrm{dist}(\bm{G},h_{\theta\bm{v}_1+(1-\theta)\bm{v}_2}(\mathcal{C}))\stackrel{(\RN{1})}{=}\nonumber\\
	&\inf_{\bm{Z}\in\mathcal{C}}\|\bm{G}-h_{\theta\bm{v}_1+(1-\theta)\bm{v}_2}(\bm{Z})\|_F\stackrel{(\RN{2})}{\le}\nonumber\\
	&\inf_{\substack{\bm{Z}_1\in\mathcal{C}\\\bm{Z}_2\in\mathcal{C}}}\|\bm{G}-\theta h_{\bm{v}_1}(\bm{Z}_1)+(1-\theta)h_{\bm{v}_2}(\bm{Z}_2)\|_F\stackrel{(\RN{3})}{\le}\nonumber\\
	&\theta \|\bm{G}-h_{\bm{v}_1}(\bm{Z}_1)\|_F+(1-\theta)\|\bm{G}-h_{\bm{v}_2}(\bm{Z}_2)\|_F\stackrel{(\RN{4})}{\le}\nonumber\\
	&\theta\mathrm{dist}(\bm{G},h_{\bm{v}_1}(\mathcal{C}))+(1-\theta)\mathrm{dist}(\bm{G},h_{\bm{v}_2}(\mathcal{C}))+\epsilon+\widetilde{\epsilon}.
	\end{align}
	Since this holds for any $\epsilon,\widetilde{\epsilon}$, $\mathrm{dist}(\bm{G},h_{\bm{v}}(\mathcal{C}))$ is a convex function. As the square of a non-negative convex function is convex, $J_{\bm{G}}(\bm{v})$ is a convex function. Finally, the function $J(\bm{v})$ is the average of convex functions, hence is convex.
	In (\ref{eq.distconvexity}), the equality $(\RN{1})$ comes from the definition of \lq\lq dist\rq\rq. $(\RN{2})$ uses the argument
	\begin{align}\label{eq.dist_2}
	&\forall \bm{Z}_1,\bm{Z}_2\in\mathcal{C}~~\exists \bm{Z}\in\mathcal{C}~~\text{such that}\nonumber\\
	&\theta h_{\bm{v}_1}(\bm{Z}_1)+(1-\theta)h_{\bm{v}_2}(\bm{Z}_2)=(\theta h_{\bm{v}_1}+(1-\theta)h_{\bm{v}_2})(\bm{Z}).
	\end{align}
	In fact, the left and right hand sides of \eqref{eq.dist_2} have the same value on $\widehat{T}:=\mathrm{supp}(h_{\bm{w}}(\bm{X}))$. To more clarify this fact, when $\bm{Z}_1,\bm{Z}_2,\bm{Z}\in \widehat{T}$, both the right and left hand sides of (\ref{eq.dist_2}), take the same value
	\begin{align}
	(\theta h_{\bm{v}_1}+(1-\theta)h_{\bm{v}_2})(\mathrm{sgn}(h_{\bm{w}}(\bm{X}))).
	\end{align}
	To verify (\ref{eq.dist_2}), it remains to prove
	\begin{align}
	&\theta h_{\bm{v}_1}(\mathcal{P}_{\widehat{T}^\perp}(\bm{Z}_1))+(1-\theta)h_{\bm{v}_2}(\mathcal{P}_{\widehat{T}^\perp}(\bm{Z}_2))=\nonumber\\
	&(\theta h_{\bm{v}_1}+(1-\theta)h_{\bm{v}_2})(\mathcal{P}_{\widehat{T}^\perp}(\bm{Z})).
	\end{align}
	To prove the above equality, we argue by contradiction. Suppose that the above \lq\lq$=$\rq\rq turns to \lq\lq$\neq$\rq\rq for all $\bm{Z}_1,\bm{Z}_2, \bm{Z}$. By setting $\bm{Z}_1=\bm{Z}_2=\bm{Z}=\bm{I}_n$, we reach a contradiction.\par
	\textit{Strict convexity}. We prove strict convexity by contradiction. If $J(\bm{v})$ were not strictly convex, there would be vectors $\bm{v}_1,\bm{v}_2\in\mathbb{R}_{+}^3$ such that
	\begin{align}\label{eq.strict}
	\mathds{E}\Big[J_{\bm{G}}(\theta \bm{v}_1+(1-\theta)\bm{v}_2)\Big]=\mathds{E}\Big[\theta J_{\bm{G}}(\bm{v}_1)+(1-\theta) J_{\bm{G}}(\bm{v}_2)\Big].
	\end{align} 
	For each $\bm{G}$ in (\ref{eq.strict}), the left-hand side is smaller than or equal to the right-hand side. Therefore, in (\ref{eq.strict}), $J_{\bm{G}}(\theta \bm{v}_1+(1-\theta)\bm{v}_2)$ and $\theta J_{\bm{G}}(\bm{v}_1)+(1-\theta) J_{\bm{G}}(\bm{v}_2)$ are almost surely equal (except at a measure zero set) with respect to Gaussian measure. Moreover, it holds that
	\begin{align}
	&J_{\bm{0}}(\theta \bm{v}_1+(1-\theta)\bm{v}_2)=\mathrm{dist}^2(\bm{0},h_{\theta \bm{v}_1+(1-\theta)\bm{v}_2}(\mathcal{C}))\stackrel{(\RN{1})}{\le}\nonumber\\
	&\inf_{\bm{Z}_1,\bm{Z}_2\in\mathcal{C}}\|\theta h_{\bm{v}_1}(\bm{Z}_1)+(1-\theta)h_{\bm{v}_2}(\bm{Z}_2)\|_F^2\stackrel{(\RN{2})}{<}\nonumber\\
	&\theta \inf_{\bm{Z}_1\in\mathcal{C}}\|h_{\bm{v}_1}(\bm{Z}_1)\|_F^2+(1-\theta)\inf_{\bm{Z}_2\in\mathcal{C}}\|h_{\bm{v}_2}(\bm{Z}_2)\|_F^2\nonumber\\
	&\stackrel{(\RN{3})}{=}\theta J_{\bm{0}}(\bm{v}_1)+(1-\theta)J_{\bm{0}}(\bm{v}_2),
	\end{align}
	where the inequality $(\RN{1})$ above, follows from (\ref{eq.dist_2}). $(\RN{2})$ stems from the strict convexity of $\|\cdot\|_F^2$. $(\RN{3})$ is due to the definition of $J_{\bm{0}}$. From (\ref{eq.dist_2}), it can be deduced that the set $h_{\bm{v}}(\mathcal{C})$ is a convex set. Since the distance to a convex set, e.g. $\mathcal{E}\subseteq \mathbb{R}^{n\times n}$ (i.e. $\mathrm{dist}(\bm{G},\mathcal{E})$) is a 1-Lipschitz function, namely
	\begin{align}
	|\mathrm{dist}(\bm{G}_1,\mathcal{E})-\mathrm{dist}(\bm{G}_2,\mathcal{E})|\le \|\bm{G}_1-\bm{G}_2\|_F~~\forall \bm{G}_1,\bm{G}_2\in\mathbb{R}^{n\times n},
	\end{align}
	and continuous with respect to $\bm{G}$, $J_{\bm{G}}(\bm{v})$ is continuous with respect to $\bm{G}$. Thus, there exists an open ball around $\bm{G}=\bm{0}\in\mathbb{R}^{n\times n}$ that we may write the following relation for some $\epsilon>0$
	\begin{align}
	&\exists \bm{U}\in\mathbb{B}_{\epsilon}^{n\times n} :\nonumber\\
	&J_{\bm{U}}(\theta\bm{v}_1+(1-\theta)\bm{v}_2)<\theta J_{\bm{U}}(\bm{v})+(1-\theta)J_{\bm{U}}(\bm{v}_2).
	\end{align}
	Since $\mathbb{B}_{\epsilon}^{n\times n}$ is a measure zero set, the above statement contradicts (\ref{eq.strict}). Hence, we have strict convexity. Continuity besides convexity of $J$ implies that $J(\bm{v})$ is convex on the whole domain $\bm{v}\in\mathbb{R}_{+}^3$.\par
	\textit{Attainment of the minimum}. Suppose that $v_{\min}:=\min\{v_1,v_2,v_3,\frac{v_2v_3}{v_1}\}>\|\bm{G}\|_F$. Then, we may write:
	\begin{align}\label{eq.attain1}
	&\mathrm{dist}(\bm{G},h_{\bm{v}}(\mathcal{C}))=\inf_{\bm{Z}\in\mathcal{C}}\|\bm{G}-h_{\bm{v}}(\bm{Z})\|_F\stackrel{(\RN{1})}{\ge}\nonumber\\
	&\inf_{\bm{Z}\in\mathcal{C}}\big(\|h_{\bm{v}}(\bm{Z})\|_F-\|\bm{G}\|_F\big)\stackrel{(\RN{2})}{=}\nonumber\\
	&\inf_{\bm{Z}\in\mathcal{C}}
	\big(\|\bm{L}\bm{B}_L^{\rm H}\bm{Z}\bm{B}_R\bm{R}^{\rm H}\|_F-\|\bm{G}\|_F\big)\stackrel{(\RN{3})}{\ge}
	v_{\min}-\|\bm{G}\|_F\ge 0,
	\end{align}
	where in \ref{eq.attain1}, the inequality $(\RN{1})$ comes from triangle inequality of Frobenius norm. The equality $(\RN{2})$ is the result of the decomposition provided in Lemma \ref{lem.decompos} and the rotational invariance of Frobenius norm. Lastly, $(\RN{3})$ is obtained by combining the facts
	\begin{align*}
	\|\bm{A}\bm{B}\bm{C}\|_F\ge \frac{\|\bm{B}\|_F}{\|\bm{A}^{-1}\|_{2\rightarrow 2}\|\bm{C}^{-1}\|_{2\rightarrow 2}}, 
	\end{align*}
	for any non-singular and conforming matrices $\bm{A},\bm{B},\bm{C}\in\mathbb{R}^{n\times n}$, 
	\begin{align*}
	\|\bm{Z}\|_F\ge 1~~\forall \bm{Z}\in\mathcal{C},
	\end{align*}
	and (\ref{eq.sigma_L}), (\ref{eq.sigma_R}). 
	By squaring (\ref{eq.attain1}), we reach
	\begin{align}\label{eq.attain3}
	J_{\bm{G}}(\bm{v})\ge \Big(v_{\min}-\|\bm{G}\|_F\Big)^2~~:~\text{when} ~~v_{\min}>\|\bm{G}\|_F.
	\end{align}
	Using the relation $\mathds{E}\|\bm{G}\|_F\ge \frac{n}{\sqrt{n^2+1}}$ (\cite[Proposition 8.
	1]{foucart2013mathematical}) and Marcov's inequality, we obtain
	\begin{align}\label{eq.attain4}
	\mathds{P}\Big(\|\bm{G}\|_F\le n\Big)\ge 1-\frac{n}{\sqrt{n^2+1}}.
	\end{align}
	Then, it holds that
	\begin{align}\label{eq.attain2}
	&J(\bm{v})\stackrel{(\RN{1})}{=}\mathds{E}\Big[J_{\bm{G}}(\bm{v})\Big|\|\bm{G}\|_F\le n\Big]\mathds{P}\Big(\|\bm{G}\|_F\le n\Big)+\nonumber\\
	&\mathds{E}\Big[J_{\bm{G}}(\bm{v})\Big|\|\bm{G}\|_F> n\Big]\mathds{P}\Big(\|\bm{G}\|_F> n\Big)\nonumber\\
	&\stackrel{(\RN{2})}{\ge} \mathds{E}\Big[J_{\bm{G}}(\bm{v})\Big|\|\bm{G}\|_F\le n\Big]\mathds{P}\Big(\|\bm{G}\|_F\le n\Big)\stackrel{(\RN{3})}{\ge}\nonumber\\
	&\Big(1-\frac{n}{\sqrt{n^2+1}}\Big)\mathds{E}\Big[\big(v_{\min}-\|\bm{G}\|_F\big)^2\Big|\|\bm{G}\|_F\le n\Big]\stackrel{(\RN{4})}{\ge}\nonumber\\
	&\Big(1-\frac{n}{\sqrt{n^2+1}}\Big)\Big(v_{\min}-n\Big)^2,
	\end{align}
	where in (\ref{eq.attain2}), $(\RN{1})$ stems from total probability theorem. $(\RN{2})$ is since $J_{\bm{G}}$ is positive. $(\RN{3})$ follows from (\ref{eq.attain3}) and (\ref{eq.attain4}). Lastly, $(\RN{4})$ is because $\big(v_{\min}-n\big)^2$ provides a lower-bound for the expression in the brackets inside the expectation.\par
	From (\ref{eq.attain2}), one can infer that when 
	\begin{align}
	v_{\min}>n\Big(1+\frac{(n^2+1)^{\frac{1}{4}}}{\sqrt{\sqrt{n^2+1}-n}}\Big).
	\end{align}
	We have that $J(\bm{v})>J(\bm{0})=n^2$. Thus any minimizer of $J$ must be in the set $\Big[0,n\Big(1+\frac{(n^2+1)^{\frac{1}{4}}}{\sqrt{\sqrt{n^2+1}-n}}\Big)\Big]^3$.\par
	\textit{Minimum is not the origin}.
Assume that $w_i=\lambda ~\forall i=1,..., 3$. Then, one may write $J(\bm{w})$ in (\ref{eq.Jv}) as
\begin{align}\label{eq.J1}
&J(\bm{w})=\mathds{E}{\rm dist}^2(\bm{G},t\lambda\partial\|\cdot\|_*(\bm{X}))\stackrel{(\RN{1})}{\le}\nonumber\\
&\inf_{t\ge 0}\mathds{E}\inf_{\bm{Z}\in\partial\|\cdot\|_*(h_{\bm{w}}(\bm{X}))}\|\bm{G}-t\lambda\bm{Z}\|_F^2\stackrel{(\RN{2})}{\le}\nonumber\\
&\mathds{E}\inf_{\bm{Z}\in\partial\|\cdot\|_*(h_{\bm{w}}(\bm{X}))}\|\bm{G}-\lambda\bm{Z}\|_F^2\stackrel{(\RN{3})}{=}\nonumber\\
&\mathds{E}\sum_{i=1}^n\inf_{\sigma_i(\bm{Z})\le 1}\bigg(\sigma_i(\bm{G})-\lambda\sigma_i(\bm{Z})\bigg)^2\stackrel{(\RN{4})}{=}\nonumber\\
&\mathds{E}\sum_{i=1}^n\big(\sigma_i(\bm{G})-\lambda\big)_{+}^2\stackrel{(\RN{5})}{=}n^2\int_{0}^{2}(u-\frac{\lambda}{\sqrt{n}})_{+}^2\frac{\sqrt{4-u^2}}{\pi}{\rm d}u\stackrel{(\RN{6})}{=}\nonumber\\
&n^2\varphi(\frac{\lambda}{\sqrt{n}})\stackrel{(\RN{7})}{<}n^2=J(\bm{0}),
\end{align} 
where $\varphi(\alpha):=\frac{-(26\alpha+\alpha^3)\sqrt{4-\alpha^2}+24(1+\alpha^2)\cos^{-1}(\frac{\alpha}{2})}{12\pi}.$\par
In (\ref{eq.J1}), $(\RN{1})$ is because the infimum of an affine function is concave and Jensen's inequality. In $(\RN{2})$, we set $t=1$. $(\RN{3})$ is because of Hoffman--Wielandt Theorem \cite[Corollary 7.3.5]{horn2013matrix}. $\RN{4}$ follows from (\ref{eq.shrinkage}). $(\RN{5})$ is the result of Lemma \ref{lem.S_vs_Sap}. $(\RN{6})$ is since 
\begin{align}
&\int_{0}^{2}(u-\alpha)_{+}^2\frac{\sqrt{4-u^2}}{\pi}{\rm d}u=\nonumber\\
&\begin{cases}
\frac{3\pi-16\alpha+3\pi\alpha^2}{3\pi}& \alpha\le 0\\
\frac{-(26\alpha+\alpha^3)\sqrt{4-\alpha^2}+24(1+\alpha^2)\cos^{-1}(\frac{\alpha}{2})}{12\pi} &0\le\alpha\le 2\\
0&\alpha>2
\end{cases}.
\end{align}
$(\RN{7})$ comes from $\varphi(\alpha)$ is a decreasing function and for sufficiently small  $\lambda>0$ is less than $1$. So, this completes the proof.
\end{proof}
\subsection{Proof of Corollary \ref{cor.sgn}}\label{proof.cor}
\begin{proof}
In relations (\ref{eq.sgn_hX}) and (\ref{eq.Pthatperp}), using MATLAB matrix notation, ${\rm sgn}(h_{\bm{w}}(\bm{X}))$ is obtained by (\ref{eq.h_wX}) and 
\begin{align}
{\rm sgn}(h_{\bm{w}}(\bm{X}))=\bm{B}_L\bm{O}_L(:,1:r)\bm{O}_R(:,1:r)^{\rm H}\bm{B}_R^{\rm H},\nonumber
\end{align}
while (\ref{eq.Pthatperp}) is the result of
\begin{align}
&\mathcal{P}_{\widehat{T}^\perp}(\bm{Z})=\bm{B}_L\bm{O}_L(:,r+1:n)\bm{O}_L(:,r+1:n)^{\rm H}\bm{B}_L^{\rm H}\bm{Z}\bm{B}_R\nonumber\\
&\bm{O}_R(:,r+1:n)\bm{O}_R(:,r+1:n)^{\rm H}\bm{B}_R^{\rm H}=\nonumber\\
&\bm{B}_L\bm{O}_L\mathcal{P}_{T_1^\perp}(\bm{I}_n)\bm{O}_L^{\rm H}\bm{B}_L^{\rm H}\bm{Z}\bm{B}_R\bm{O}_R\mathcal{P}_{T_1^\perp}(\bm{I}_n)\bm{O}_R^{\rm H}\bm{B}_R^{\rm H}=\nonumber\\
&\bm{B}_L\bm{O}_L\mathcal{P}_{T_1^\perp}(\bm{O}_L^{\rm H}\bm{B}_L^{\rm H}\bm{Z}\bm{B}_R\bm{O}_R)\bm{O}_R^{\rm H}\bm{B}_R^{\rm H}.
\end{align}
\end{proof}
\subsection{Proof of Lemma \ref{lem.S_vs_Sap}}\label{proof.lem.S_vs_Sap}
\begin{proof}
Define a random vector ${h}\in\mathbb{R}^{n_1}$ where its elements $h_i$s are randomly chosen without replacement from the set $\{f_1,\dots, f_{n_1}\}$. Since $\sigma_i$s and $f_i$s are all positive and non-increasingly ordered, it is straightforward to check that
\begin{align}\label{eq.rel11}
&\frac{1}{n_1}\sum_{i=1}^{n_1}\mathds{E}_{{\sigma}_i}(\sigma_i(\frac{\bm{G}}{\sqrt{n_2}})-f_i)_{+}^2\le\nonumber\\
& \frac{1}{n_1}\sum_{i=1}^{n_1}\mathds{E}_{{\sigma}_i,{h}_i}(\sigma_i(\frac{\bm{G}}{\sqrt{n_2}})-h_i)_{+}^2.
\end{align}
 Due to the Fact \ref{fact.me} and the conditional expectation, we may write:
\begin{align}\label{eq.rel2}
&\mathds{E}_{{\sigma}_i,{h}_i}({\sigma}_i(\frac{\bm{G}}{\sqrt{n_2}})-{h}_i)_{+}^2=\mathds{E}_{\sigma_i}\mathds{E}_{h_i}({\sigma}_i(\frac{\bm{G}}{\sqrt{n_2}})-{h}_i)_{+}^2=\nonumber\\
&\frac{1}{n_1}\sum_{j=1}^{n_1}\mathds{E}_{\sigma_i}({\sigma}_i(\frac{\bm{G}}{\sqrt{n_2}})-f_j)_{+}^2.
\end{align}
Use \eqref{eq.rel11} and \eqref{eq.rel2} to reach:
\begin{align}\label{eq.rel3}
&\frac{1}{n_1}\sum_{i=1}^{n_1}\mathds{E}_{{\sigma}_i,{h}_i}(\sigma_i(\frac{\bm{G}}{\sqrt{n_2}})-h_i)_{+}^2=\nonumber\\
&\frac{1}{n_1}\sum_{j=1}^{n_1}\frac{1}{n_1}\sum_{i=1}^{n_1}\mathds{E}_{\sigma_i}(\sigma_i(\frac{\bm{G}}{\sqrt{n_2}})-f_j)_{+}^2=\nonumber\\
&\frac{1}{n_1}\sum_{j=1}^{n_1}\mathds{E}_{\sigma}(\sigma(\frac{\bm{G}}{\sqrt{n_2}})-f_j)_{+}^2,
\end{align}
where we used the fact that $\mathds{P}\{h_i=f_j\}=\frac{1}{n_1}$ in the first equality. Now, we use the relation
\begin{align}\label{eq.marchen_rel}
&\mathds{E}_{{\sigma}}({\sigma}(\frac{\bm{G}}{\sqrt{n_2}})-f_j)_{+}^2=\nonumber\\
&\int_{l_b(s)}^{u_b(s)}(u-f_j)_{+}^2\frac{\sqrt{(u_b(s)^2-u^2)(u^2-l_b(s)^2)}}{\pi u s}{\rm{d}}u,
\end{align}
which comes from the fact that the distribution of singular values of a Gaussian matrix tends to the Mar\v{c}enko--Pastur law \cite[Theorem 3.6]{bai2010spectral} with probability one. By substituting \eqref{eq.marchen_rel} into \eqref{eq.rel3}, we shall have that
\begin{align}
&\frac{1}{n_1}\sum_{i=1}^{n_1}\mathds{E}_{{\sigma}_i,{h}_i}(\sigma_i(\frac{\bm{G}}{\sqrt{n_2}})-h_i)_{+}^2=\nonumber\\
&\frac{1}{n_1}\sum_{j=1}^{n_1}\int_{l_b(s)}^{u_b(s)}(u-f_j)_{+}^2\frac{\sqrt{(u_b(s)^2-u^2)(u^2-l_b(s)^2)}}{\pi u s}{\rm{d}}u,
\end{align}
which concludes the result.

\end{proof}
\ifCLASSOPTIONcaptionsoff
  \newpage
\fi

\end{document}